\documentclass{aa}  
\usepackage{color} 
\usepackage[squaren, Gray, cdot]{SIunits}
\usepackage{graphicx}
\usepackage{txfonts}
\usepackage{natbib}
\bibpunct{(}{)}{;}{a}{}{,} 
\usepackage{amsmath}
\usepackage{tikz}
\usepackage{hyperref}
\usepackage{ulem} 
\usepackage{comment}
\usepackage{booktabs}
\usepackage{placeins}

\definecolor{col3}{RGB}{245,170,7}
\definecolor{col2}{RGB}{217,82,57}
\definecolor{col1}{RGB}{111,18,111}

\newcommand*\diff{\mathop{}\!\mathrm{d}}

\def\ltsima{$\; \buildrel < \over \sim \;$}
\def\simlt{\lower.5ex\hbox{\ltsima}}
\def\gtsima{$\; \buildrel > \over \sim \;$}
\def\simgt{\lower.5ex\hbox{\gtsima}}
\def\deg{^{\circ}}

\begin{document} 

   \title{Is the mm/submm dust polarization a robust tracer of the magnetic field topology in protostellar envelopes ? A model exploration}
  \titlerunning{Protostellar $B$-fields traced with dust polarization}

   \author{Valeska Valdivia \inst{\ref{inst1},\ref{inst2}}
          \and
          Ana\"{e}lle Maury \inst{\ref{inst1}}
          \and
          Patrick Hennebelle \inst{\ref{inst1}}
        }

        \institute{Laboratoire AIM, Paris-Saclay, CEA/IRFU/SAp - CNRS - Universit\'e Paris Diderot, 91191 Gif-sur-Yvette Cedex, France
        \label{inst1}
        \and %
        Department of Physics, Nagoya University, Furo-cho, Chikusa-ku, Nagoya, Aichi 464-8602, Japan \\ \email{valdivia.valeska@g.mbox.nagoya-u.ac.jp}\label{inst2}
             }

   \date{Received Month dd, yyyy; accepted Month dd, yyyy}


  \abstract
   {High-resolution millimeter and submillimeter (mm and submm) polarization observations have opened a new era in the understanding of how magnetic fields are organized in star forming regions, unveiling an intricate interplay between the magnetic fields and the gas in protostellar cores. However, to assess the role of the magnetic field in the process of solar-type star formation,  it is important to understand to what extent these polarized dust emission are good tracers of the magnetic field in the youngest protostellar objects.}
   {In this paper, we present a thorough investigation of the fidelity and limitations of using dust polarized emission to map the magnetic field topologies in low-mass protostars.}
   {To assess the importance of these effects, we performed an analysis of magnetic field properties in 27 realizations of magnetohydrodynamics (MHD) models following the evolution of physical properties in star-forming cores. Assuming a uniform population of dust grains the sizes of which follow the standard MRN size distribution, we analyzed the synthetic polarized dust emission maps produced when these grains align with the local $B$-field because of radiative torques (B-RATs).}
   {We find that mm and submm polarized dust emission is a robust tracer of the magnetic field topologies in inner protostellar envelopes and is successful at capturing the details of the magnetic field spatial distribution down to radii $\sim 100~\mathrm{au}$. Measurements of the line-of-sight-averaged magnetic field line orientation using the polarized dust emission are precise to $<15\deg$ (typical of the error on polarization angles obtained with observations from large mm polarimetric facilities such as ALMA) in about $75\%-95 \%$ of the independent lines of sight that pass through protostellar envelopes. Large discrepancies between the integrated $B$-field mean orientation and the orientation reconstructed from the polarized dust emission are mostly observed in (i) lines of sight where the magnetic field is highly disorganized and (ii) those that probe large column densities. Our analysis shows that the high opacity of the thermal dust emission and low polarization fractions could be used to avoid using the small fraction of measurements affected by large errors.}
   {}


   \maketitle
%

\section{Introduction}

{The fate of the evolution of protostellar cores is a result of the interplay between their gravity, turbulence, magnetic fields, and chemical composition through coupled processes that control the dynamics of the gas as well as its thermo-dynamical state. In order to better understand the star formation process and to untangle the different contributions, we rely on observations on one hand and theoretical models on the other. The quality of the modeling and the quality of the interpretation of actual observations depend on how well we can connect these two approaches, making the study of how observed data actually traces the physico-dynamical state of the gas a key topic in modern astrophysics.} 

\noindent Particularly, magnetic fields ($B$-fields) are a fundamental component of the star formation process at all scales and evolutionary stages. They are key agents that regulate the star formation at scales of molecular clouds, by providing a support against gravitational collapse \citep{Mouschovias1999, Li2014}, and are thought to channel the mass accretion to form interstellar filaments (the preferred sites of star formation). At core scales ($\simlt 3000$ au), if $B$-fields are efficiently coupled with the protostellar envelope made of dust and gas, they may also play a key role in regulating the final outcome of the star formation process, by transporting a significant fraction of the core angular momentum outward, hence allowing the formation of stars, setting the pristine disk size and shaping the properties of the upcoming star and planets.

While observing the $B$-fields in star-forming environments cannot be done directly, various techniques have been developed to probe the magnetic properties in astrophysical structures. Among them, the most widely used method for the dense molecular conditions consists in interpreting the polarized emission of ISM dust grains. Indeed, several theoretical works suggest that nonspherical dust grains spinning around their axis of maximal inertia (the shortest axis of the dust grain) tend to align with the magnetic fields through dissipative processes producing a polarized dust emission that is perpendicular to the magnetic field lines \citep{Davis1951, Purcell1979, Roberge2004}. The mechanisms able to align the dust grains in such a manner include the radiative alignment torques (B-RATs, \citealt{LazarianHoang2007MNRAS, Andersson2015}), a new variant of RATs including super-paramagnetic dust grains (MRAT, \citealt{Hoang2016}), the supersonic mechanical grain alignment (or Gold effect, \citealt{Gold52}), the mechanical alignment torques (MATs, \citealt{DavisGreenstein1951,LazarianHoang2007ApJ, Reissl2022}), and other mechanisms (not discussed in this paper) operating at specific conditions that can produce a polarized signal but do not necessarily trace the magnetic field. For instance, the dust self-scattering \citep{Kataoka2015} caused by the scattered light of anisotropic continuum emission by large dust grains (with sizes comparable to the observed wavelengths) can explain the polarization levels observed in the innermost regions of protostellar disks independently of the magnetic field, as well as another variant of the RAT theory (k-RATs, \citealt{LazarianHoang2007ApJ, Tazaki2017}), where dust grains in regions near a strong radiation source precess around the radiation anisotropy vector instead of precessing around the magnetic field lines. 
Current observations with single-dish instruments (such as Planck, BLASTPOL, and the JCMT POL2), as well as interferometric facilities (such as ALMA, CARMA and SMA), are providing a wealth of polarimetric data that, if properly interpreted, can deepen our insight into the different processes occurring in different environments. For instance, recent interferometric observations with the SMA allowed \citet{Galametz2020} to highlight the potentially crucial role of the relative orientation between the magnetic field orientation inferred from the polarized dust emission and the rotation axis inferred from the direction of the outflows and jets in the fragmentation and multiplicity properties of protostars.   
Recently, a few works used numerical models and radiative transfer methods to characterize the dust polarization expected from physical models of clouds, cores, and disks \citep{Kuffmeier20, Brunngraber21, Lam21} and qualitatively compare them to observations. Most conclude that, considering the known mechanisms that generate dust polarization at millimeter wavelengths, such polarization 
likely originates from magnetically aligned grains in inner envelopes and dust scattering in disks. 
However, while the ability to reconstruct large-scale magnetic field structures from the polarized signal arising from magnetically aligned dust grains is well established for diffuse interstellar gas, hardly any work \citep[see][ focusing on bridge structures]{Kuffmeier20} has yet been dedicated to assessing the robustness of polarization 
as a tracer of the magnetic field topology in the case of objects displaying a higher degree of complexity such as young protostars.
 
Specifically, polarimetric observations aiming at probing the magnetic fields of protostellar cores face several challenges. As the collapse proceeds, the magnetic field lines are dragged, bent, and distorted, making the definition of a true mean orientation difficult to obtain. The increasing gas density subsequently rises the opacity to levels where even at millimeter (mm) wavelengths it is hard to trace the full information along the line of sight. On the other hand, the properties of polarized dust emission heavily rely on the ability of dust grains to align with the magnetic fields: local irradiation and its properties may affect how accurately the polarized dust emission represents the mean integrated orientation of the magnetic field lines.
Understanding how well and under which conditions the magnetic field morphologies are constrained by polarimetric observations can help us to understand the relative importance of the physical processes shaping the star formation process.


\begin{table*}
\caption{Parameters of the protostellar models}              
\label{table:1}      
\centering                                      
\begin{tabular}{l c c c c | r r r r }          
\hline\hline                        
 \noalign{\smallskip}
ID & $\mu$ & $\beta_\mathrm{rot}$  & $\theta $ & $\mathcal{M}$  & $t_\mathrm{form}$ [yr]  &  $t_\mathrm{0.1M_\odot}$ [yr] &  $t_\mathrm{0.2M_\odot}$ [yr] & $t_\mathrm{0.3M_\odot}$ [yr]\\    
 \noalign{\smallskip}
\hline                                  
 \noalign{\smallskip}
    R1 &        3.33    &       0.01            &       $30^\circ$      &       0  &      55428   &       2493    &       5991            &       13484\\      
    R2 &        3.33    &       0.04            &       $30^\circ$      &       0  &      60328   &       2472    &       8760            &       19509\\
    R3 &        10.0    &       0.04            &       $30^\circ$      &       0  &      50987   &       5179    &       14816   &       30282\\
    R4 &        6.67    &       0.04            &       $30^\circ$      &       0  &      51206   &       3693    &       8927            &       17943\\
    R6 &        3.33    &       0.01            &       $30^\circ$      &       1  &      66942   &       3618    &       12321   &       30486\\ 
    R7 &        3.33    &       0.04            &       $90^\circ$      &       0  &      56006   &       2461    &       6001            &       13532\\
    R8 &        6.67    &       0.01            &       $30^\circ$      &       1  &      57132   &       2460    &       7331            &       16413\\
    R9 &        10.0    &       0.04            &       $90^\circ$      &       0  &      49636   &       5112    &       15843   &       42005\\
    R10&        3.33    &       0.0025  &       $30^\circ$      &       0  &      52577   &       2828    &       7015            &       12982\\
\hline                                             
\end{tabular}
\tablefoot{
Selected numerical models from \citet{Hennebelle2020}. $\mu$ is the mass-to-flux ratio normalized by the critical value, $\beta_\mathrm{rot}$ is the ratio between rotational energy and gravitational energy, $\theta$ is the inclination angle of the initial rotation axis with respect to the magnetic field axis, and $\mathcal{M}$ is the initial Mach number.
$t_\mathrm{form}$ is the time at which the sink forms in the simulation, while $t_\mathrm{0.1M_\odot}$ , $t_\mathrm{0.2M_\odot}$ and $t_\mathrm{0.3M_\odot}$ are the ages of the sink at which it reaches $0.1$, $0.2$, and $0.3~\mathrm{M_\odot}$, respectively.
}
\end{table*}

\section{Methods} \label{SectMethods}

\subsection{Non-ideal MHD models of protostellar cores}\label{NumSim}

Protostellar cores are expected to display a wide variety of physical conditions, especially in terms of masses, level of turbulence, magnetization degree, and level of rotation and relative orientation between the magnetic field and the rotation axis. While an exploration of the complete parameter space is out of the scope of this study, we attempt to use realistic models of Class~0 protostars to investigate whether polarized dust emission is a reliable tracer of the magnetic field topology inside protostellar envelopes. We therefore use a subset of the physically motivated numerical simulations by \citet{Hennebelle2020}, which were initially run to understand how the interplay between the turbulence, magnetization degree, initial rotation, and the initial angle between rotation axis and the  magnetic field orientation define the outcome of the evolution of protostellar cores. Even though most stars are formed in multiple stellar systems \citep[see][and references therein]{Offner2022} whose dynamics is still influenced by their environment, these isolated cores are believed to {grasp the complexity} of the physical process at work at envelope scales, and hence can be trusted as realistic 3D models of the density, temperature, gas kinematics, and magnetic field structures in prototypical solar-type Class~0 protostars. 

The numerical simulations were performed using the \textsc{Ramses} code \citep{Teyssier2002, Fromang2006}, which is able to solve the MHD equations, including the ambipolar diffusion, to treat the gas dynamics. The thermal evolution of the gas is computed using an equation of state (EoS) that mimics the opacity effects on the radiative transfer in dense gas \cite[see Eq.~1 of][]{Hennebelle2020}. A sink particle approach is used to follow the formation and evolution of the protostar \citep{Bate1995, Bleuler2014}.  
The code benefits from the adaptive mesh refinement (AMR) technique to adapt the resolution. In these simulations, the refinement criterion is the local Jeans mass, 
for which we impose that the Jeans length is resolved by at least 20 cells. The structure of the data is a fully threaded ocTree. \\
The setup of the simulations is the same as that of \citet{Hennebelle2016} and \citet{Masson2016}, with the only difference being that it includes sink particles  and covers a wider range of initial conditions. \\

All the model cores have an {initial} mass of $1~\mathrm{M_\odot}$, and the magnetic field is initially uniform and aligned with the $z$ direction. The ratio of thermal to gravitational energy is $\alpha = 0.4$ for all the simulations. 
The selected models cover three levels of rotation (set by the ratio of rotational to gravitational energy $\beta_\mathrm{rot} = E_\mathrm{rot}/E_\mathrm{grav}= 0.0025, 0.01, 0.04$), three levels of magnetization (given by the mass-to-flux ratio normalized to the critical mass-to-flux ratio $\mu = (M/\Phi)/(M/\Phi)_\mathrm{crit}  = 3.33, 6.67, 10.0$). Most of the models have a Mach number $\mathcal{M}=0$ and an inclination between the initial rotation axis and the initial magnetic field orientation $\theta = 30^\circ$, but special cases including turbulence ($\mathcal{M}=1$) and cases using an extreme tilt of $\theta = 90^\circ$ are also included. The run identifiers (ID), as well as their parameters, are summarized in the first part of Table~\ref{table:1}.\\

\begin{figure}
\centering
\includegraphics[width=0.48\textwidth, trim={1.9cm 1.3cm 1.0cm 1.8cm},clip]{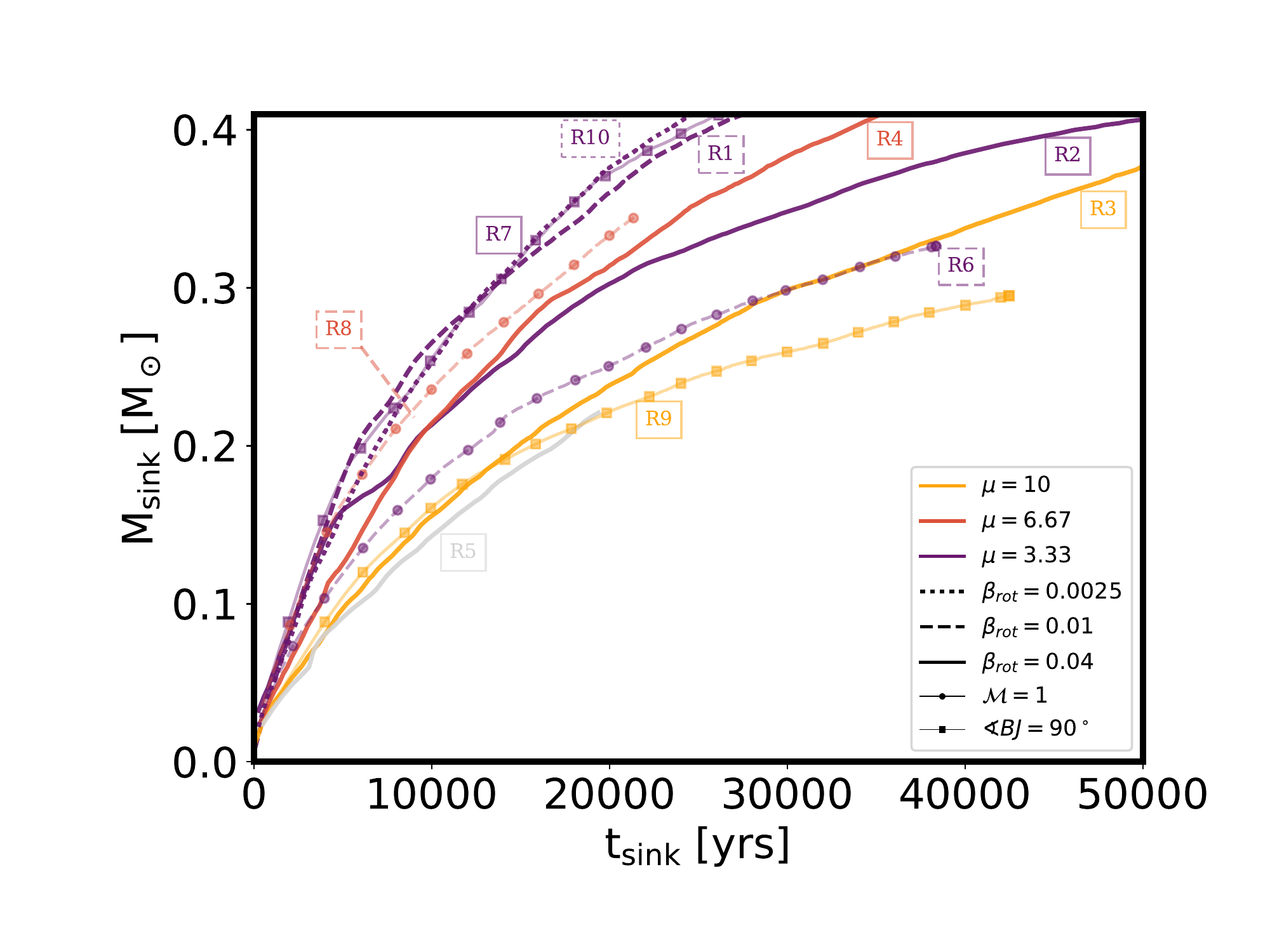}
\caption{Sink-mass evolution as a function of the time since the formation of the sink $t_\mathrm{age}$.The color identifies the normalized mass-to-flux ratio, the line-style identifies the rotational to gravitational energy ratio, and the symbols indicate the presence of turbulence (circles) or an extreme inclination of the rotation with respect to the magnetic field (squares). We keep the same identifiers in the following figures.}  
\label{msinkevo}
\end{figure}

\subsection{Protostar mass evolution and output selection}
\label{PSmassev}

\noindent As the gravitational collapse proceeds, the density increases towards the center of the simulation box.  When the local density reaches a threshold of $n_\mathrm{thres}=3\times10^{13}~\mathrm{cm^{-3}}$, a sink particle is created at the highest resolution level, at the peak of the local clump formed by the neighboring cells whose density is higher than $n_\mathrm{thres}/10$. The time at which the sink forms in the simulation, shown in the sixth column of Table~\ref{table:1}, indicates how fast these conditions are reached. The sink particle mimics the formation and evolution of the protostellar embryo, allowing us to follow its mass evolution and dynamics. 
The sink formation happens faster for the less magnetized models (R9 and R3), while the slowest formation corresponds to the R6, which in addition to being more magnetized ($\mu=3.33$) and having a moderate rotation ($\beta_\mathrm{rot}=0.01$), is turbulent. The importance of the additional support against collapse provided by the rotation itself is reflected on the delayed formation time of the sink in models R1, R2, and R10 that, in spite of having the same initial parameters except for the rotation, increases with $\beta_\mathrm{rot}$. 

The sink particle can further increase its mass by accreting gas from neighboring cells within a radius of four times the size of the most refined grid nearby the sink position. At each time-step and  from each of its   neighboring cells,  
 the sink accretes $10\%$ of the gas whose density is above $n_\mathrm{thres}/3$. 
The sink mass evolution shown in Fig.~\ref{msinkevo} seems to indicate that the most important parameter in terms of the accretion is the mass-to-flux ratio. But contrary to the trend seen for the sink formation time, the initial accretion proceeds faster in the more magnetized cases as a consequence of the formation of more massive disks when the magnetic field is weaker and the rotation stronger. 
All in all, an increasing $B$-field delays the formation of the protostellar embryo at the center of the collapsing core, but allows the early mass accretion on the protostar to proceed faster. Although other processes may also be playing a role, we note that the two least magnetized simulations (R9 and R3) have the earliest sink formation, and show the lowest mass-accretion rates. 
The strength and geometry of the magnetic field can also play a more episodic role during the protostellar formation process: for example simulation R2 exhibits a reduced mass-accretion rate when its sink particle has reached roughly $0.15~\mathrm{M_\odot}$. This can be explained by the development of a magnetic cavity (the loop-like structure seen in the $x-y$ map of Fig.~\ref{Ncolmaps}), which is a consequence of the interchange instability \citep{Krasnopolsky2012} often seen in simulated disks \citep{Matsumoto2017, Machida2014, Joos2012, Seifried2011}.

%
In order to analyze and compare simulations with such different evolution, we use the fraction of the initial-mass core accreted onto the stellar embryo (traced by the sink mass) as an indicator of the evolutionary stage of the protostar--envelope system. For each simulation, we selected three snapshots that correspond to the moment when $10\%$, $20\%,$ and $30\%$ of the initial core mass has been transferred to the embryo, or equivalently, the sink mass, $m_\mathrm{sink}$, has reached $0.1$, $0.2$, and $0.3~\mathrm{M_\odot}$, respectively, thus sampling the Class 0 stage ($M_\mathrm{env}>M_\mathrm{star}$).  
The time at which the sink forms in each simulation ($t_\mathrm{form}$), as well as the age at which the sink reaches the selected masses relative to its formation time are shown in the right-hand side of Table~\ref{table:1}. 
The general evolution of the sink mass for all the simulations is shown in Fig.~\ref{msinkevo}. 
We decided not to use the very weakly magnetized run R5 since it fragments early in the evolution 
 and none of the fragments reach $0.3~\mathrm{M_\odot}$.   
 

\subsection{Magnetic field mean orientation using bidirectional statistics}\label{bidirstat}

As dust grains with opposite spinning directions produce an identical polarized pattern, the polarized dust emission is insensitive to the direction of the magnetic field, and, as a consequence, the magnetic field orientation can be considered as a bidirectional quantity. When the data are diametrically bimodal, the mean angle obtained by computing the mean projected sinus and cosinus is orthogonal to the actual mean orientation due to canceling effects. For example, toroidal magnetic field lines will produce a poloidal field geometry when they are computed edge-on because lines in the foreground and background have opposite directions that will cancel out.

Even though the usual integrated magnetic field in simulations of interstellar gas seems to satisfactorily trace the mean orientation, it is not well adapted to describing the mean orientation in regions where the magnetic field lines are particularly tangled, as in the case of protostellar cores because of the aforementioned canceling effects. As this effect extends into the envelope of protostellar cores, it is necessary to find a suitable description of the magnetic field mean orientation that is compatible with the polarized dust emission.  

Here, and in the following, we compute the orientation angle, and any other angle, as the position angle with respect to the northern (vertical) direction, in the sense that increases anti-clockwise, as recommended by the IAU\footnote{\url{https://www.iau.org/static/archives/announcements/pdf/ann16004a.pdf}}, in the range $-90\deg$ to $+90\deg$. To compute the mean orientation of the magnetic field, we borrow a statistical technique widely used in geography to treat bimodal orientation data \citep{Jones1969, mardia1975}. 
In order to calculate the mean orientation of the magnetic field lines along the line of sight regardless of the direction of the 
$B$-vectors, we proceed to double the individual angles and to compute the density-weighted mean sinus and cosinus of $2\phi_B$, where $\phi_B$ is the local magnetic field angle projected onto the plane of the sky,  as follows:
\begin{eqnarray}
\langle \sin  (2\phi_B)\rangle &=& \frac{\int_{los}  \rho \sin (2\phi_B) \, \diff l}{\int_{los}  \rho \, \diff l}  ,\\
\langle \cos (2\phi_B)\rangle &=& \frac{\int_{los}  \rho \cos(2\phi_B) \, \diff l}{\int_{los}  \rho \, \diff l,} 
,\end{eqnarray}

\noindent where $\rho$ is the local gas number density, and the averages are computed along the line of sight \textit{los}. The mean orientation of the magnetic field $\langle\phi_B\rangle$ is then calculated as:
\begin{equation}
 \langle \phi_B \rangle  = 0.5 \arctan ( \langle \sin(2\phi_B)\rangle, \langle \cos(2\phi_B)\rangle).
 \label{Eq_mean_phiB}
\end{equation}

The angle-doubling procedure is consistent with analytical formulae to compute the Stokes parameters $Q$ and $U$, such as described in \citet{Landi2004} or \citet{MartinezGonzalez2012}. To estimate the intrinsic level of organization of the magnetic field lines along the line of sight we define the circular variance $V_\phi$ as follows:
\begin{equation}\label{eqnVphi}
V_\phi = 1- \sqrt{ \langle \sin (2\phi_B) \rangle^2 +  \langle \cos (2\phi_B) \rangle^2 }.
\end{equation}
Clustered values of $\phi_B$ will produce circular variance values close to 0, while an increasing dispersion of angles (or in other words an increasing level of disorganization) will produce values closer to $1$. Figure \ref{Vphi2sigma} shows the relation between the circular variance $V_\phi$ and the typical standard deviation $\sigma_\phi$ in degrees. 

\begin{table}
\caption{Summary of the parameters used in \textsc{Polaris} for all the models described in Table~\ref{table:1}. 
}              
\label{table:2}      
\centering                                      
\begin{tabular}{l c l}          
\hline\hline                        
Quantity & Symbol & Value \\
\hline\hline      
 \noalign{\smallskip}
        \multicolumn{3}{c}{\textit{Radiation sources}}                  \\ 
\hline                                  
 \noalign{\smallskip}
Central source luminosity       &       $L_\mathrm{star}$               &       $1~\mathrm{L_\odot}$\\      
Central source radius   &       $R_\mathrm{star}$               &
$1~\mathrm{R_\odot}$\\
Photon packages per $\lambda$   &       $N_\mathrm{ph, star}$           &       $500000$\\
{External radiation field} & $G_0$ & $1$ (\citealt{Mathis1983})\\
Photon packages per $\lambda$   &       $N_\mathrm{ph, ISRF}$           &       $10000$\\
\hline  
 \noalign{\smallskip}
        \multicolumn{3}{c}{\textit{Synthetic observation parameters}}                   \\ 
\hline                                  
 \noalign{\smallskip}
Distance to observer    &       $D$     &       $250~\mathrm{pc}$\\
Wavelengths     &       $\lambda$       &       $0.8, 1.3, 3.0 ~\mathrm{mm}$\\
Map size        &       $L_\mathrm{map}$        &       $8000~\mathrm{au}$\\
Resolution & $\diff x$ & $5~\mathrm{au}$ \\
\hline                                             
        \multicolumn{3}{c}{\textit{Dust parameters}}    \\ 
\hline                                  
 \noalign{\smallskip}
Gas to dust ratio & $\rho_\mathrm{g}/\rho_\mathrm{d}$ & 100\\
Minimum grain size      &       $a_\mathrm{min}$                &       $5~\mathrm{nm}$\\      
Maximum grain size      &       $a_\mathrm{max}$                &       $20~\mathrm{\mu m}$\\
Aspect ratio               &    $\epsilon$                           &      $0.5$ \qquad(oblate)\\
Power-law index $^{(a)}$   & $\alpha$ & -3.5 \\
Astronomical silicates  &       $f_\mathrm{sil}$        &       $62.5~\%$\\
Graphite                                &       $f_\mathrm{graph}$      &       $37.5~\%$\\
Suprathermal dust $^{(b)}$ & $f_{\mathrm{high}-J}$ & $0.25$ \\
\hline                                             
\end{tabular}
\tablefoot{$^{(a)}$ Power-law index of the size distribution $\diff n(a)\propto a^{\alpha}$ \citep{Mathis1977}, $^{(b)}$ Fraction of dust grains spinning suprathermally. }
\end{table}

\subsection{Polarized dust emission with the dust radiative transfer code POLARIS }\label{methodpolar}

\subsubsection{\textsc{Ramses} datacubes}\label{secdatacubes}

We use the subset of the simulations presented in \citet{Hennebelle2020} summarized in Table~\ref{table:1}. 
Each \textsc{Ramses} output provides a datacube containing the gas density, the gas pressure (or equivalently the gas temperature), the three components of the velocity field, as well as the three components of the magnetic field.   
These gas properties are extracted and put in a suitable format to be post-processed with \textsc{Polaris}.
To facilitate the propagation of photons and avoid missed photon packages from the central source due to long computational times, we artificially emptied a small central sphere of radius $4~\mathrm{au}$ around the sink particle {as in \cite{Valdivia2019}}. This procedure (only performed for the radiative transfer in post processing) is a pragmatic choice to allow a faster radiative transfer despite the very optically thick layers surrounding the sink, minimizing computational artifacts and thus improving the quality of the radiative transfer at the scales studied in this paper ($50-4000~\mathrm{au}$). As most of the reprocessing of the protostellar photons into infrared (IR) photons happens at radii $4-10~\mathrm{au}$, this procedure does not noticeably affect the temperature and irradiation spectrum at the scales we are investigating in this study. 

To describe the evolution of the simulated collapse, we provide integrated maps of the total column density ($N$), total mean magnetic pressure ($P_\mathrm{mag}$), and mean magnetic field orientation ($\phi_B$), as well as the associated circular variance ($V_\phi$) at key time-steps in Sect.~\ref{resnumsim}. 
We point out that the advantage of providing the mean magnetic pressure ($P_\mathrm{mag}$) rather than the mean magnetic field is that it reflects the strength of the magnetic field without canceling out the contributions of vectors with opposite directions.
The mean magnetic pressure is computed as the density-weighted average of the local magnetic pressure, $P_\mathrm{mag}$, which is given by: 
\begin{equation}
P_\mathrm{mag} = \frac{B^2}{8\pi} =\left(\frac{B}{5~\mathrm{\mu G}}\right)^2\times 10^{-12}~\mathrm{erg~cm^{-3}},
\end{equation}
\noindent where $B = \sqrt{B_x^2 + B_y^2 + B_z^2}$ is the local magnetic field strength in $\mathrm{\micro G}$.

\subsubsection{Radiation sources and dust properties}

As the protostellar luminosity as well as the efficiency of the conversion from gravitational energy into accretion luminosity are not well constrained and can vary significantly during the Class 0 stage \citep{Offner2011, Cheng2022, Kuffmeier2018, Jensen2018}, we adopt a simple description.  
For all the simulations, and regardless of the protostar mass, we model the central radiation source as a blackbody of $1~\mathrm{L}_\odot$ of radius of $1~\mathrm{R}_\odot$ located at the sink position. For a more detailed study investigating the dependency of dust polarized emission with varying accretion luminosity, and associated radiation field penetrating the envelopes of Class 0 protostars, see for example \citet{LeGouellec2020} and Le Gouellec et al. in prep.
We include an external isotropic radiation field of strength $G_0 = 1$ following the \citet{Mathis1983} description. \\
The dust properties are assumed uniform throughout the simulation box and are the same as those in \citet{Valdivia2019}, namely a gas-to-dust ratio of $100$, with a composition of $62.5$ per cent astronomical silicates and $37.5$ per cent graphite grains, which matches the Galactic extinction curve \citep{Mathis1977}. The dust grains are assumed to be oblate with an aspect ratio of $0.5$ \citep{Hildebrand95} and following a power-law size distribution \citep{Li2001} of 
\begin{equation}
\diff n(a)\propto a^{-3.5}\diff a ,  
\end{equation}
\noindent where $a$ is the effective dust radius of a spherical grain of equivalent volume, and $n(a)$ is the number of dust grains of effective radius $a$. The lower and upper cutoff of the size distribution are set as $a_\mathrm{min}=5 ~\mathrm{nm}$ and $a_\mathrm{max}=20 ~\mathrm{\micro m}$, respectively.
 {The choice of $a_\mathrm{max}$ stems from a compromise between our previous study \citep{Valdivia2019}, suggesting that the polarization fractions observed towards protostellar cores can only be approached when including a population of large grains, and the need to avoid the regime of very large grains where the DDSCAT \citep{Draine2000DDSCAT} introduces artifacts on the absorption coefficients \cite[see][]{Reissl2016}.All the parameters are summarized in Table~\ref{table:2} and are the same used for all the selected snapshots.}
 


\subsubsection{Temperatures and dust alignment with \textsc{Polaris} }\label{dustalig}

We post-processed each one of the snapshots described in Sect.~\ref{NumSim} with the \textsc{Polaris} code \citep{Reissl2016}. The photon propagation is based on a Monte Carlo scheme and, along with the local values of the density and the magnetic field, it is used to compute the dust temperature and the grain alignment efficiency on a cell-by-cell basis. As the gas temperature computed by \textsc{Ramses}  using an EoS is the result of an approximation and not the result of an actual radiative transfer, we recompute the gas temperature as well.\\
In addition to the gas and dust heating, the radiation field induces a radiation torque on dust grains, spinning them up to suprathermal rotation velocities, at which they can align with the magnetic field \citep{Barnett1915, Dolginov76, LazarianHoang2007MNRAS, Hoang2014}. 
The grain alignment efficiency is computed according to the B-RATs theory as implemented in \textsc{Polaris} \citep{Reissl2016}. In principle, the fraction of dust grains spinning at suprathermal velocities ($f_{\mathrm{high}-J}$, which is the stable population of aligned grains)  depends on the dust grain properties such as the specific shape, composition, and magnetic properties. As this quantity cannot be calculated exactly, it is set to $f_{\mathrm{high}-J}=0.25$. This rather conservative choice is able to produce a detectable amount of polarization. Recently, \citet{Herranen2021} found that the alignment efficiency might be higher than predicted, which in the case of aggregates can reach $f_{\mathrm{high}-J}\sim 0.35-1$. Also, the presence of iron inclusions can increase this quantity to nearly unity \citep{Hoang2016}.   
 Additionally, we take into account the fact that, due to internal thermal fluctuations, nonspherical dust grains are not expected to be perfectly aligned, but precessing around their rotation axes; we do this by enabling the imperfect internal alignment mechanism (II, \citealt{Barnett1915, Purcell1979}), as implemented in \textsc{Polaris}. To include the effect of random gas collisions, we include the imperfect Davis-Greenstein alignment (IDG, or paramagnetic relaxation \citealt{Davis1951, Jones1967, Purcell1979, Spitzer1979}), which despite being rather inefficient \citep{Reissl2017} can contribute to aligning a part of the small grains. Silicate grains and graphite grains have very different paramagnetic properties, which differ by roughly six orders of magnitude \citep{Draine1996ASPC, Hoang2014}. This results in a very randomized distribution of graphite grain orientations, preventing them from contributing to the polarized emission. Finally, we do not include the effects of the self-scattering \citep{Kataoka2015}, because it operates in the innermost regions of the disk (typically within a radius of $\sim 10~\mathrm{au}$), which are excluded from our analysis, or the k-RAT mechanism (alignment with the radiation anisotropy), which requires radiation conditions much stronger than those considered in this work to be efficient \citep{LazarianHoang2007ApJ, Tazaki2017}.\\
 
 \subsubsection{Synthetic observations of the Stokes maps with \textsc{Polaris} }\label{SOPolaris}
 {For each one of the selected snapshots, we set our model at a distance of $250~\mathrm{pc}$ and observe a squared region of $8000~\mathrm{au}$ on a side centered at the position of the protostellar embryo (materialized by the sink particle) at a resolution of $5~\mathrm{au}$ per pixel. Each snapshot is observed synthetically along each one of the main axes $x$, $y$, and $z$ ($y$--$z$, $x$--$z$, and $x$--$y$ maps respectively) at three wavelengths: $0.8$, $1.3,$ and $3.0~\mathrm{mm}$. } \\

For each synthetic observation, \textsc{Polaris} provides perfectly observed maps of the four components of the Stokes vector $S = (I, Q, U, V)^T$, where $I$ is the total dust emission, $Q$ and $U$ are the linearly polarized dust emission components, and $V$ is the circularly polarized emission. \\

Finally, we compute the total linearly polarized intensity maps ($PI$) and the polarization fraction $p_\mathrm{frac}$ from the Stokes maps as follows:
\begin{eqnarray}
PI &=& \sqrt{Q^2 + U^2}\\ 
p_\mathrm{frac} &=& PI/I.
\end{eqnarray}
At far-infrared (FIR) to millimiter wavelengths, the magnetic field orientation inferred from the polarized emission ($\phi_\mathrm{pol}$) is obtained by rotating the polarization vector $\psi = \frac{1}{2}\arctan(U,Q) 
 $ by $90^\circ$. \\

\begin{figure*}[htp!]
\centering
    \includegraphics[width=0.99\textwidth, trim={2.65cm 1.25cm 2.72cm 4.1cm},clip]{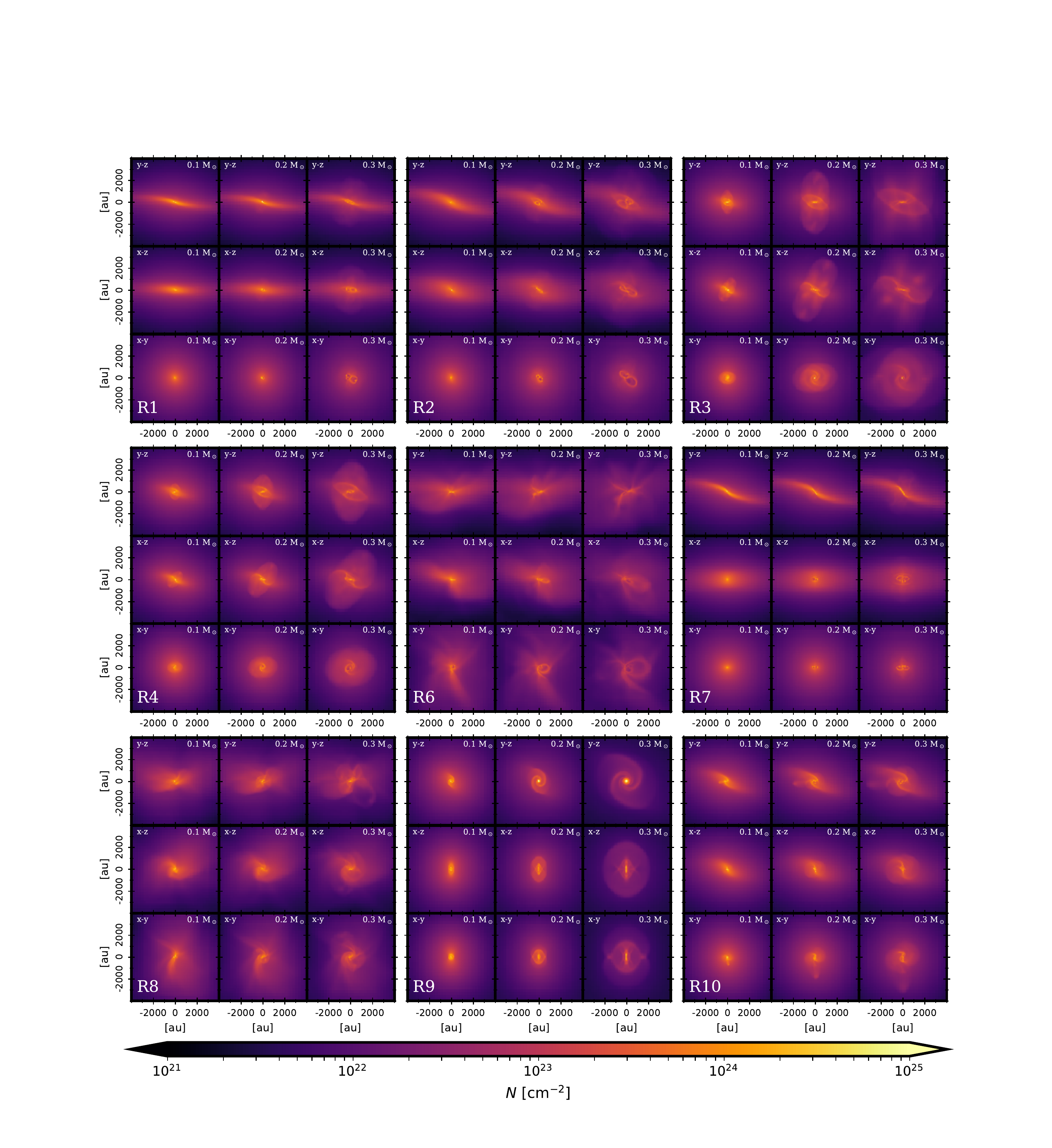}
\caption{Column density maps of all the selected outputs described in Table~\ref{table:1}. Each block corresponds to a given simulation (the identifier is indicated in the lower left corner of the block). Each block displays column density maps integrated along the $x$, $y,$ and $z$ axes (from top to bottom: $y$-$z$, $x$-$z$, and $x$-$y$ maps, respectively), and for a snapshot at $m_\mathrm{sink} = 0.1, 0.2,$ and $0.3~\mathrm{M_\odot}$(from left to right). The projection and evolutionary step are shown at the top of each individual plot. Higher resolution figures for individual simulations are available online.}
\label{Ncolmaps}
\end{figure*}

\begin{figure*}
\centering
 \includegraphics[width=0.99\textwidth, trim={2.65cm 1.195cm 2.72cm 4.1cm},clip]{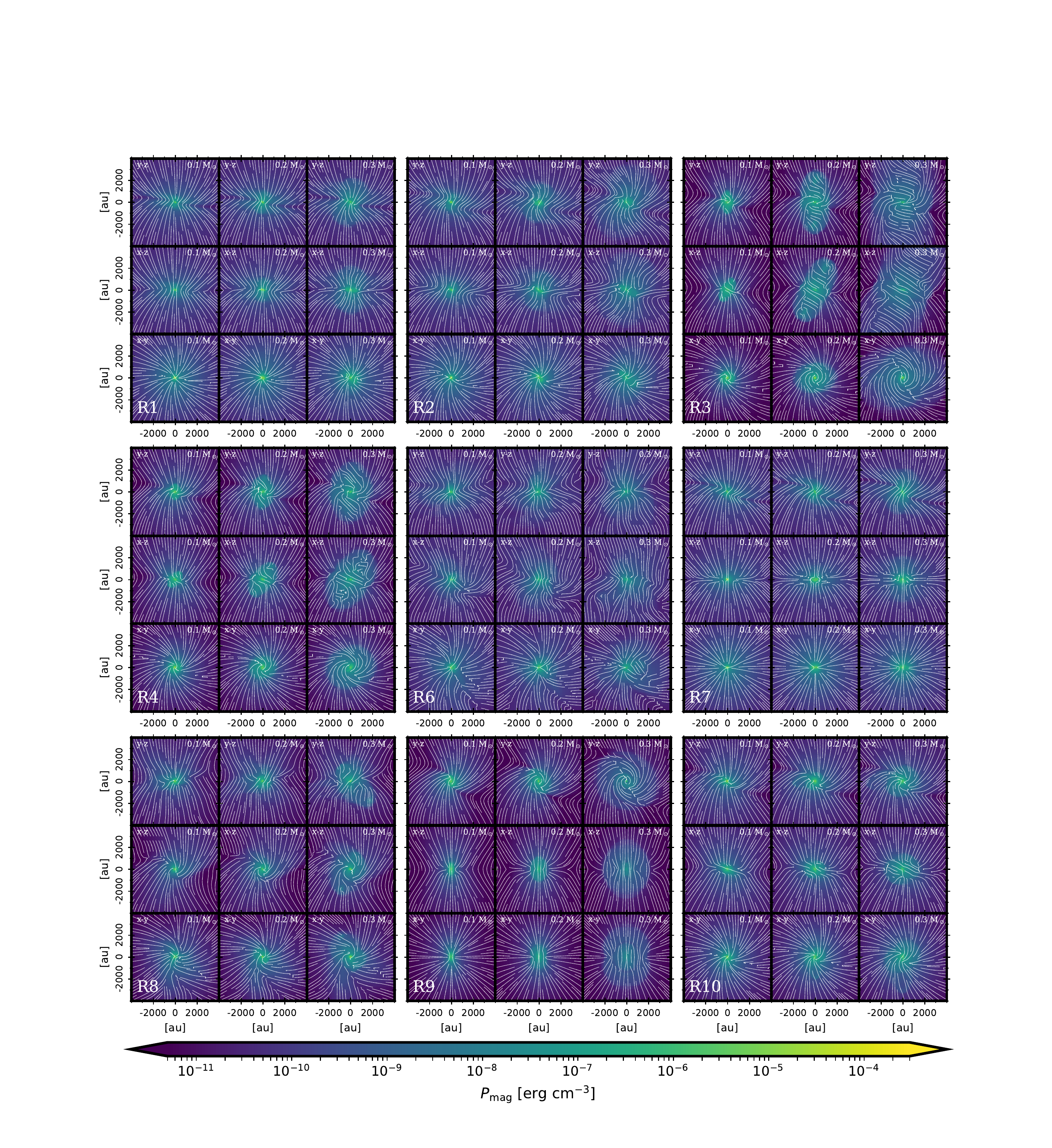}
\caption{Mean magnetic pressure $P_\mathrm{mag}$ maps and the orientation of the magnetic field lines along each projection for all the simulations. Each block corresponds to a given simulation and the plots are organized using the same structure as in Fig.~\ref{Ncolmaps}. Higher resolution figures for individual simulations are available online.} 
\label{Pmagmaps}
\end{figure*}

\begin{figure*}
\centering
  \includegraphics[width=0.99\textwidth, trim={2.65cm 1.35cm 2.72cm 4.1cm},clip]{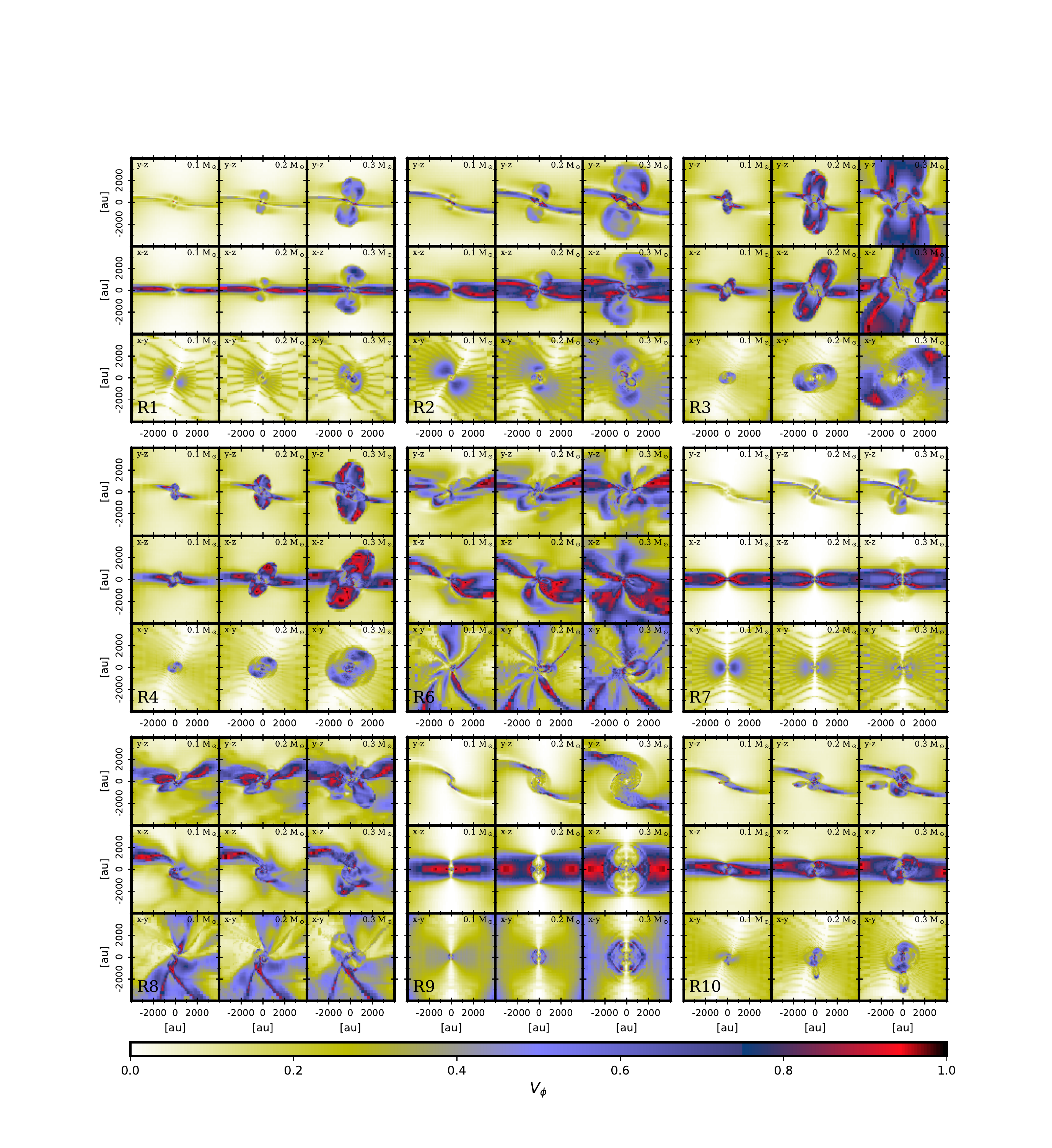}
\caption{Circular variance $V_\phi$ of the $B$-field lines in the \textsc{Ramses} data cubes (See Sect.~\ref{bidirstat}). The figure is organized using the same structure as in Figs.~\ref{Ncolmaps} and \ref{Pmagmaps}. The simulation identifier is given in the bottom left corner of the block, while the projection and evolutionary step are shown at the top of each individual plot.}
\label{Vphimaps}
\end{figure*}

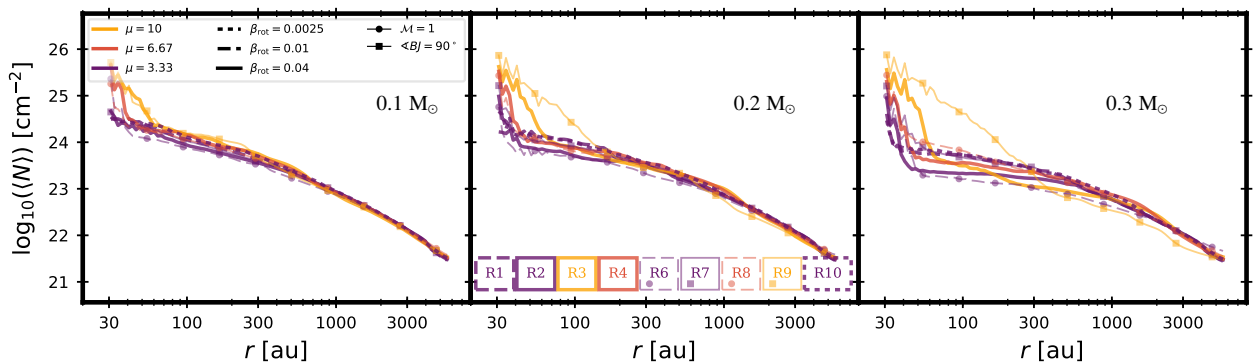
\begin{figure*}
\centering
\begin{tikzpicture}
\node[above right] (img) at (0,0) {
  \includegraphics[width=0.9\textwidth, trim={1.7cm 0.1cm 2.0cm 1.2cm},clip]{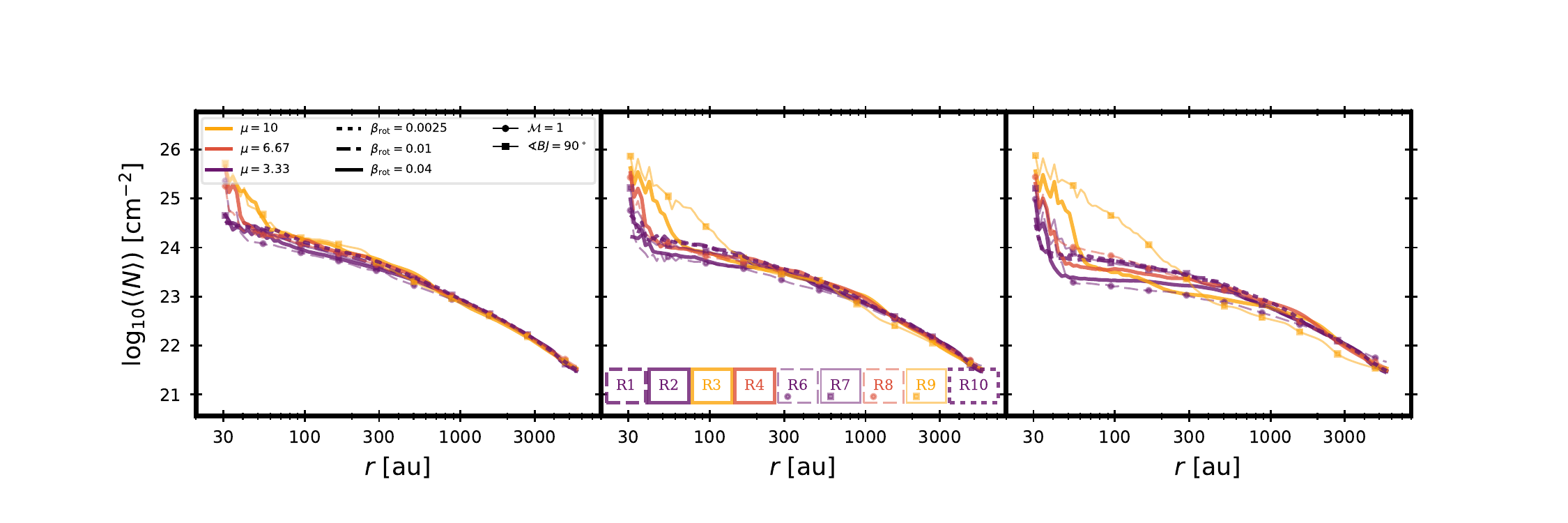}
  };
    \node [rotate=0, scale=0.8] at (5.4,  4.0) {\fontfamily{phv}\selectfont $0.1~\mathrm{M_\odot}$};
    \node [rotate=0, scale=0.8] at (10.1,  4.0) {\fontfamily{phv}\selectfont $0.2~\mathrm{M_\odot}$};
    \node [rotate=0, scale=0.8] at (15.0,  4.0) {\fontfamily{phv}\selectfont $0.3~\mathrm{M_\odot}$};        
\end{tikzpicture}  
\caption{Radial mean column density profiles. From left to right: profiles at $m_\mathrm{sink} = 0.1, 0.2,$ and $0.3~\mathrm{M_\odot}$. }
\label{Nvsdist}
\end{figure*}

\begin{figure*}
\centering
\begin{tikzpicture}
\node[above right] (img) at (0,0) {
  \includegraphics[width=0.9\textwidth, trim={1.7cm 0.1cm 2.0cm 1.2cm},clip]{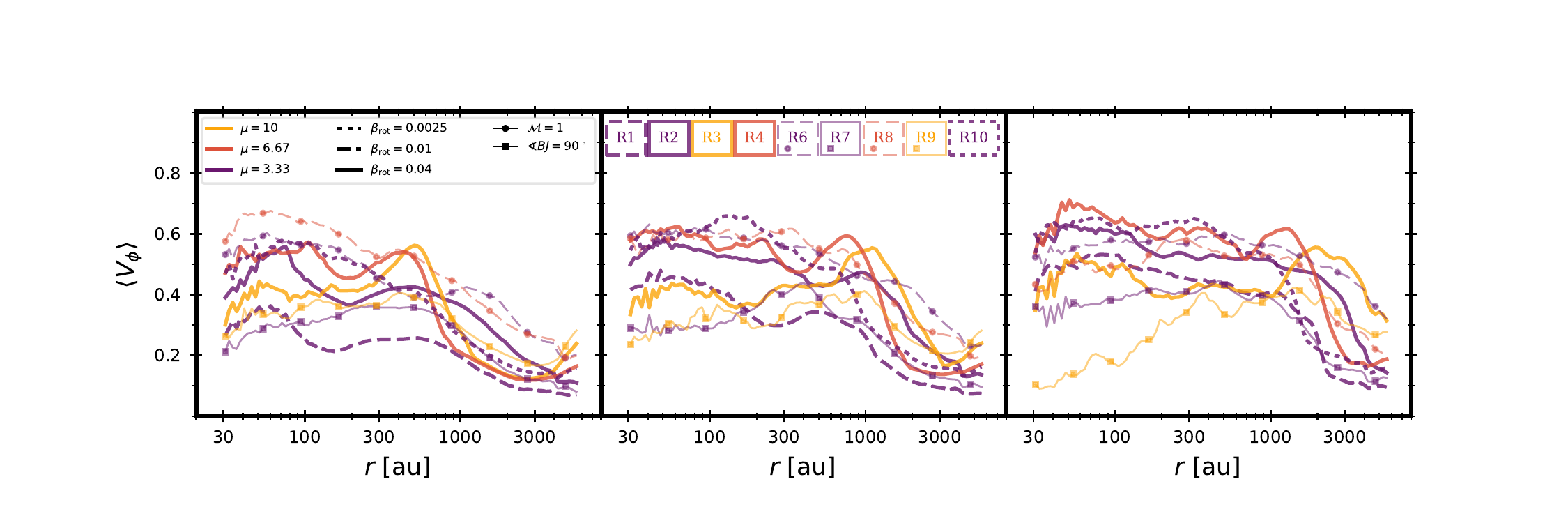}
  };
    \node [rotate=0, scale=0.8] at (5.4,  4.0) {\fontfamily{phv}\selectfont $0.1~\mathrm{M_\odot}$};
    \node [rotate=0, scale=0.8] at (10.1,  4.0) {\fontfamily{phv}\selectfont $0.2~\mathrm{M_\odot}$};
    \node [rotate=0, scale=0.8] at (15.0,  4.0) {\fontfamily{phv}\selectfont $0.3~\mathrm{M_\odot}$};        
\end{tikzpicture}  
\caption{Radial mean profiles of the circular variance $V_\phi$. From left to right: profiles at $m_\mathrm{sink} = 0.1, 0.2,$ and $0.3~\mathrm{M_\odot}$.}
\label{Vphivsdist}
\end{figure*}


\section{From 3D models to 2D maps: evolution of the integrated properties in the models} 
\label{resnumsim}

The general evolution of the numerical simulations and the dependence on the initial conditions are detailed in \citet{Hennebelle2020}. In the present paper, we focus on the description of the properties relevant to the polarized dust emission, particularly the protostar evolution (represented by the sink particle), the column densities, and the integrated mean orientation of the magnetic field lines as well as its intrinsic degree of disorganization (quantified by the variance of the angle orientation along the line of sight).

In this section we present integrated maps along the $x$, $y,$ and $z$ axes for the selected simulations and snapshots described in Table~\ref{table:1}. We chose to present these maps at the same resolution and scale as those of the synthetically observed ones in order to allow a direct comparison between both\footnote{Higher resolution figures for individual panels of Figs.~\ref{Ncolmaps}, \ref{Pmagmaps}, \ref{pfrac_maps_2}, \ref{pfrac_maps_1}, and \ref{pfrac_maps_3} are available online.}. In Fig.~\ref{Ncolmaps}, we present the total column density maps, which correspond to the total number of hydrogen atoms integrated along the line of sight. Figure~\ref{Pmagmaps} shows the density-weighted mean magnetic pressure ($P_\mathrm{mag}$, as described in Sect.~\ref{secdatacubes}) with overlaid mean orientation ($\phi_B$, computed following Eq.~\ref{Eq_mean_phiB}), aiming at providing a description of the magnetic field strength and mean orientation. Figure~\ref{Vphimaps} shows the circular variance ($V_\phi$, as defined in Eq.~\ref{eqnVphi}), which quantifies the level of disorganization of the magnetic field lines along the line of sight. Finally, we summarize these results by computing the average radial profiles of the three axes of $N$ and $V_\phi$ in Figs.~\ref{Nvsdist} and \ref{Vphivsdist}, respectively. 

\subsection{Integrated 2D maps}
The column densities shown in Fig.~\ref{Ncolmaps} display a wide variety of gas distribution features including outflows, spiral structures, loops, and more complex density distributions depending on the initial physical conditions. More precisely, the simulations including turbulence display very asymmetrical features, while those of the remaining simulations appear more uniform.
Figure~\ref{Pmagmaps} shows how the magnetic field lines are dragged after the initial gravitational collapse,  forming the characteristic hourglass shape, but the further evolution of the gas distribution and the final morphology of the magnetic field {are} controlled by the interplay of the specific physical conditions. 
As in the case of the evolution of the sink particle, the most important parameter is the mass-to-flux ratio ($\mu$). Figure~\ref{Pmagmaps}, and more particularly simulations R2, R3, and R4, shows that as the magnetic field lines are more prone to bending and distortion in the less magnetized cases ($\mu=10$), the magnetic field lines evolve into a helicoidal morphology in the outflow, which develops more easily. As the magnetization level increases, the outflow remains more confined and the helicoidal feature is less evident. The effect of the magnetization level can also be seen in simulations R7 and R9. Both simulations have the same rotation parameters ($\beta=0.04$ and  $\theta=90^\circ$), but the outcome of the evolution is very different. Simulation R9 ($\mu=10$) displays a well-developed spiral structure seen in the $y-z$ projection (nearly face-on projection)
of the column density map shown in Fig.~\ref{Ncolmaps} as well as in the projected magnetic field lines (streamplot) 
of Fig.~\ref{Pmagmaps}. On the other hand, in simulation R7 the stronger magnetic tension inhibits the development of high-contrast density waves seen as spiral arms, and the associated warped $B$-field lines.
Simulations R6 and R8 also share the same rotation ($\beta=0.04$ and  $\theta=30^\circ$) and turbulent ($\mathcal{M}=1$) parameters, but in this case it seems that the turbulence dominates the evolution of the gas dynamics as well as the evolution of the magnetic field lines. Nevertheless, the circular variance (Fig.~\ref{Vphimaps}) seems to be higher and more extended for R6. 

The circular variance maps (Fig.~\ref{Vphimaps}) describe the level of organization of the projected orientation of the magnetic field lines onto the plane of the sky, while traveling along the line of sight. 
As all the magnetic field lines are initially set parallel to the $z$ direction, all the $V_\phi$ maps are initially uniform and equal to zero. As the gas collapses inwards, dragging the magnetic field lines, the projected orientations of the magnetic field are no longer parallel and the $V_\phi$ starts to increase. 
At larger scales ($r>4000~\mathrm{au}$, not shown in the figure), the magnetic field orientation remains extremely well organized and the hourglass pattern is barely affected by the rotation or the outflow.  

At smaller scales, as those shown in the figure, the magnetic field remains relatively ordered along the lines of sight exhibiting yellow to light blue colors, which corresponds to a typical angular dispersion of the orientation of the magnetic field of $\sigma_\phi = 20^\circ-30^\circ$ (see Appendix~\ref{sec:sigmaV}), while regions with purple to red have larger angular dispersions ($\sigma_\phi>30\deg$). These regions of larger $V_\phi$ are ubiquitously seen towards the midplanes of the nearly edge-on projections, as well as in regions where the early hourglass structure of the magnetic field is perturbed either by the development of the outflow (very prominent in simulations R2 and R3, for example) or by the turbulent gas motions (simulations R6 and R7).   

We note that the projection that displays the most organized $B$-field is $z$ ($x-y$ maps on the bottom row of each block,) which is the direction of the initial magnetic field lines ($V_\phi$ remains small even for the cases R7 and R9 where the rotation axis is tilted by $90^\circ$ with respect to the $z$ direction). 

\subsection{Radial profiles}
To summarize the global behavior of the simulations and to compare the time evolution of both the collapse and the stress it produces on the magnetic field topology, we compare the mean radial profiles of the total column density and of the circular variance of the $B$-field orientation (Figs.~\ref{Nvsdist} and \ref{Vphivsdist}, respectively). For each evolutionary step and simulation, we stack the integrated maps of the three projections, and compute the mean value of the variable in radial logarithmic bins. 
From the  evolution of the column density radial distribution (Fig.~\ref{Nvsdist}), we clearly distinguish the usual development of a high-density inner region peaking around the protostar (column densities higher than  $\sim2\times10^{24}~\mathrm{cm^{-2}}$ within a radius of  $r\lesssim60~\mathrm{au}$). 
We also notice a flattening of the density profile at intermediate radii with time, which is  due to the development of a flattened envelope sometimes also referred to as the "pseudo-disk" \citep{Galli1993, Li1996,Hennebelle2009}, while the outer layers of the protostellar envelopes ($r>1000~\mathrm{au}$) remain mostly unaffected.  All the simulations, except for the simulation with a lower degree of magnetization, present a comparable global evolution. Simulation R9, and to a lesser extent simulation R3 (low magnetization cases and $\theta = 90\deg$ and $\theta = 30\deg$, respectively),  display a wider overdensity region due to the development of the spiral structures and a thicker equatorial region with respect to the rotation axis.

The behavior of the mean circular variance radial profiles of Fig.~\ref{Vphivsdist} seems quite different from that of the mean column density radial profile. The $V_\phi$ profiles have slightly lower values in the inner region, remaining more uniform with typical values of $V_\phi \sim 0.2-0.6$ at intermediate radii, while in the outer region they decrease to unperturbed values. 
The regions displaying the largest variance are located at intermediate distances, following the development of the outflow. 
In these regions, the development of a helicoidal magnetic field, along with the projection effects, induces a higher circular variance in the cavity walls of the outflow, as remarkably shown in the case of R3.
However, even at advanced Class 0 evolution ($0.3~\mathrm{M_\odot}$ in the central embryo), the stress applied by the gravitational collapse and outflowing motions to the $B$-field only result in rather low $V_\phi$ with values $< 0.6$ (equivalent to typical dispersion of the $B$-field angle along the line of sight $\simlt 30\deg$).
A particular case is the simulation R9, which displays a decreasing $V_\phi$ with time in the inner $\sim500~\mathrm{au}$ with time due to the development of a more organized spiral pattern guided by the rotation due to a lower magnetic tension, and that remains unperturbed due to the lack of outflow. 
There is not a clear dependence between the mean $V_\phi$ and the degree of magnetization, but the presence of turbulence seems to produce a more uniform and more stable level of disorganization of the magnetic field lines.

 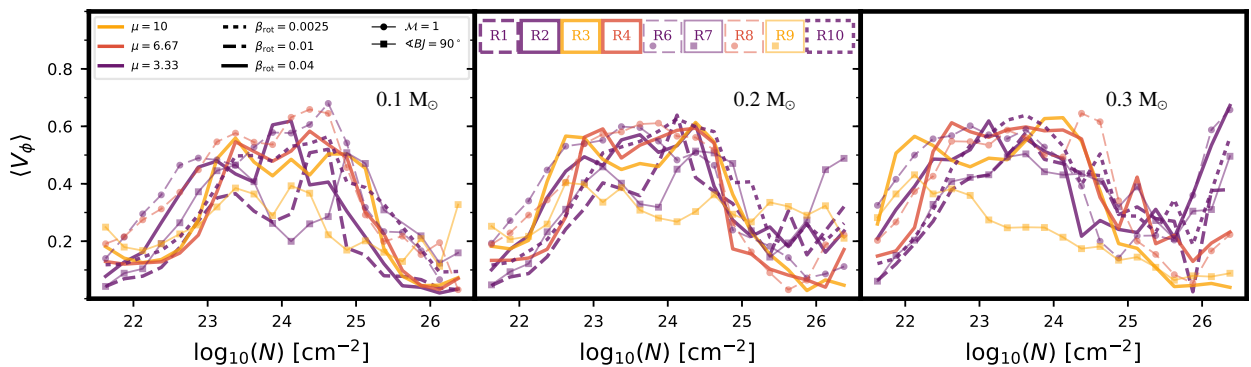
\begin{figure*}
\centering
\begin{tikzpicture}
\node[above right] (img) at (0,0) {
  \includegraphics[width=0.9\textwidth, trim={1.6cm 0.1cm 2.0cm 1.2cm},clip]{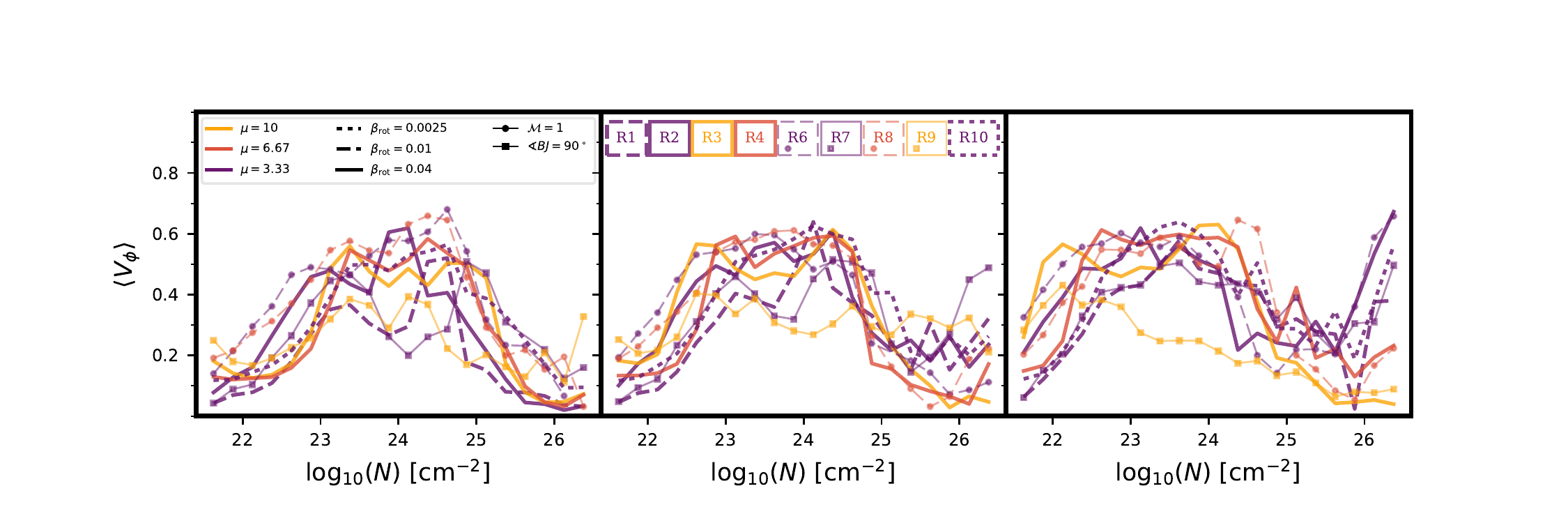}
  };
    \node [rotate=0, scale=0.8] at (5.4,  4.0) {\fontfamily{phv}\selectfont $0.1~\mathrm{M_\odot}$};
    \node [rotate=0, scale=0.8] at (10.1,  4.0) {\fontfamily{phv}\selectfont $0.2~\mathrm{M_\odot}$};
    \node [rotate=0, scale=0.8] at (15.0,  4.0) {\fontfamily{phv}\selectfont $0.3~\mathrm{M_\odot}$};          %
\end{tikzpicture}
\caption{Mean circular variance $V_\phi$ per logarithmic bin of column density. From left to right: results at $m_\mathrm{sink} = 0.1, 0.2,$ and $0.3~\mathrm{M_\odot}$.}
\label{VphivsN}
\end{figure*}

To better understand how the circular variance is related to the column density, we compute the mean value of $V_\phi$ in logarithmic bins of total column density. Figure~\ref{VphivsN} shows these results for the stacked data. We notice that initially only the intermediate column densities present high levels of disorganization of the magnetic field lines. These regions are related to the equatorial region, where the column densities increase due to the collapse and the further accretion is guided along the magnetic field lines, at the same time as the lines are distorted by the gas rotation. At later times, as the gas becomes denser, regions with high $V_\phi$ develop at higher column densities for the simulations with a higher degree of magnetization (purple lines indicating $\mu = 3.33$), for all the cases (with turbulence, or rotation orthogonal to the magnetic field). It is interesting to notice that simulations R3 and R9 (yellow lines), despite having the same level of magnetization and rotation, differing only in the initial rotation angle, display a very different behavior at intermediate column densities, particularly at around $N\sim10^{24}~\mathrm{cm^{-2}}$, where the simulation R3 have values of $V_\phi > 0.6$, while simulation R9 barely surpasses $0.2$.

\begin{figure*}
\centering
    \includegraphics[width=0.99\textwidth, trim={2.65cm 1.3cm 2.72cm 4.1cm},clip]{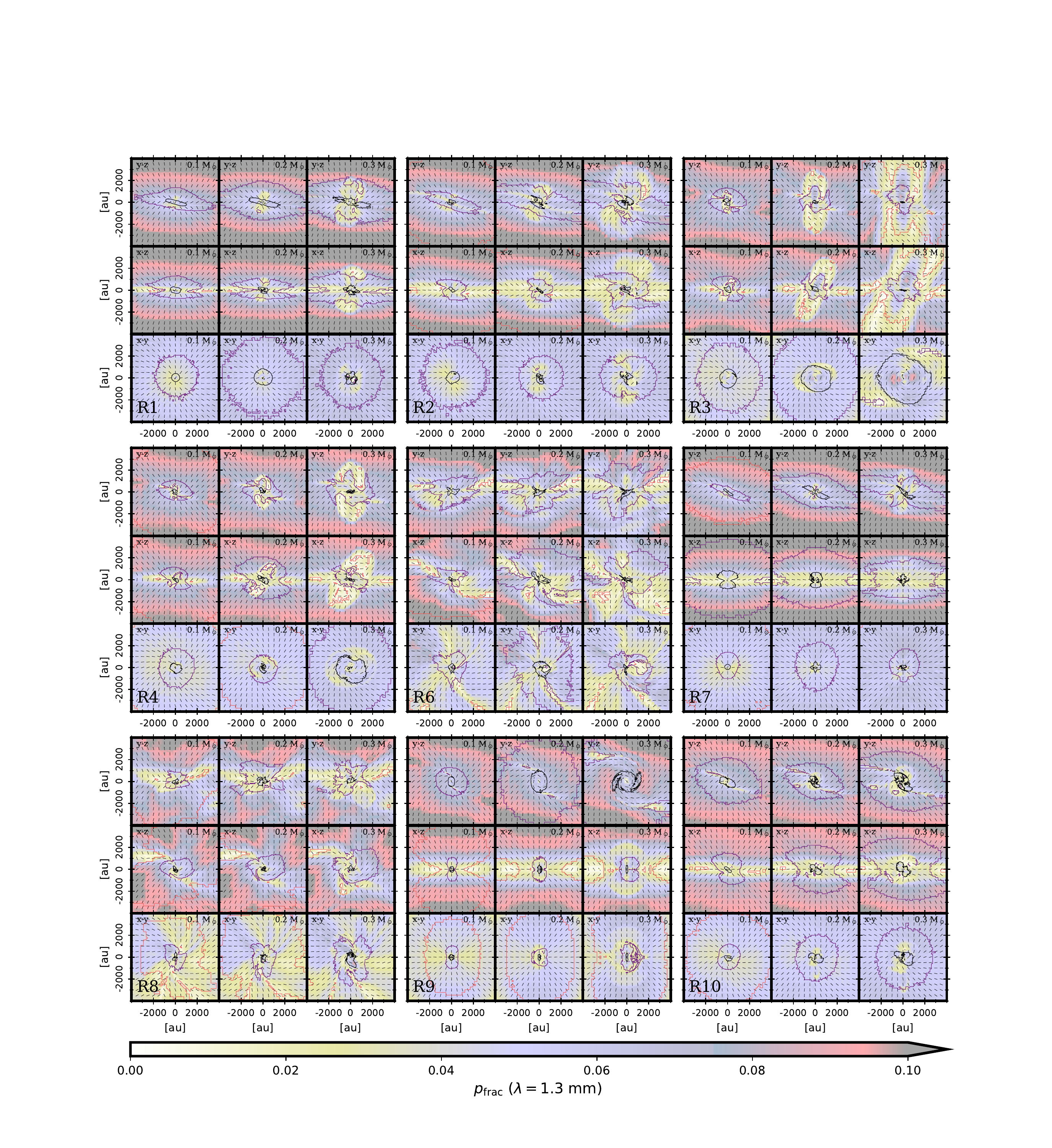}
\caption{Results from the synthetic observation at $\lambda = 1.3~\mathrm{mm}$ showing the polarization fraction (background image), the polarized intensity contours, and the inferred magnetic field orientation vectors. The contours correspond to the linearly polarized intensity normalized to the peak value at levels $10^{-2}$ (black), $10^{-3}$ (purple), $10^{-4}$ (red), and $10^{-5}$ (yellow). Higher resolution figures for individual simulations are available online. }
\label{pfrac_maps_2}
\end{figure*}  
  

\begin{figure*}
\centering
\begin{tikzpicture}
\node[above right] (img) at (0,0) {
  \begin{tabular}{@{}l@{}}
  \includegraphics[width=0.9\textwidth, trim={1.2cm 1.65cm 2.2cm 1.7cm},clip]{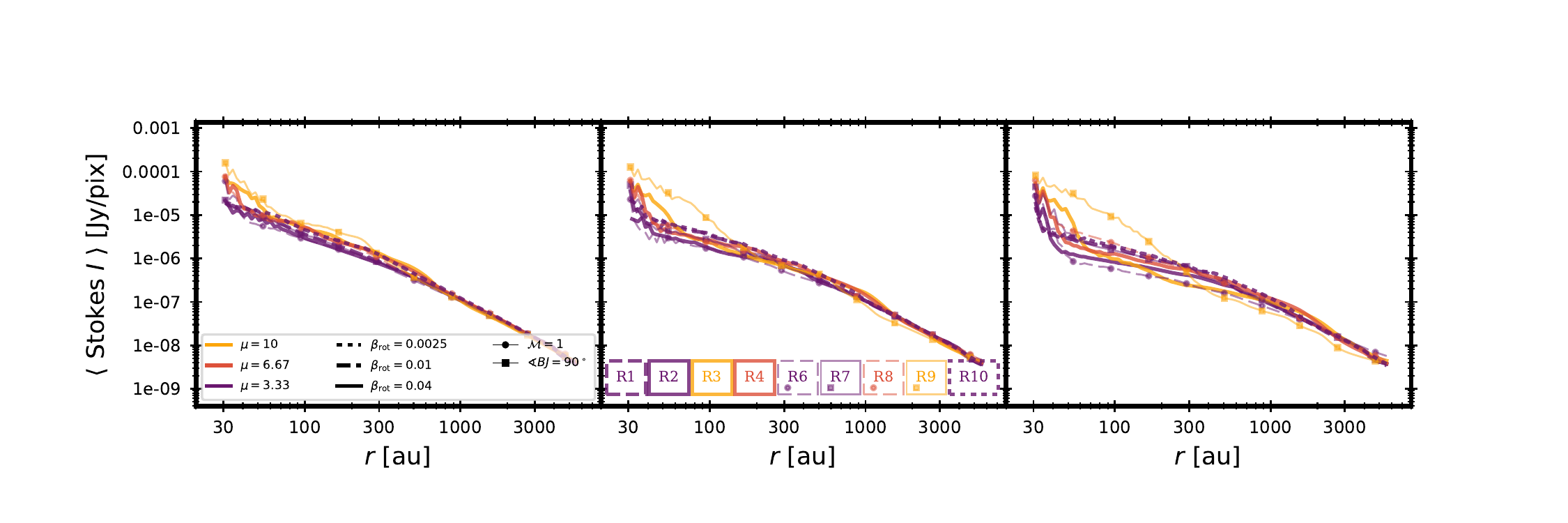}\\
  \includegraphics[width=0.9\textwidth, trim={1.2cm 1.65cm 2.2cm 1.7cm},clip]{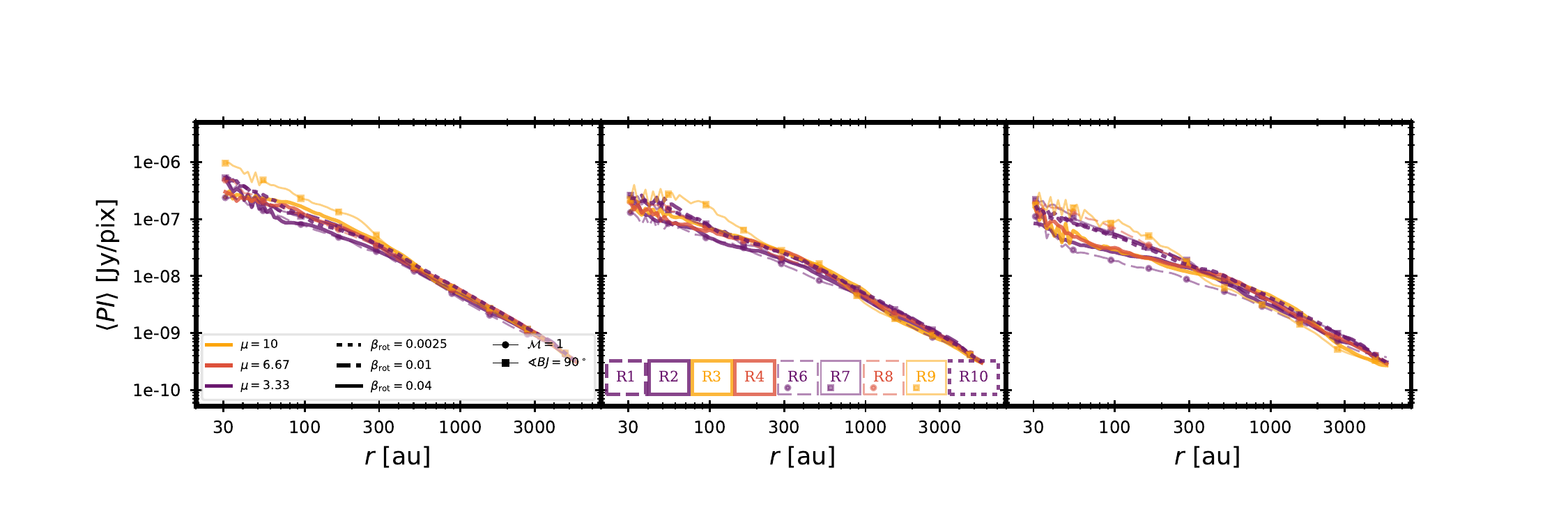}\\
  \includegraphics[width=0.9\textwidth, trim={1.2cm 0.8cm 2.2cm 1.7cm},clip]{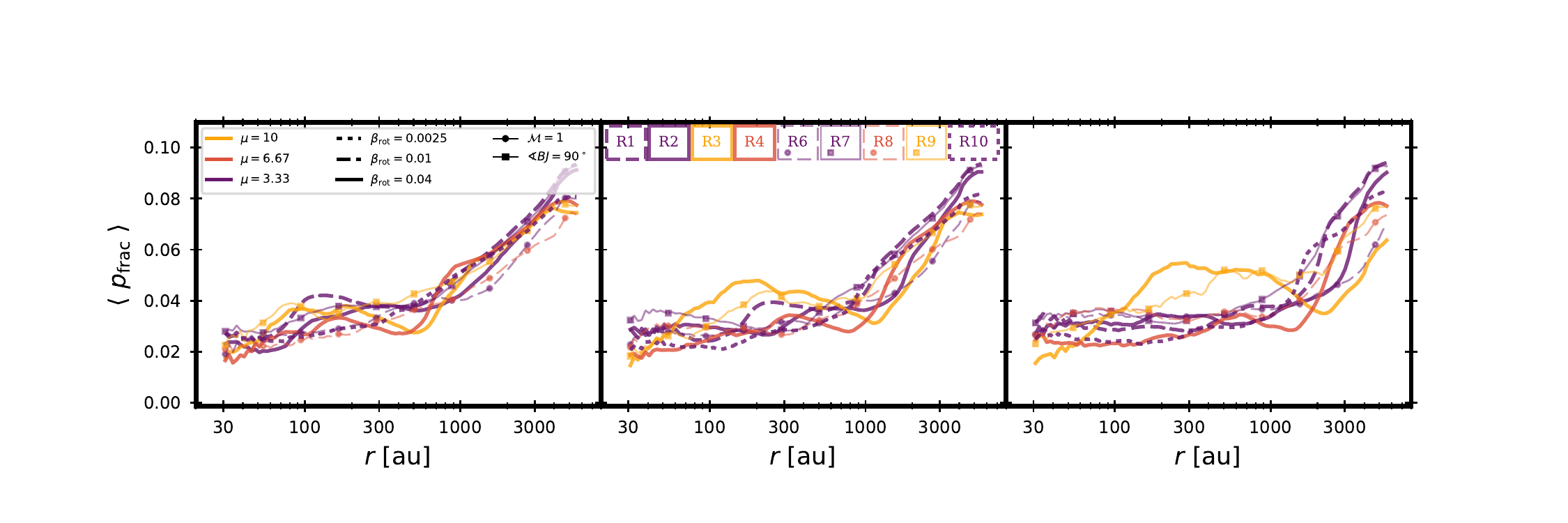}   
  \end{tabular}};
\node [rotate=0, scale=0.8] at (5.4,  11.5) {\fontfamily{phv}\selectfont $0.1~\mathrm{M_\odot}$};
\node [rotate=0, scale=0.8] at (10.1, 11.5) {\fontfamily{phv}\selectfont $0.2~\mathrm{M_\odot}$};
\node [rotate=0, scale=0.8] at (15.0, 11.5) {\fontfamily{phv}\selectfont $0.3~\mathrm{M_\odot}$};        
\end{tikzpicture}  
\caption{Radial profiles of the synthetic observation at $\lambda=1.3~\mathrm{mm}$. From top to bottom: Total dust emission (Stokes $I$), total linearly polarized dust emission ($PI$), and polarization fraction ($p_\mathrm{frac}$). From left to right: Results at $m_\mathrm{sink} = 0.1, 0.2,$ and $0.3~\mathrm{M_\odot}$.}
\label{radial_Stokes_RAT_2}
\end{figure*}

\section{Synthetic maps of the polarized dust emission}\label{sectSynthObs}
We produced perfectly integrated synthetic observations of the polarized dust emission for the central $8000~\mathrm{au}$, observed at a distance of $250~\mathrm{pc}$, and at a resolution of $5~\mathrm{au}$ per pixel (see Table~\ref{table:2}).  Fig.~\ref{pfrac_maps_2}  presents synthetic maps of the polarization fraction 
observed at $1.3~\mathrm{mm}$  (background color maps), with overlaid polarized intensity contours (normalized to the peak value) and the magnetic field vectors (inferred from the polarized emission as described in Sect.~\ref{methodpolar}). 
To summarize and simplify the comparison between all the simulations, we computed the radial profiles, stacking the data for the integrated quantities along the three projections.
We provide the radial profiles of the Stokes $I$, the intensity of the total linearly polarized dust thermal emission $PI,$ and the polarization fraction $p_\mathrm{frac}$ in Fig.~\ref{radial_Stokes_RAT_2}.
We present similar maps, but for $\lambda = 0.8$ and $3.0~\mathrm{mm}$ (Figs.~\ref{pfrac_maps_1} and \ref{pfrac_maps_3}), as well as similar radial profiles (Figs.~\ref{radial_Stokes_RAT_1} and \ref{radial_Stokes_RAT_3}) in the Appendix. \\
The total dust emission maps (Stokes $I$ maps, not shown here) peak towards the central region neighboring the protostar, closely following the column density as expected. This is also seen by comparing the radial profiles of the Stokes $I$ maps (top rows of Fig.~\ref{radial_Stokes_RAT_2} and \ref{radial_Stokes_RAT_1} and \ref{radial_Stokes_RAT_3}) to the total column density maps (Fig.~\ref{Nvsdist}), which show good agreement between these two quantities, but show slight departures at high column densities due to the large associated opacities.\\ 
The total linearly polarized intensity (contour lines of Figs.~\ref{pfrac_maps_2}, ~\ref{pfrac_maps_1} and ~\ref{pfrac_maps_3}, and middle row panels of Figs.~\ref{radial_Stokes_RAT_2}, ~\ref{radial_Stokes_RAT_1} and ~\ref{radial_Stokes_RAT_3}) shows higher values toward the central region (due to higher densities and temperatures), but the associated polarization fractions are the lowest in these regions (see the bottom panels of Figs.~\ref{radial_Stokes_RAT_2}, ~\ref{radial_Stokes_RAT_1} and ~\ref{radial_Stokes_RAT_3}).    
The highest polarization fractions are found in the external regions, typically reaching $5\%$ to $10\%$, and gradually decrease in a nonmonotonical manner toward the central regions. More precisely, Fig.~\ref{pfrac_maps_2} shows that $p_\mathrm{frac}$ drops more strongly in the midplane and the outflow regions, particularly near the cavity walls. Even though the outflow regions show lower polarization levels, the simulation R3 shows a rather high $p_\mathrm{frac}$ in the central part of the outflow region. This is likely due to the fact that the outflow cavity is well developed, allowing the transversal component of the  helicoidal magnetic field to dominate the polarization pattern.
A qualitative comparison of Fig.~\ref{Vphivsdist} and Fig.~\ref{pfrac_maps_2} suggests that the polarization fraction drops in the lines of sight where the $B$-field happens to have a large variance ($V_\phi$). In the following section, we quantify and discuss this behavior, and its consequences as to our ability to trace the $B$-field in low-mass protostellar envelopes with dust polarization observations.

It is interesting to note that, even though the simulations analyzed in this paper correspond to an initial core mass of only $1~\mathrm{M_\mathrm{\odot}}$, the less magnetized cases, and particularly simulation R3, produce a polarization pattern remarkably similar to the one observed by \citet{Hull2017} for Ser-emb 8, particularly in the $x-z$ maps (see Fig.~\ref{pfrac_maps_1}) at the time when the protostellar mass reached $0.3~\mathrm{M_\mathrm{\odot}}$, consistent with the scenario proposed by \citet{Hull2017}, but without invoking the presence of initial turbulence. 

\section{Do polarized dust emission maps capture the underlying topology of protostellar magnetic fields?}

\subsection{Fidelity of the inferred magnetic field orientation}
In this work, we aim to assess whether or not (sub)mm polarized dust emission maps of protostellar envelopes ---which are routinely used to discuss the affects of the magnetic field during the collapse of protostellar cores into a star--disk system--- are indeed robust tracers of the magnetic field lines threading those cores. To do so, we compared the two quantities introduced in Sect.~\ref{SectMethods}: 
the magnetic field orientation reconstructed from the synthetic observation of polarized dust emission ($\phi_\mathrm{pol}$, Sect.\ref{SOPolaris}) and the mean orientation of the magnetic field in the model integrated along the line of sight and computed using the circular statistics ($\phi_B$, Eq.~\ref{Eq_mean_phiB}). We compute the difference between model and observations as the following difference:
$\Delta \phi = \phi_\mathrm{pol} - \phi_B$, and the discrepancy as the absolute value of the difference, $|\Delta \phi|$. 

\subsubsection{Dust polarized emission: an overall good tracer of $B$-fields in protostellar environments}

\begin{figure}
\centering
\includegraphics[width=0.49\textwidth, trim={2.0cm 1.3cm 1.35cm 0.5cm},clip]{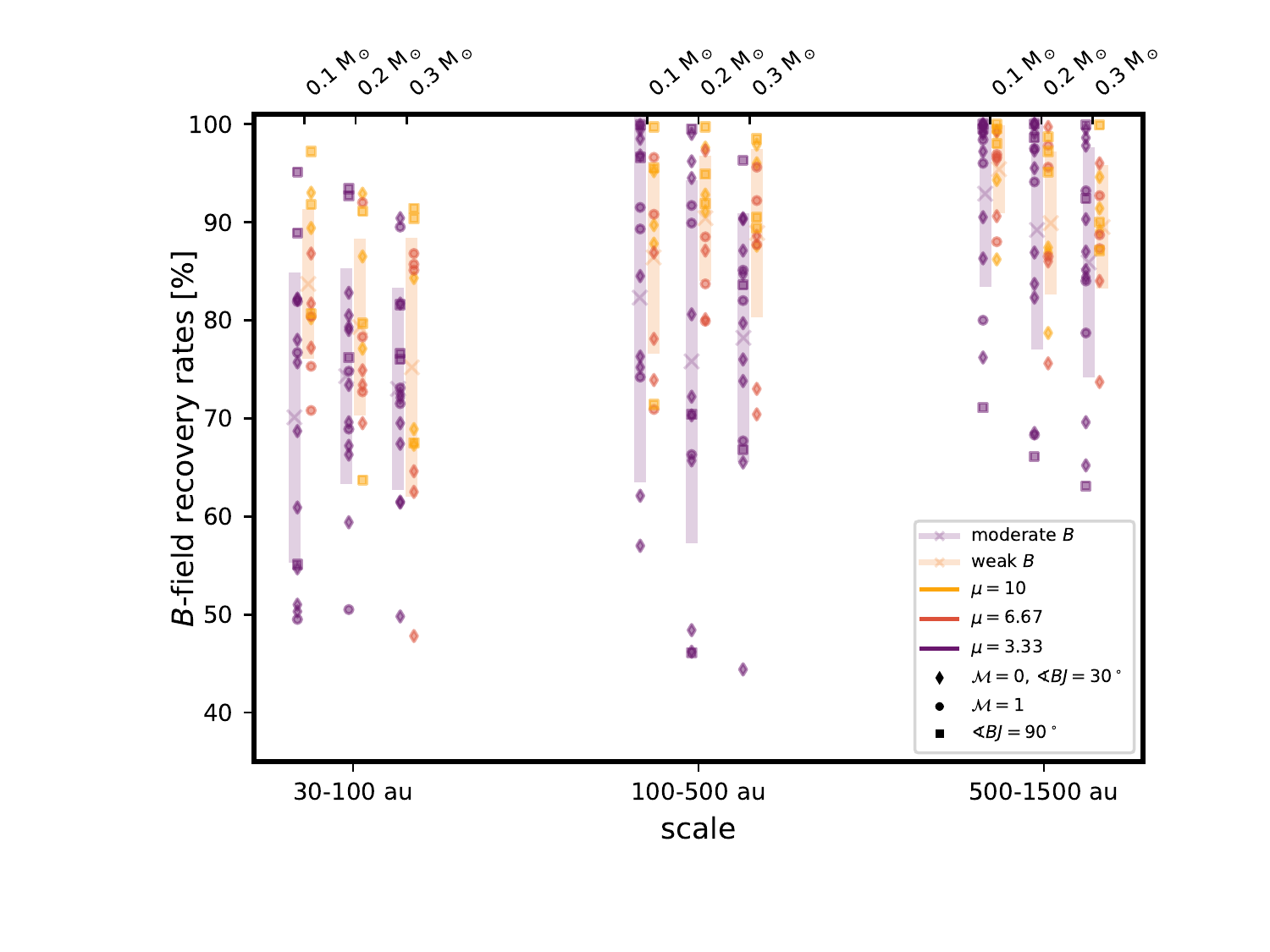}
\caption{$B$-field recovery rate for all the synthetic observations at $\lambda=0.8~\mathrm{mm}$. Individual points correspond to those of Table~\ref{table:Ap}, while the cross symbols and vertical bars  correspond to the the averaged values and their standard variation for the two cases defined in Table~\ref{table:3}, respectively: moderate ($\mu = 3.33$) and weak ($\mu > 6$) magnetic fields. The values are separated in three spatial scales (indicated at the bottom of the figure), and slightly shifted for the sake of clarity for the three evolutionary stages given by the protostellar mass  indicated at the top of the figure.}  
\label{RecFull}
\end{figure}

\begin{table}
\caption{Fidelity of $B$-field mapping from polarized dust emission at $\lambda=0.8~\mathrm{mm}$ in protostellar envelopes at different scales.}              
\label{table:3}      
\centering                                      
\begin{tabular}{c c c c c }          
\hline\hline                        
 \noalign{\smallskip}
  $m_\mathrm{sink}$ & \multicolumn{4}{c}{\textit{$B$-field recovery rates and $\sigma$ $[\%]$   }}\\
 \tiny{$[\mathrm{M_\odot}]$} & \tiny{All scales}  & \tiny{$30-100$ au} & \tiny{$100-500$ au} & \tiny{$500-1500$ au} \\    
\hline\hline      
 \noalign{\smallskip}
        \multicolumn{5}{c}{\textit{$\mu = 3.33$}}                       \\ 
\hline                                  
 \noalign{\smallskip}
%
0.1 &    96.9~ \tiny{3.1} &      70.1~ \tiny{14.8} &     82.3~ \tiny{18.8} &        92.9~~ \tiny{9.5} \\
0.2 &    95.6~ \tiny{4.2} &      74.3~ \tiny{11.0} &     75.8~ \tiny{18.5} &        89.2~ \tiny{12.2} \\
0.3 &    93.3~ \tiny{5.3} &      73.0~ \tiny{10.3} &     78.2~ \tiny{12.7} &        85.9~ \tiny{11.7} \\
\hline
 \noalign{\smallskip}
        \multicolumn{5}{c}{\textit{$\mu = 6.67, 10.$}}                  \\ 
\hline                                  
 \noalign{\smallskip}
0.1 &    97.3~ \tiny{2.1} &      83.7~~ \tiny{7.6} &     86.4~ \tiny{9.8} &        95.4~ \tiny{4.5} \\
0.2 &    95.7~ \tiny{2.7} &      79.3~~ \tiny{9.0} &     90.4~ \tiny{6.4} &        89.9~ \tiny{7.3} \\
0.3 &    90.7~ \tiny{6.3} &      75.2~ \tiny{13.2} &     88.9~ \tiny{8.6} &        89.5~ \tiny{6.3} \\
\hline                                             
\end{tabular}
\tablefoot{
The recovery rate corresponds to the fraction of lines of sight where measurement of the $B$-field orientation from polarized dust emission recovers the integrated $B$-field orientation within an error of $<15\deg$ in the modeled Class 0 envelopes. The percentages shown in this table correspond to the averaged recovery rate, while the standard deviation of each measurement is given in a smaller font next to each mean value. The results are separated in two cases: moderate magnetic field ($\mu = 3.33$, simulations R1, R2, R6, R7, and R10) and weak magnetic field ($\mu = 6.67, 10.0$, simulations R3, R4, R8, and R9). Percentages have been averaged over all the corresponding simulations and projections at a given evolutionary step (given by the protostar mass $m_\mathrm{sink}$).
The scales correspond to the whole map including all the scales (second column), and in concentric rings containing all the lines of sight with radial distances to the protostar in the interval indicated in the header of the table. All the individual recovery rates are given in Table~\ref{table:Ap}. 
}
\end{table}

\begin{table}
\caption{Fidelity of $B$-field mapping from polarized dust emission at $\lambda=0.8~\mathrm{mm}$ in protostellar envelopes as a function of the polarized intensity.}              
\label{table:4}      
\centering                                      
\begin{tabular}{c c c c }          
\hline\hline                        
 \noalign{\smallskip}
  $m_\mathrm{sink}$ & \multicolumn{3}{c}{\textit{$B$-field recovery rates and $\sigma$ $[\%]$   }}\\
    \tiny{$[\mathrm{M_\odot}]$} & \tiny{$PI_\mathrm{norm} \geq 10^{-2}$} & \tiny{$PI_\mathrm{norm} \geq 10^{-3}$} & \tiny{$PI_\mathrm{norm} \geq 10^{-3}$} \\    
\hline\hline      
 \noalign{\smallskip}
        \multicolumn{4}{c}{\textit{$\mu = 3.33$}}                       \\ 
\hline                                  
 \noalign{\smallskip}
0.1 &    84.4~ \tiny{17.7} &     93.3~ \tiny{8.8} &      98.4~ \tiny{2.0} \\
0.2 &    83.7~ \tiny{14.2} &     92.7~ \tiny{8.4} &      97.7~ \tiny{2.5} \\
0.3 &    86.6~~ \tiny{8.9} &     92.7~ \tiny{6.3} &      96.5~ \tiny{3.3} \\
\hline
 \noalign{\smallskip}
        \multicolumn{4}{c}{\textit{$\mu = 6.67, 10.$}}                  \\ 
\hline                                  
 \noalign{\smallskip}
0.1 &    92.9~ \tiny{4.3} &      95.6~ \tiny{3.7} &      99.0~ \tiny{1.0} \\
0.2 &    93.6~ \tiny{4.4} &      94.0~ \tiny{3.8} &      98.0~ \tiny{1.6} \\
0.3 &    92.2~ \tiny{6.0} &      93.8~ \tiny{4.7} &      95.4~ \tiny{3.6} \\
\hline                                             
\end{tabular}
\tablefoot{
The recovery rates are defined as in Table~\ref{table:3} for an error of $<15\deg$. The recovery rates shown in this table correspond to the mean values found for regions with a total linear polarized intensity normalized to the peak value ($PI_\mathrm{norm} = PI/PI_\mathrm{peak}$) higher than $10^{-2}$, $10^{-3}$, and $10^{-4}$, respectively. These values correspond to the first three contours of Fig.~\ref{pfrac_maps_1}. As in Table~\ref{table:3}, the corresponding standard deviations are given in a smaller font.
}
\end{table}

\noindent We built difference maps between the model magnetic field line orientations and the reconstructed magnetic field orientations from polarized dust emission observed at $\lambda = 0.8, 1.3$, and $3.0~\mathrm{mm}$ (see Figs.~\ref{dphi_1}, \ref{dphi_2} and \ref{dphi_3}, respectively). 
We note that there is generally good agreement between the inferred ($\phi_\mathrm{pol}$) and the actual ($\phi_{B}$) mean orientations, with localized regions displaying large differences. 
From these maps, we can quantify ---in a statistical way--- the ability of the polarized dust emission to recover the true magnetic field topology. We define the recovery rate as the fraction of sight lines in a map for which the magnetic field orientation is retrieved within a given error, or in other words, have a discrepancy $|\Delta\phi|$ lower than a given threshold. This recovery rate is found to be similar at the three wavelengths for each map, slightly increasing with the wavelength.
At $\lambda=0.8~\mathrm{mm}$, which is the frequency that displays the lowest recovery rates, the magnetic field is typically recovered within an error of less than $15\deg$ (typical of most observations using dust polarized emission as a probe for magnetic topologies) for more than $95 \%$ of the lines of sight over the full maps, 
with only some cases with no clear dependence on the magnetization level, rotation, or inclination, recovering less than $90\%$. These cases correspond to late evolutionary steps and projections along the $y$-axis ($x-z$ maps). 
The worst recovery rate ($79.7\%$) is obtained for simulation R3, $m_\mathrm{sink} = 0.3~\mathrm{M_\odot}$ and projected along the $y$-axis. For this map, most of the lines of sight for which the orientations were not recovered within an error of $15^\circ$ are associated to higher circular variances (the recovery rate for the lines of sight with $V_\phi < 0.5$ is $99.7\%$).  Table~\ref{table:3} summarizes the mean recovery rates for the synthetic maps at $\lambda=0.8~\mathrm{mm}$ at different evolutionary stages and at different scales, averaged for all the projections and simulations.  The detailed results are given in Table~\ref{table:Ap}. Additionally, we provide a synthesis of these tables in Fig.~\ref{RecFull}. 
 These tables show that the magnetic field orientation reconstructed from polarized dust emission observations is remarkably good at large scales and that even at intermediate scales the recovery rate is typically higher than $70\%$ (see also Fig.~\ref{RecFull}). Table~\ref{table:ap.1.3} and Table~\ref{table:ap.3.0} show even better results at longer wavelengths ($\lambda=1.3~\mathrm{mm}$ and $\lambda=3.0~\mathrm{mm}$, respectively), consistent with the findings of \citet{Kuffmeier20}.  
To check whether or not the polarized emission that is  susceptible to being observed recovers the magnetic field topology, we computed the recovery rates for different levels of total linearly polarized dust emission normalized to the peak ($PI_\mathrm{norm} = PI/PI_\mathrm{peak}$) at $\lambda = 0.8~\mathrm{mm}$. We use the same values of the contour levels of Fig.~\ref{pfrac_maps_1}. We computed the mean recovery rates over all the simulations and projections separated in the same two cases as in Table~\ref{table:3}. We find that even for the first contour, which corresponds to the most intense emission and which is  therefore the most likely to be detected, the recovery rates are found to be higher than $83\%$ for the moderate magnetic field simulations, and higher than $92\%$ for the less magnetized simulations. As the contour level decreases, the recovery rate increases, consistently with the results of Table~\ref{table:3} seen for larger scales. 
\\

Finally, note that the models analyzed here implement nonideal MHD processes under ionization conditions that may not be prototypical at the smallest scales around protostars: for example, \citet{Cabedo2022} estimated cosmic-ray ionization rates in B335 from observations probing the $\sim 100-500$ au scales which are up to two orders of magnitude higher than the predicted ionization generated by external cosmic rays alone. If the ionization was indeed to be much stronger at small envelope radii due to locally accelerated cosmic rays, that would make the coupling of B fields to the local gas much more efficient and models that implement both ideal and nonideal MHD with realistic local ionization conditions would need to be explored in the future.

%

\subsubsection{Physical and observational effects responsible for poor measurements in disks and outflows}
\begin{figure*}
\centering
\begin{tikzpicture}
\node[above right] (img) at (0,0) {
  \includegraphics[width=0.9\textwidth, trim={1.7cm 0.65cm 2.0cm 1.cm},clip]{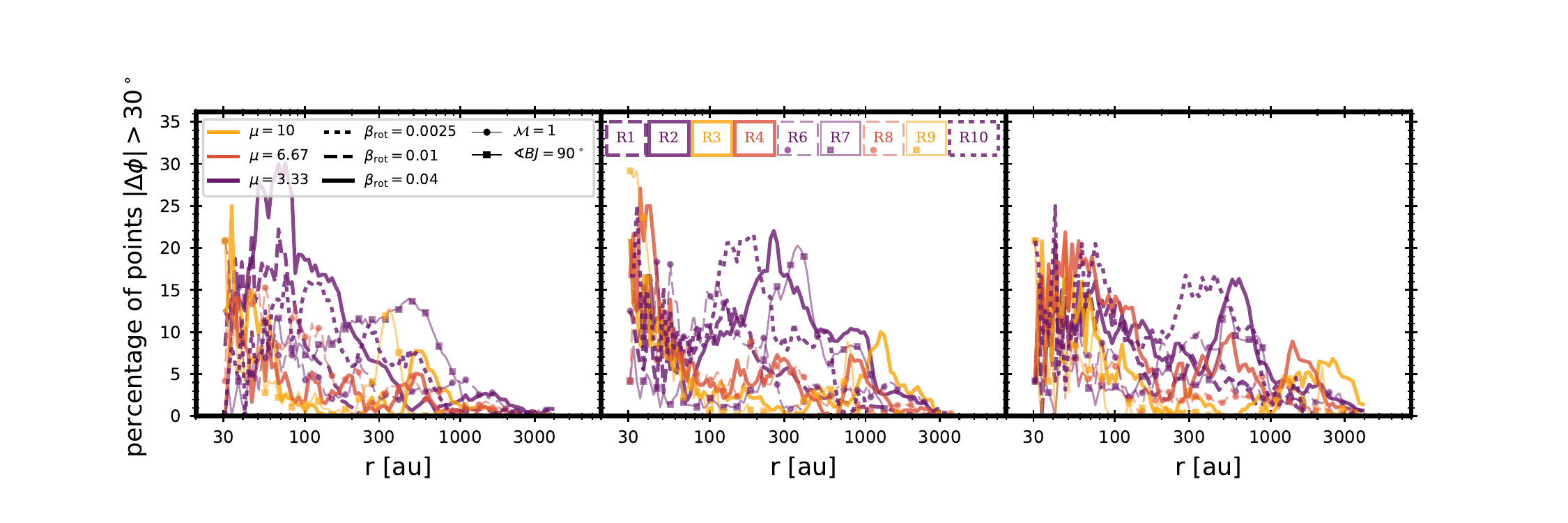}
  };
    \node [rotate=0, scale=0.8] at (5.4,  4.0) {\fontfamily{phv}\selectfont $0.1~\mathrm{M_\odot}$};
    \node [rotate=0, scale=0.8] at (10.1,  4.0) {\fontfamily{phv}\selectfont $0.2~\mathrm{M_\odot}$};
    \node [rotate=0, scale=0.8] at (15.0,  4.0) {\fontfamily{phv}\selectfont $0.3~\mathrm{M_\odot}$};        
\end{tikzpicture}  
\caption{Percentage of pixels with $|\Delta\phi| > 30^\circ$ at $\lambda = 1.3~\mathrm{mm}$ in logarithmic radial bins.}
\label{radial_error30deg}
\end{figure*}
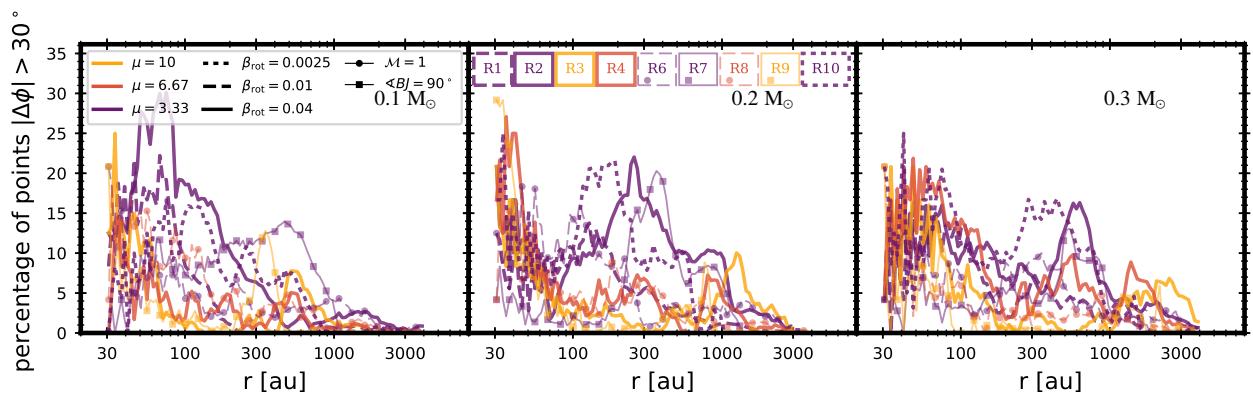

We now present our investigation of the lines of sight associated to large errors in the reconstructed magnetic field maps, where our goal is to identify the local physical properties 
responsible for the mismatch. As the local gas density and its temperature both scale with radii in protostellar envelopes, we first looked for the distribution of inaccurate measurements as a function of distance to the central protostellar embryo. Figure ~\ref{radial_error30deg} shows the percentage of lines of sight that show a discrepancy (expressed in terms of the absolute value of the difference $|\Delta\phi|$) of higher or equal to $30^\circ$ as a function of radius (using $100$ logarithmic bins). 
This figure shows distinctive radial distances at which the points with large discrepancies represent an important {contribution}. These regions develop outwards in time, diluting this fraction over a larger volume. The more magnetized simulations ($\mu=3.33$) display percentages of higher than $10\%$ over larger regions. Simulations with $\theta = 90^\circ$ start to show regions of high discrepancy at a much larger radius. Once the sinks in the simulations have reached $0.3~\mathrm{M_\odot}$, most of the simulations show a region of high discrepancy within a distance of $30-100~\mathrm{au}$ with respect to the sink position. Even though Figs.~\ref{radial_error30deg} and \ref{Vphivsdist} do not seem to display the same behavior, these regions of higher discrepancy seem to be connected to larger levels of disorganization of the field line orientations along the line of sight.  

\begin{figure*}
\centering
\begin{tikzpicture}
\node[above right] (img) at (0,0) {
  \begin{tabular}{@{}l@{}}
  \includegraphics[width=0.85\textwidth, trim={0.95cm 4.1cm 0.0cm 4.15cm},clip]{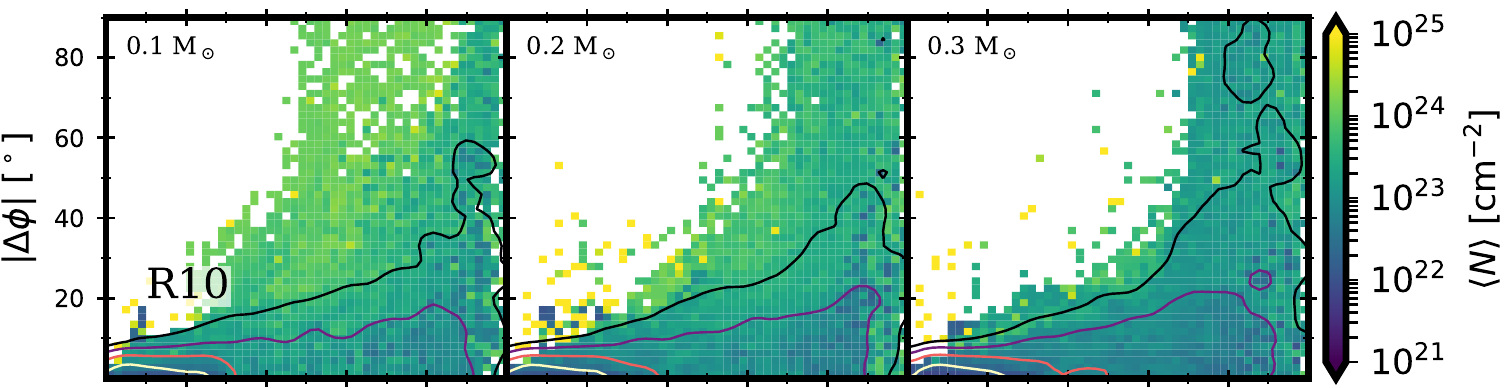}\\
  \includegraphics[width=0.85\textwidth, trim={0.95cm 4.1cm 0.0cm 4.15cm},clip]{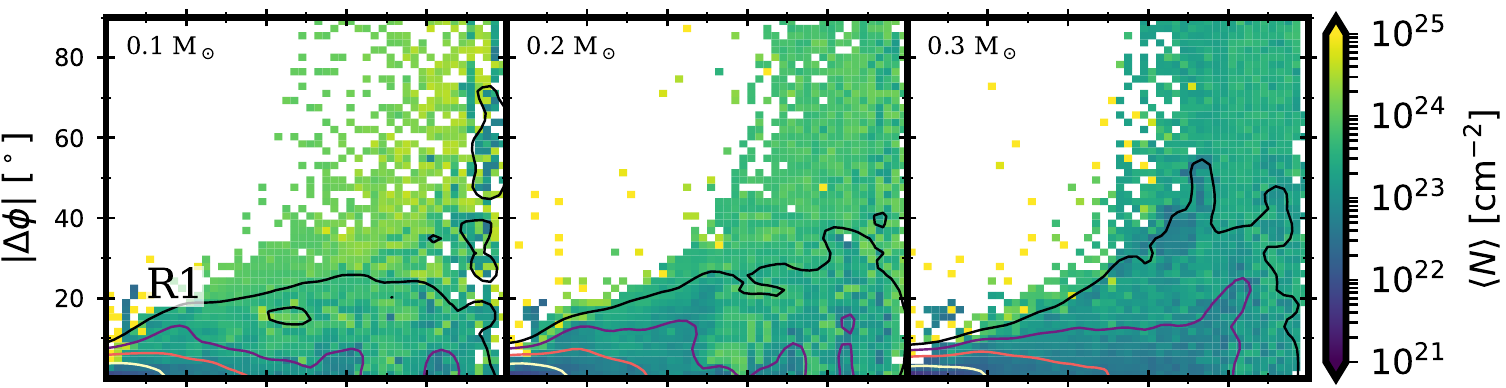}\\
  \includegraphics[width=0.85\textwidth, trim={0.95cm 4.1cm 0.0cm 4.15cm},clip]{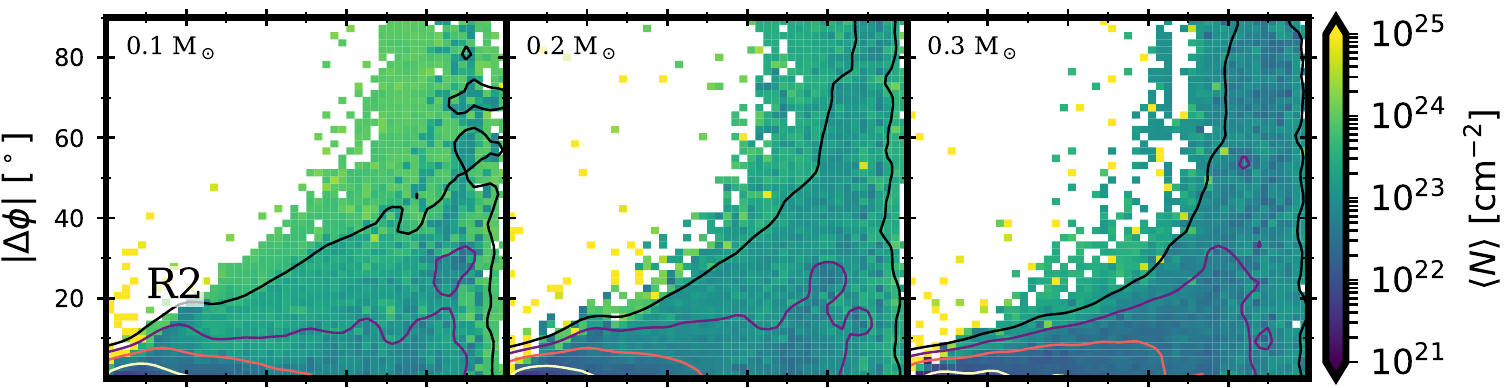}\\
  \includegraphics[width=0.85\textwidth, trim={0.95cm 4.1cm 0.0cm 4.15cm},clip]{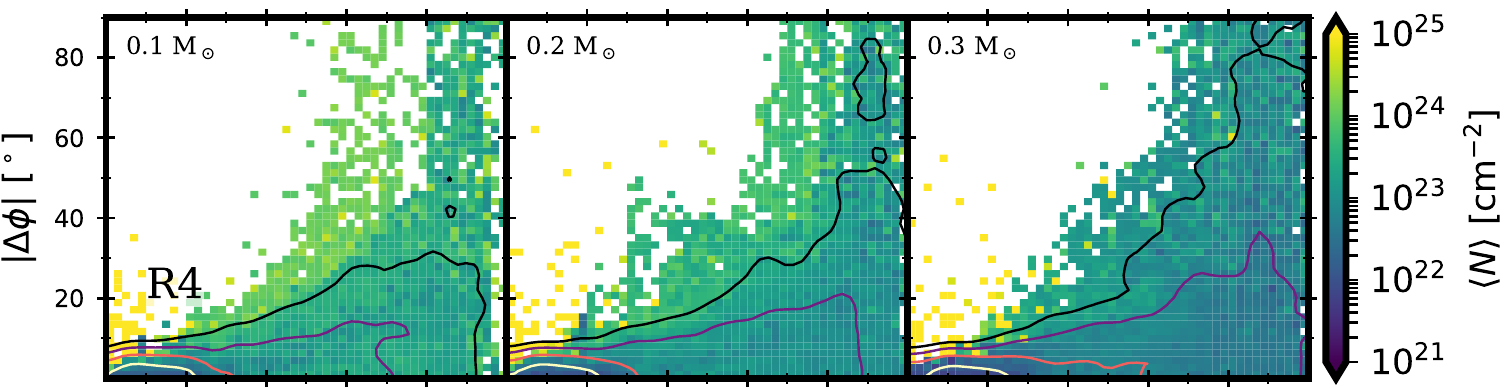}\\
  \includegraphics[width=0.85\textwidth, trim={0.95cm 3.00cm 0.0cm 4.15cm},clip]{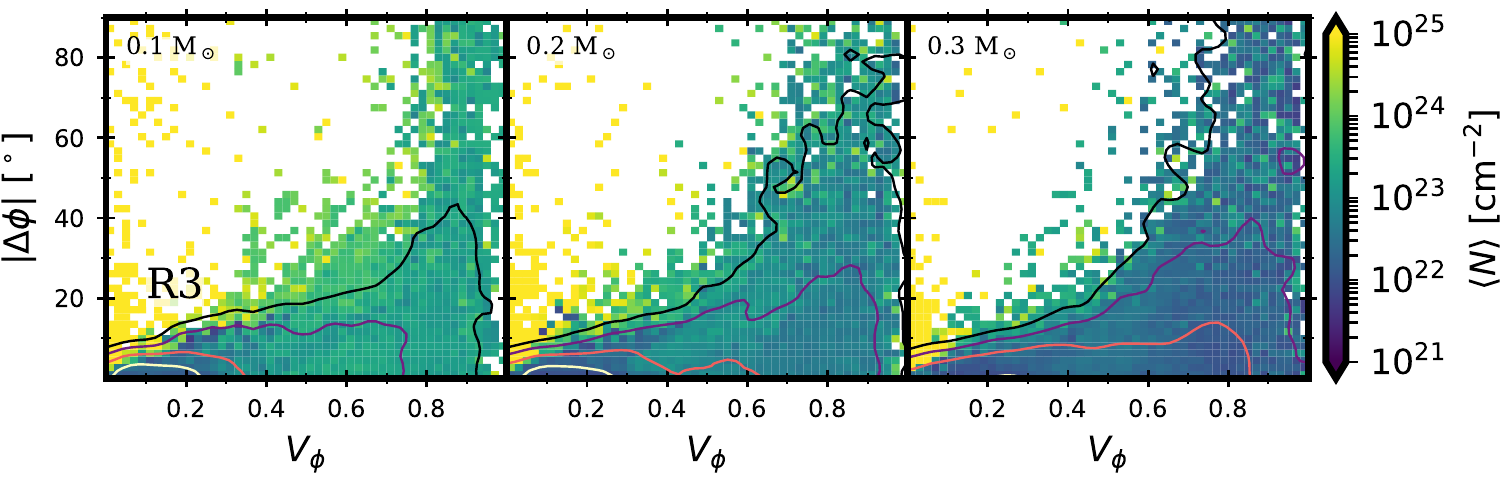}   
  \end{tabular}};
  \draw [|-|, line width=1.3pt, color=col1] (0.05,9.721) -- (0.05,21.827);
  \draw [|-|, line width=1.5pt, color=col2] (0.05,5.576) -- (0.05,9.397);
  \draw [|-|, line width=1.5pt, color=col3] (0.05,1.433) -- (0.05,5.252);
  \node [rotate=90, scale=1.5, color=col1] at (-0.2,15.8) {\fontfamily{phv}\selectfont $\mu = 3.33$};
  \node [rotate=90, scale=1.5, color=col2] at (-0.2,7.5) {\fontfamily{phv}\selectfont $\mu = 6.67$};
  \node [rotate=90, scale=1.5, color=col3] at (-0.2,  3.2) {\fontfamily{phv}\selectfont $\mu = 10$};
  \node [rotate=0, scale=0.9] at (2.1,20.90) {\fontfamily{phv}\selectfont $\beta = 0.0025$};
  \node [rotate=0, scale=0.9] at (2.0,16.90) {\fontfamily{phv}\selectfont $\beta = 0.01$};
  \node [rotate=0, scale=0.9] at (2.0,12.80) {\fontfamily{phv}\selectfont $\beta = 0.04$};
  \node [rotate=0, scale=0.9] at (2.0,  8.60) {\fontfamily{phv}\selectfont $\beta = 0.04$};
  \node [rotate=0, scale=0.9] at (2.0,  4.50) {\fontfamily{phv}\selectfont $\beta = 0.04$};
\end{tikzpicture}
\caption{Distribution of the discrepancy between the $B$ angle inferred from the synthetic polarized dust emission at $\lambda=1.3~\mathrm{mm}$ and the
mean orientation of the $B$ lines in the simulation as a function of the circular variance $V_\phi$  for the simulations with standard conditions. All these simulations do not include any initial turbulence ($\mathcal{M} = 0$) and have an initial inclination angle between the initial rotation axis and the magnetic field $\theta$ of $30^\circ$. The contour lines show the smoothed 2D histogram contour levels at $10^5$ (yellow), $10^4$ (red), $10^3$  (purple), and $10^2$ (black) counts. The color-coded background corresponds to the mean column density.}
\label{DphiVstdcond2}
\end{figure*}

\begin{figure*}
\centering
\begin{tikzpicture}
\node[above right] (img) at (0,0) {
  \begin{tabular}{@{}l@{}}
  \includegraphics[width=0.85\textwidth, trim={0.95cm 4.1cm 0.0cm 4.15cm},clip]{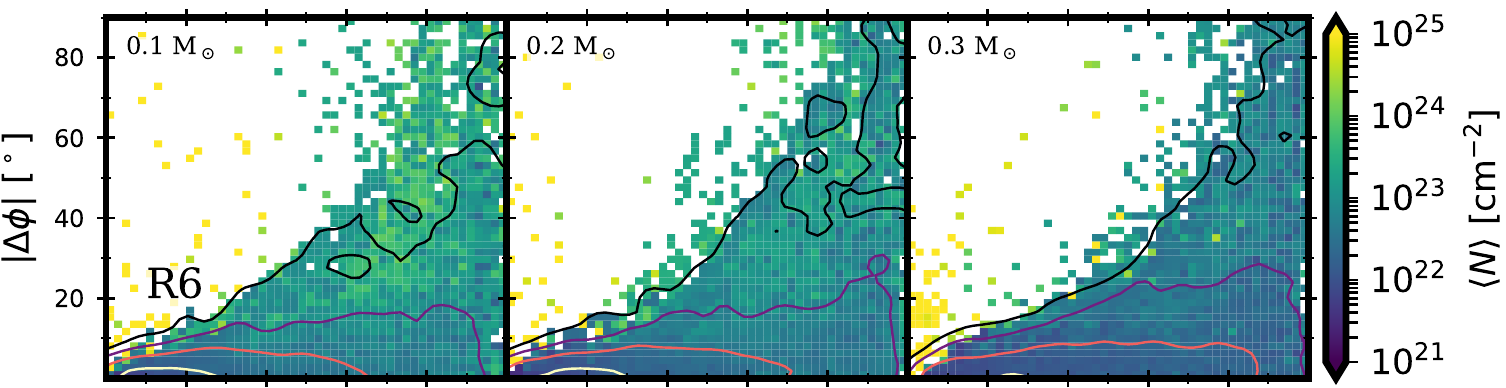}\\
  \includegraphics[width=0.85\textwidth, trim={0.95cm 4.1cm 0.0cm 4.15cm},clip]{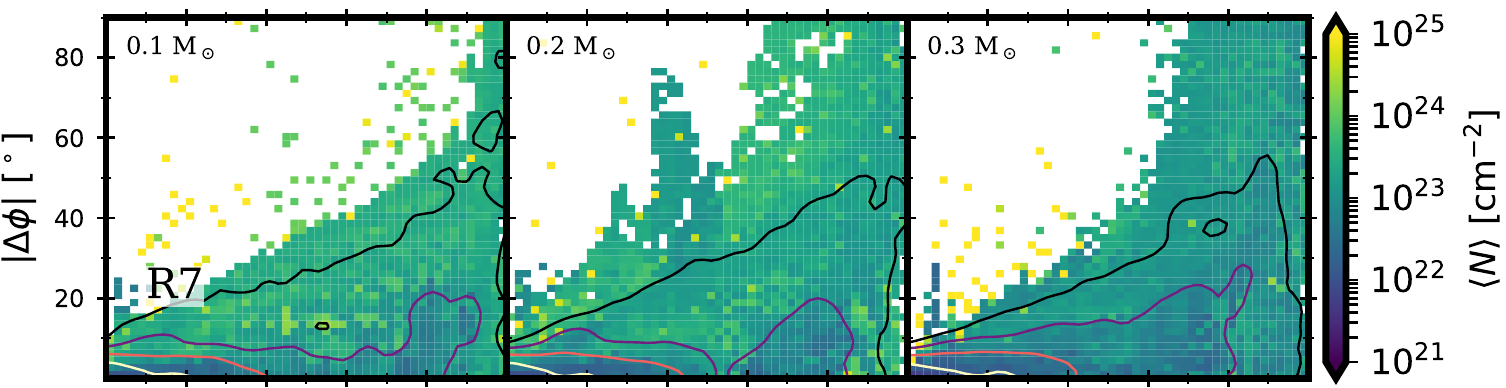}\\
  \includegraphics[width=0.85\textwidth, trim={0.95cm 4.1cm 0.0cm 4.15cm},clip]{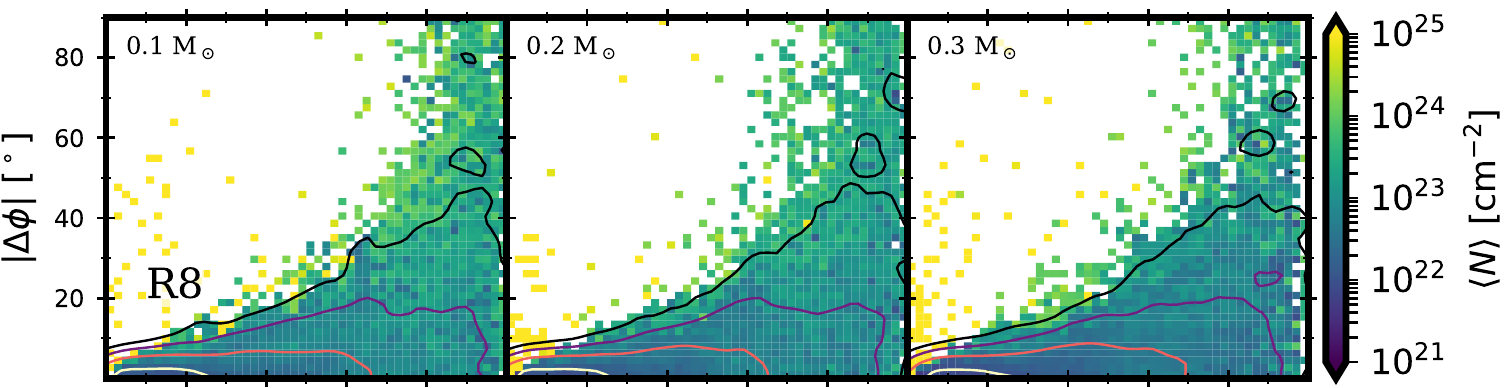}\\
  \includegraphics[width=0.85\textwidth, trim={0.95cm 3.00cm 0.0cm 4.15cm},clip]{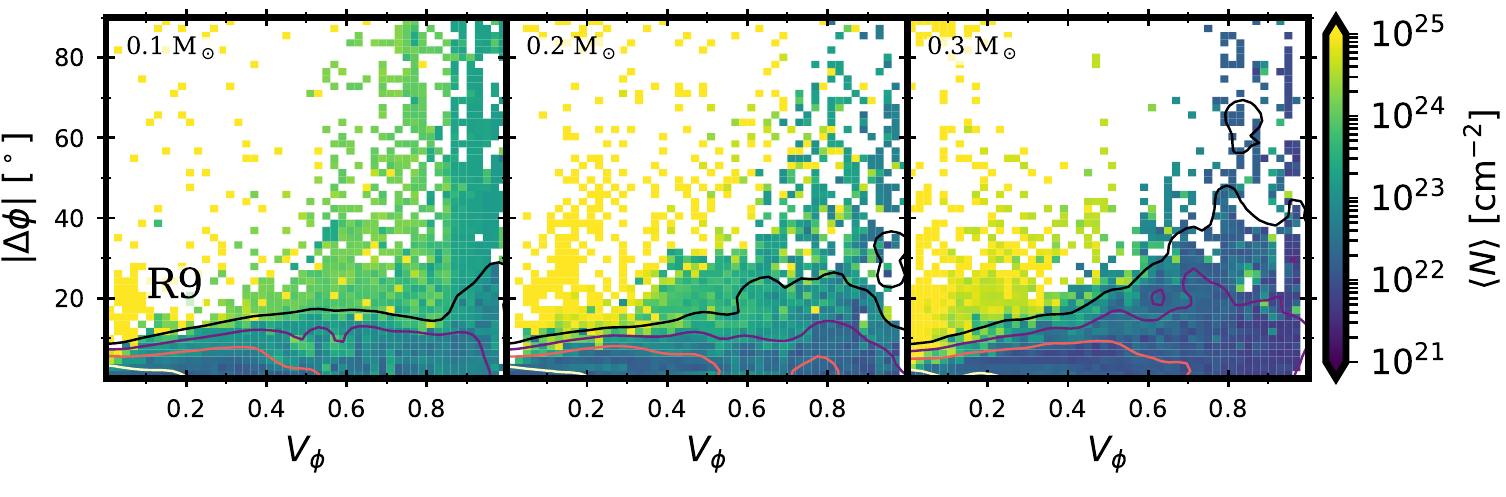}   
  \end{tabular}};
  \draw [|-|, line width=1.3pt, color=col1] (0.05,9.721) -- (0.05,17.684);
  \draw [|-|, line width=1.5pt, color=col2] (0.05,5.576) -- (0.05,9.397);
  \draw [|-|, line width=1.5pt, color=col3] (0.05,1.433) -- (0.05,5.252);
  \node [rotate=90, scale=1.5, color=col1] at (-0.2,14.0) {\fontfamily{phv}\selectfont $\mu = 3.33$};
  \node [rotate=90, scale=1.5, color=col2] at (-0.2,7.5) {\fontfamily{phv}\selectfont $\mu = 6.67$};
  \node [rotate=90, scale=1.5, color=col3] at (-0.2,  3.2) {\fontfamily{phv}\selectfont $\mu = 10$};
  \node [rotate=0, scale=0.9] at (2.5,16.90) {\fontfamily{phv}\selectfont $\beta = 0.01, \mathcal{M} = 1$};
  \node [rotate=0, scale=0.9] at (2.5,12.80) {\fontfamily{phv}\selectfont $\beta = 0.04, \theta = 90^\circ$};
  \node [rotate=0, scale=0.9] at (2.5,  8.60) {\fontfamily{phv}\selectfont $\beta = 0.01, \mathcal{M} = 1$};
  \node [rotate=0, scale=0.9] at (2.5,  4.450) {\fontfamily{phv}\selectfont $\beta = 0.04, \theta = 90$};  
\end{tikzpicture}
\caption{Distribution of $|\Delta \phi|$ at $\lambda=1.3~\mathrm{mm}$ as a function of the circular variance $V_\phi$ for the simulations with nonstandard conditions. The contours and the background image are the same as in Fig.~\ref{DphiVstdcond2}. Simulations R6 and R8 include turbulence ($\mathcal{M}=1$), while the simulations R7 and R9 have an initial angle between the rotation axis and the magnetic field axis of $\theta=90^\circ$.}
\label{DphiVnonstd2}
\end{figure*}

To better understand the role of the intrinsic degree of organization of the magnetic field lines and the reliability of the orientation inferred from dust polarization observations, we present 2D histograms showing the absolute value of the difference $|\Delta \phi|$ as a function of the circular variance $V_\phi$ at $\lambda=1.3~\mathrm{mm }$  in Figs.~\ref{DphiVstdcond2}
and \ref{DphiVnonstd2}. Analogous results for $\lambda=0.8~\mathrm{mm}$ and $\lambda=3.0~\mathrm{mm}$ are shown in the Appendix in Figs.~\ref{DphiVstdcond1} and \ref{DphiVnonstd1}, and Figs.~\ref{DphiVstdcond3} and \ref{DphiVnonstd3}, respectively. All the results correspond to the stacked values for the three projections ($y-z$, $x-z$ and $x-y$) of Fig.~\ref{dphi_2} (Fig.~\ref{dphi_1} and Fig.~\ref{dphi_3} for $\lambda=0.8~\mathrm{mm}$ and $\lambda=3.0~\mathrm{mm}$). The contour lines corresponds to the 2D histogram contour levels of the distribution of the lines of sight at $10^5$ (yellow), $10^4$ (red), $10^3$ (purple), and $10^2$ (black) counts, while the color-coded background displays the mean column density for a given bin of $V_\phi$ and $|\Delta\phi|$. 
We present in Fig.~\ref{DphiVstdcond2} the distribution for the more standard conditions (no initial turbulence and $\theta=30^\circ$, namely simulations R1, R2, R3, R4 and R10), while Fig.~\ref{DphiVnonstd2} shows the same results but for the simulations including turbulence (R6, R8), or with a rotation axis perpendicular to the initial magnetic field orientation (R7, R9). In both cases, the results have been organized by blocks of decreasing level of magnetization (increasing $\mu$), and increasing level of rotation (increasing values of $\beta_\mathrm{rot}$). 
In all the simulations, most of the points in the map (contours higher or equal to $10^4$ counts) are characterized by low levels of $|\Delta \phi|$ (typically $\leq 10^\circ$) and low to moderate values of $V_\phi$, with low values of the mean column density. These points corresponds to regions that, other than the hourglass, have not been significantly disturbed by the infall, the turbulence, or the outflows and cover a great surface of the maps (yellowish regions of Fig.~\ref{Vphimaps} for example). 
Figures~\ref{DphiVstdcond2} and \ref{DphiVnonstd2} show a general behavior for points at low to moderate column densities ($\langle N \rangle \leq 10^{25}~\mathrm{cm^{-3}}$), where the discrepancy between the inferred magnetic field orientation and the actual mean orientation increases with the intrinsic degree of disorganization of the field lines and is consistent with the expected angular dispersion (see Fig.~\ref{Vphi2sigma}).

On the other hand, at higher column densities, the behavior differs from what is expected, displaying $|\Delta \phi|$ values higher than what can be attributed to the dispersion only. Particularly, points with high $|\Delta\phi|$ ($ \geq 30^\circ$) and low values of $V_\phi$ ($\leq 0.2$) are puzzling, but they remain extremely rare. Even the abundance of points with $|\Delta \phi| > 30^\circ$ and $V_\phi \leq 0.5$ remain extremely low at the three wavelengths (typically less than $0.01\%$, with only the very particular case of the simulation R7, with projection $y-z$ at $m_\mathrm{sink} = 0.3~\mathrm{M_\odot,}$ reaching $0.1\%$). These points can be only partially explained by a lower efficiency of the grain alignment (see Appendix~\ref{lowVhighdelta} for a comparison with the case of perfectly aligned dust grains), a higher opacity of the layers surrounding the protostar ---which only allows us to recover the foreground information---, and a dust temperature gradient, which are not included in the computation of the mean orientation of the magnetic field. 
To assess whether or not the opacity is responsible of the lower recovery rates, we stack the maps at $\lambda=0.8~\mathrm{mm}$ for all the simulations, time-steps, and projections and compute the recovery rate in three opacity ranges. For all the lines of sight with $\tau_\mathrm{0.8~mm} \geq 1,$ we obtain a recovery rate of only $42.5\%$. For $\tau_\mathrm{0.8~mm}$ in the range $0.1-1$, we obtain a recovery rate of $79.4\%$, while for $\tau_\mathrm{0.8~mm}$ in the range $0.01-0.1$ the recovery rate reaches $81.7\%$. This clearly shows that the reconstructed $B$-fields of regions with high column densities most likely suffer from opacity effects, and that these regions might benefit from multi-wavelength observations. 
\\
Figure~\ref{DphiVstdcond2} suggests that, in otherwise similar initial conditions ($\beta=0.04$, $\mathcal{M}=0$, $\theta=30^\circ$), the more magnetized case (R2) has initially (at $m_\mathrm{sink} = 0.1~\mathrm{M_\odot}$) higher levels of $|\Delta\phi|$, but as the simulations evolve, the overall levels of discrepancy become comparable (similar contour levels). 
Nevertheless, the most puzzling measurements being the sight lines where the magnetic field is relatively well organized (low variance), but displaying large discrepancies (high $|\Delta\phi|$), are more important for the least magnetized simulation (R3) at all evolutionary stages  (shown in Fig.~\ref{DphiVstdcond2}), and they are mostly associated with high-column-density lines of sight which correspond to the very small radii (disk scales) where radiative transfer effects cannot easily be neglected, even at mm wavelengths.\\ 
For the nonstandard configurations shown in Fig.~\ref{DphiVnonstd2}, those including turbulence
show relatively similar levels of discrepancy (simulations R6 and R8), indicating that the relative importance of the magnetization level is reduced in the presence of turbulence. Nevertheless, the presence of turbulence does not seem to be the main driver of the low recovery rates. For instance, Table~\ref{table:turb} shows that simulations R1 and R6 display similar recovery rates, within the error bars, at different scales.\\ 
On the other hand, simulations with $\theta=90^\circ$ (R7 and R9) show an increased importance of the magnetization level. Simulation R9 (the least magnetized one) exhibits a relatively peculiar behavior, where although the overall $|\Delta\phi|$ values remain low, it develops a region where the discrepancy is maximal (even for values of $V_\phi$ lower than $0.2$), reaching values closer to $90^\circ$.
\\
Figure~\ref{DphiVstdcond2} also shows that the accuracy of polarized dust emission maps in tracing the magnetic field topology in protostars has weak and rather unclear dependence on the initial values of the rotational energy $\beta_\mathrm{rot}$ in the more magnetized case ($\mu=3.33$, first three rows), displaying overall a larger level of discrepancy between the intrinsic magnetic field mean orientation and the reconstructed orientation from the polarized dust emission for the simulation with larger rotational energy (R2), and the lowest one for the simulation with intermediate rotational energy level (R1).

\begin{figure*}
\centering
\begin{tikzpicture}
\node[above right] (img) at (0,0) {
  \includegraphics[height=0.27\textwidth, trim={2.2cm 1.5cm 4.8cm 1.6cm},clip]{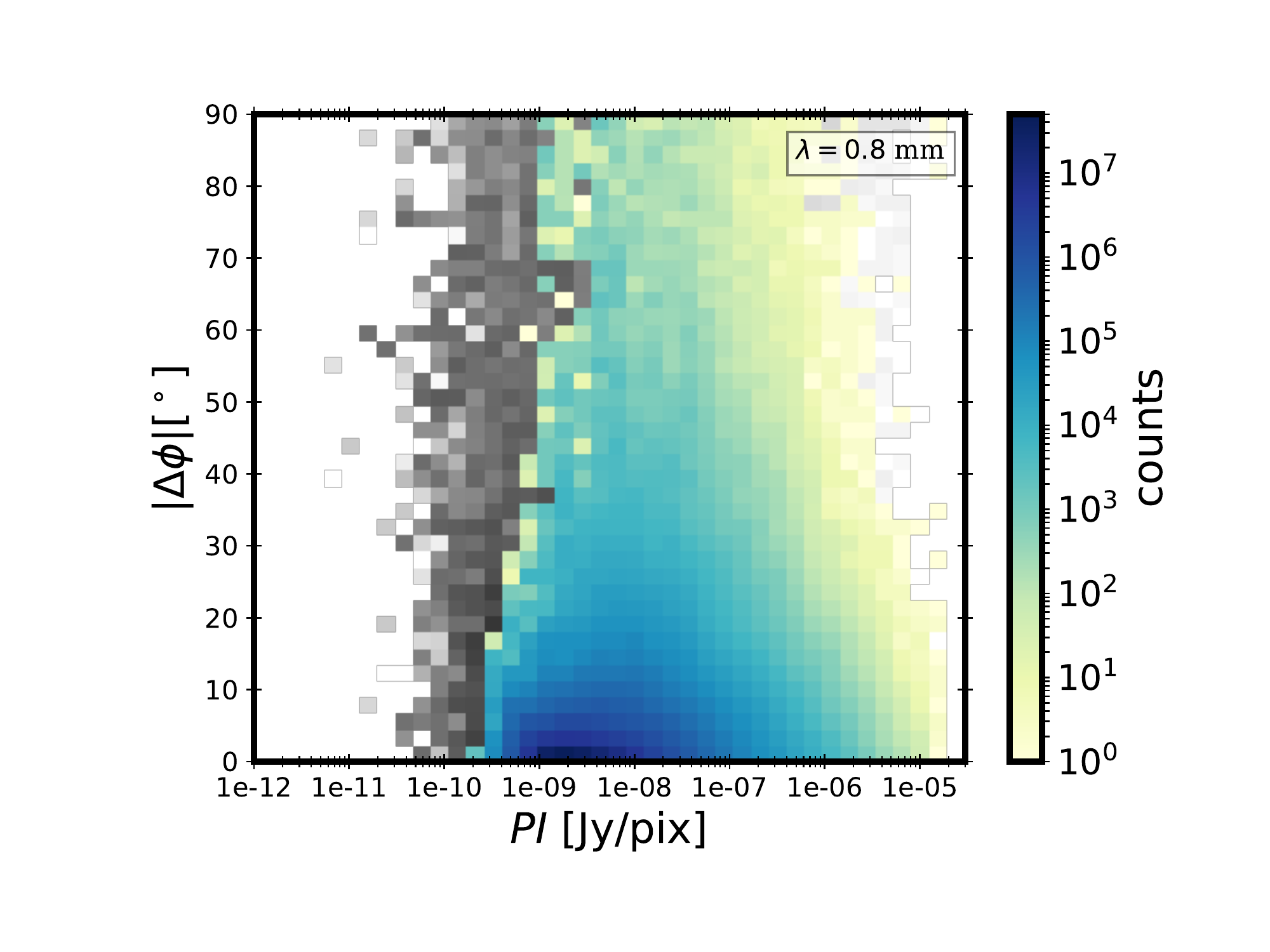}
  \includegraphics[height=0.27\textwidth, trim={2.2cm 1.5cm 4.8cm 1.6cm},clip]{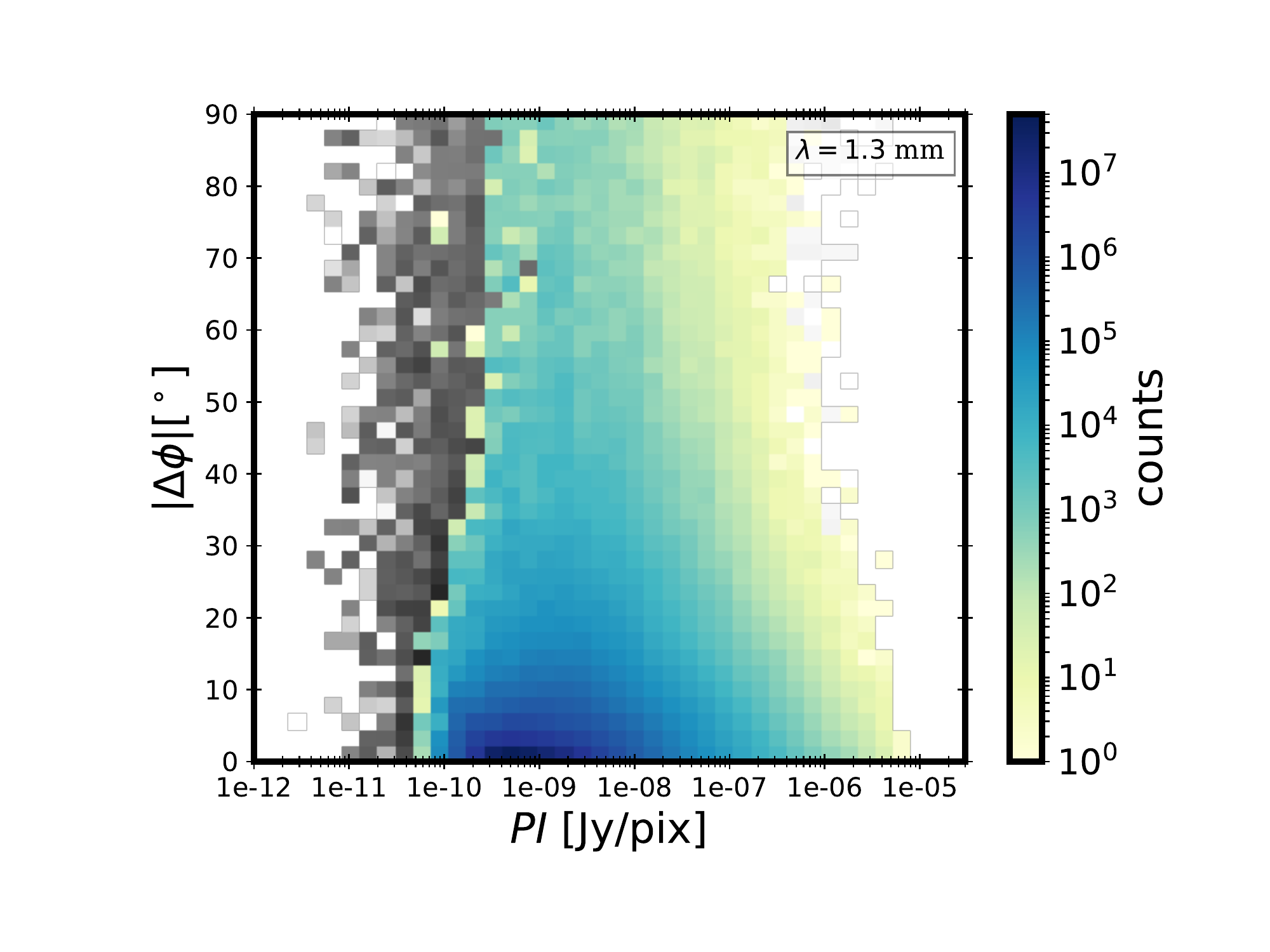}
  \includegraphics[height=0.27\textwidth, trim={2.2cm 1.5cm 1.cm 1.6cm},clip]{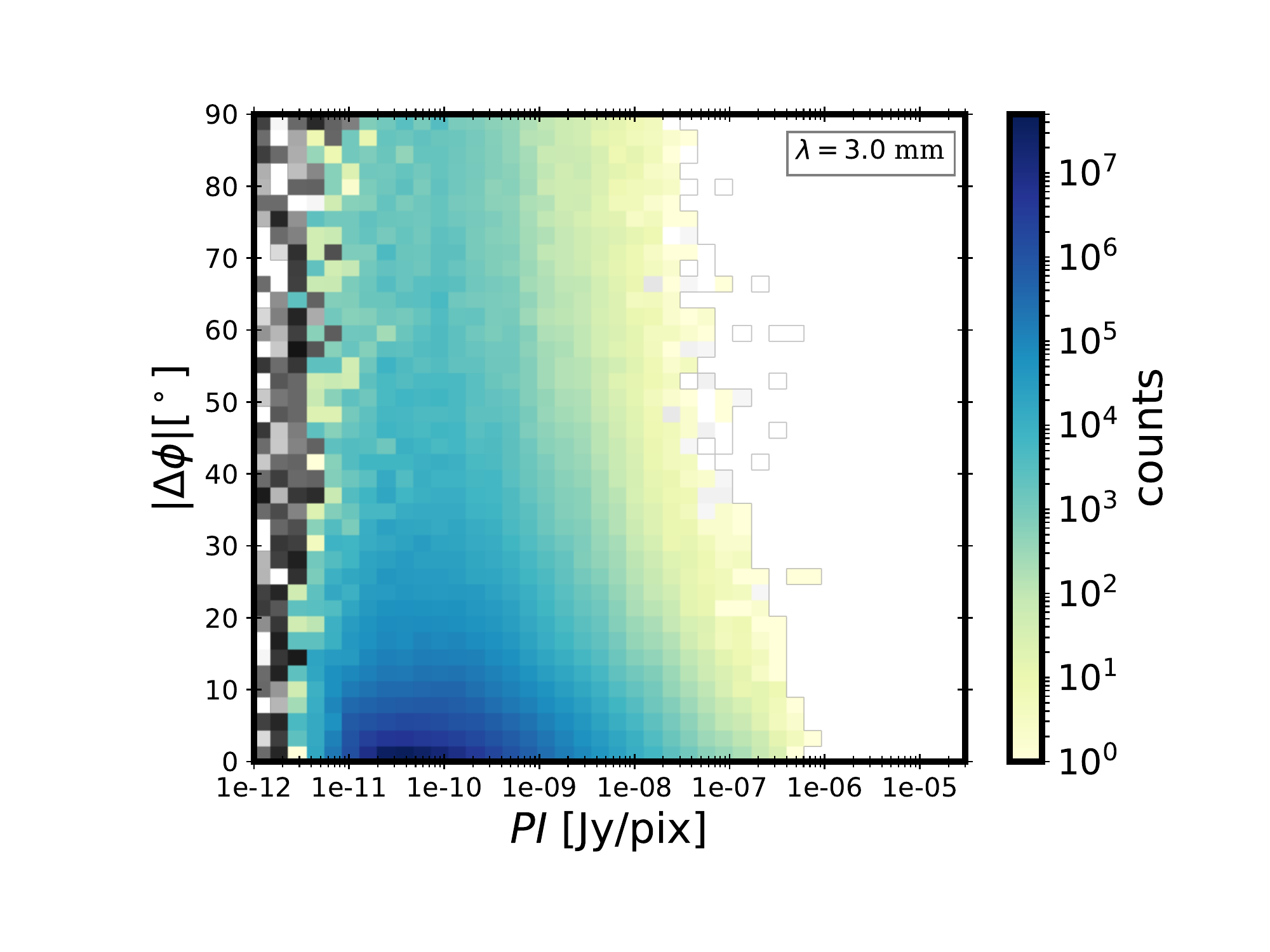}   
  };
\end{tikzpicture}  
\caption{Two-dimensional histograms of the discrepancy between the mean $B$-field orientation and the  orientation inferred from the polarized dust emission, and the intensity of the polarized dust emission at three wavelengths. From left to right $\lambda=0.8, 1.3$, and $3.0~\mathrm{mm}$. Each histogram has been constructed using the data stacked for all simulations, time-steps, and projections. The color histograms show the results using only the lines of sight that have a polarization fraction above $1\%$, while the gray-scale part corresponds to the rest of the points.}
\label{hist2d_dphi_poli}
\end{figure*}

\begin{figure*}
\centering
\begin{tikzpicture}
\node[above right] (img) at (0,0) {
  \begin{tabular}{@{}l@{}}
  \includegraphics[height=0.27\textwidth, trim={1.9cm 1.5cm 4.8cm 1.6cm},clip]{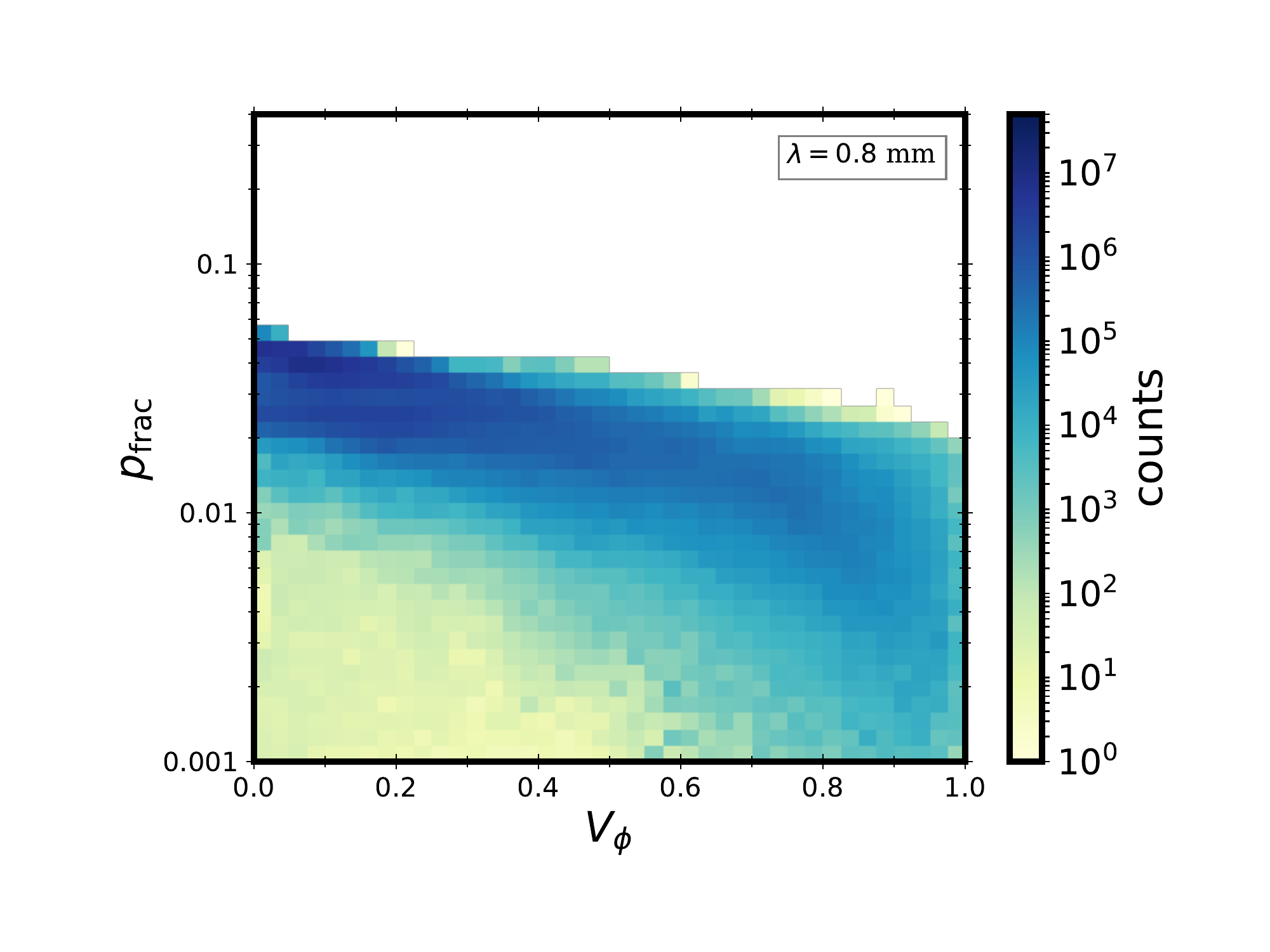}
  \includegraphics[height=0.27\textwidth, trim={2.5cm 1.5cm 4.8cm 1.6cm},clip]{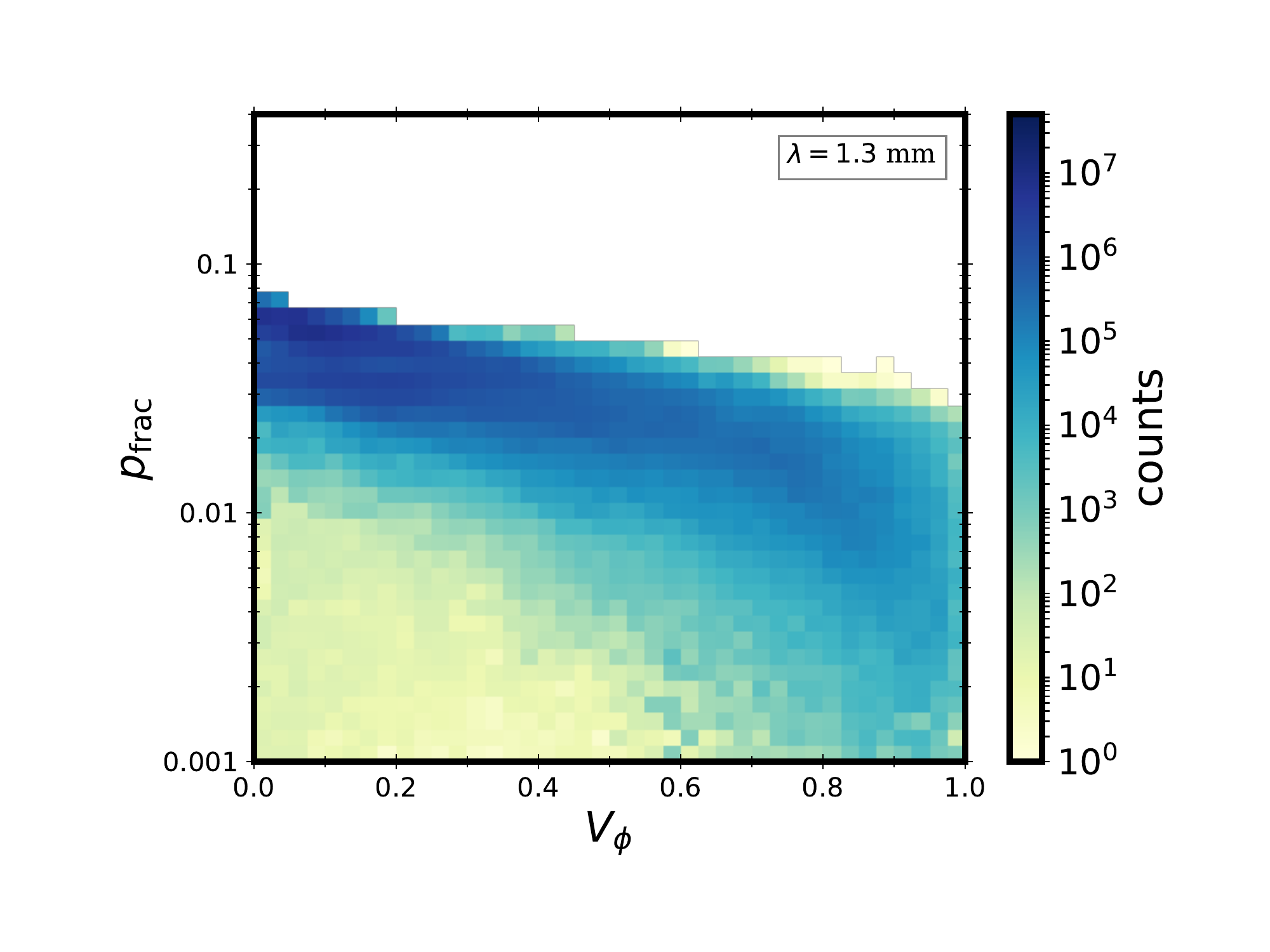}
  \includegraphics[height=0.27\textwidth, trim={2.5cm 1.5cm 1.cm 1.6cm},clip]{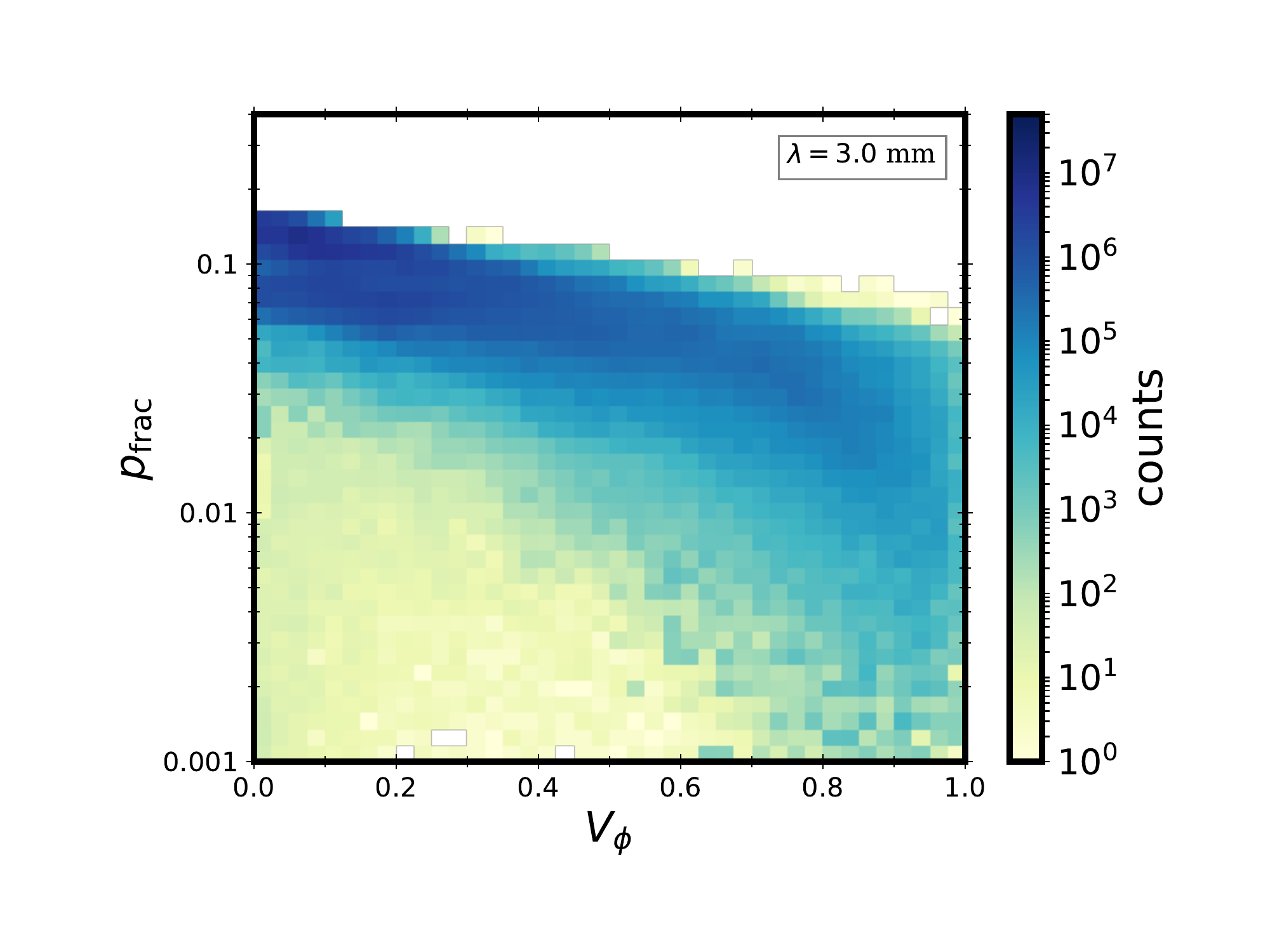}\\
    \includegraphics[height=0.27\textwidth, trim={1.9cm 1.5cm 4.8cm 1.6cm},clip]{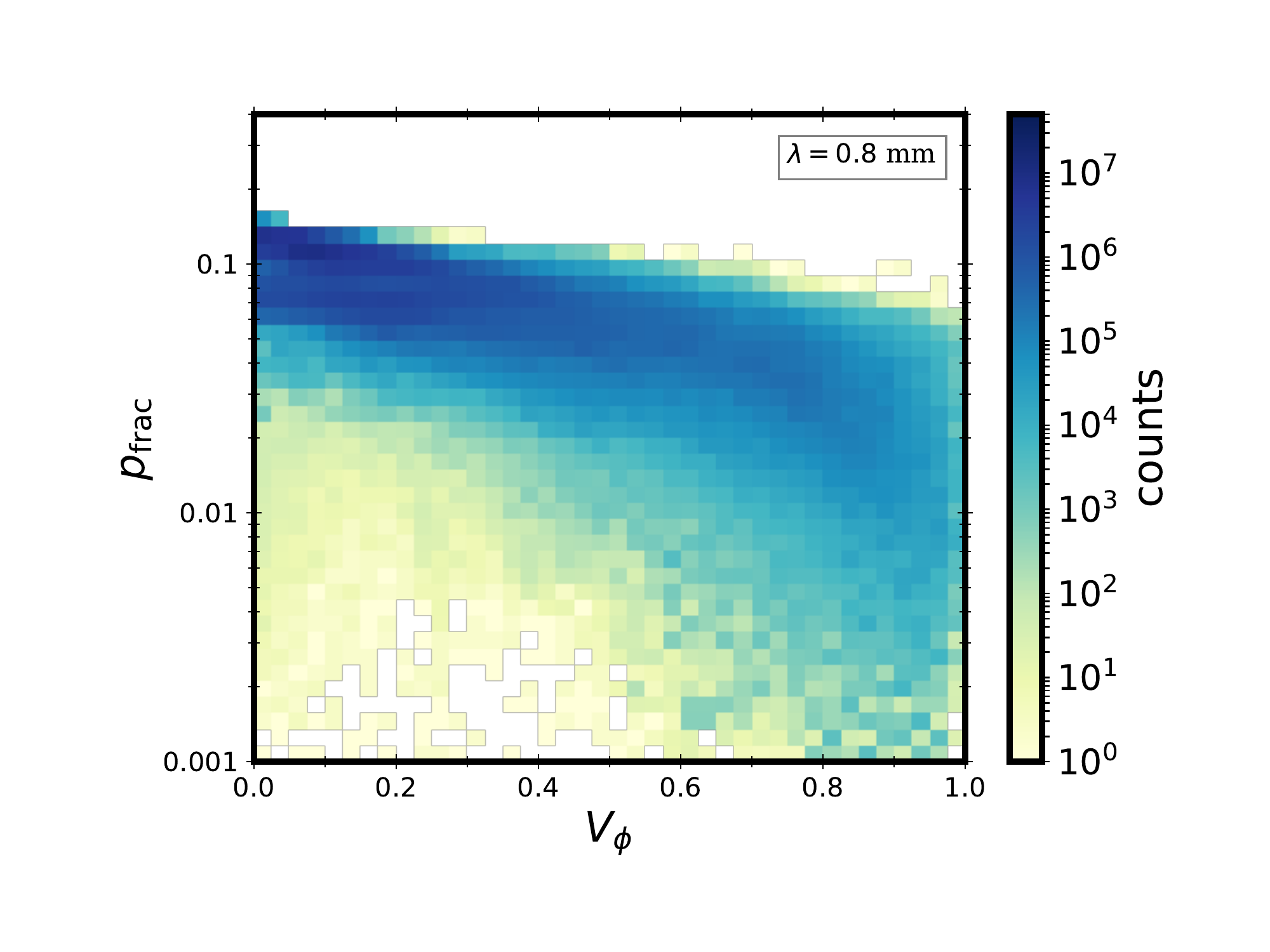}
  \includegraphics[height=0.27\textwidth, trim={2.5cm 1.5cm 4.8cm 1.6cm},clip]{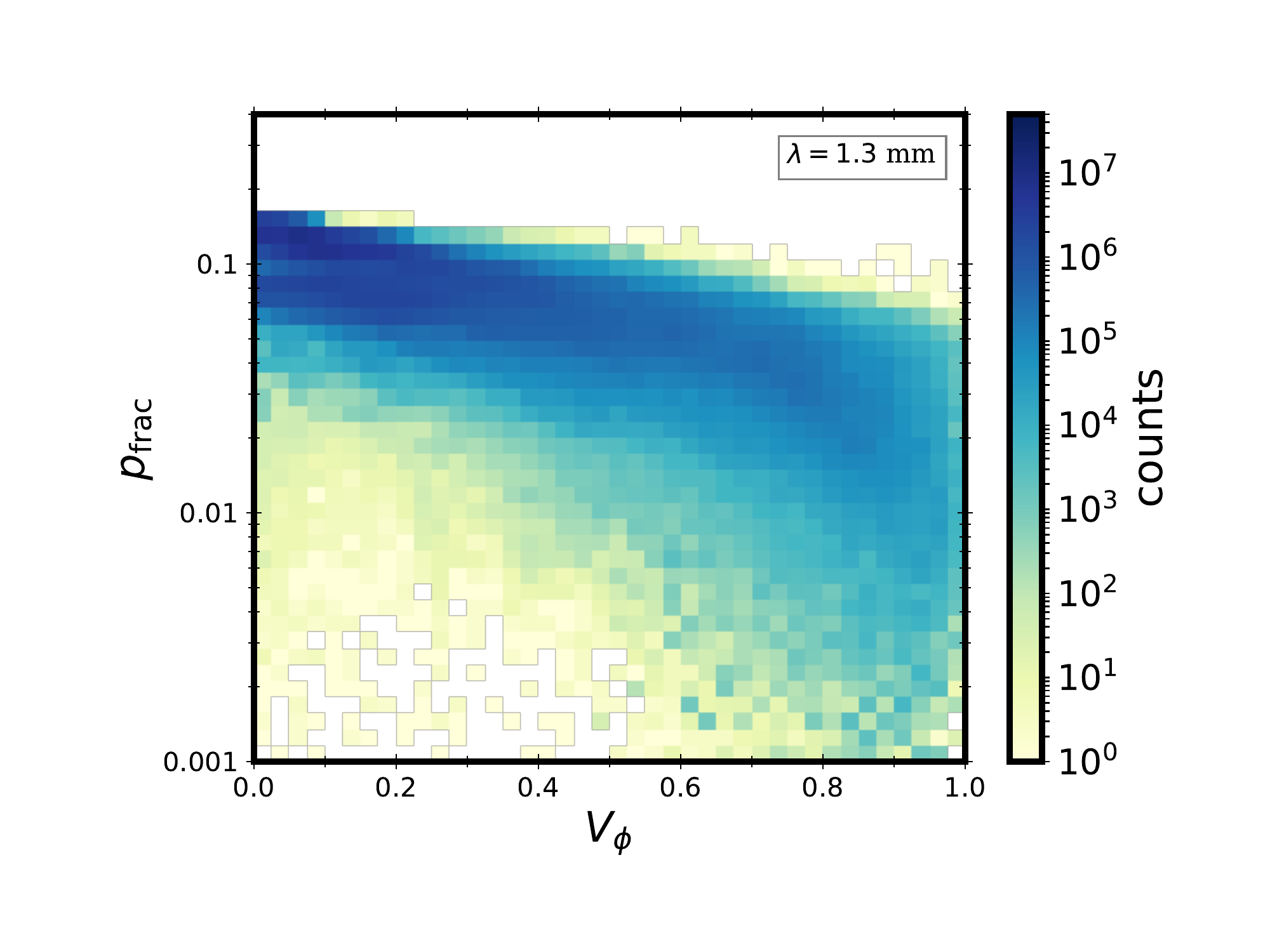}
  \includegraphics[height=0.27\textwidth, trim={2.5cm 1.5cm 1.cm 1.6cm},clip]{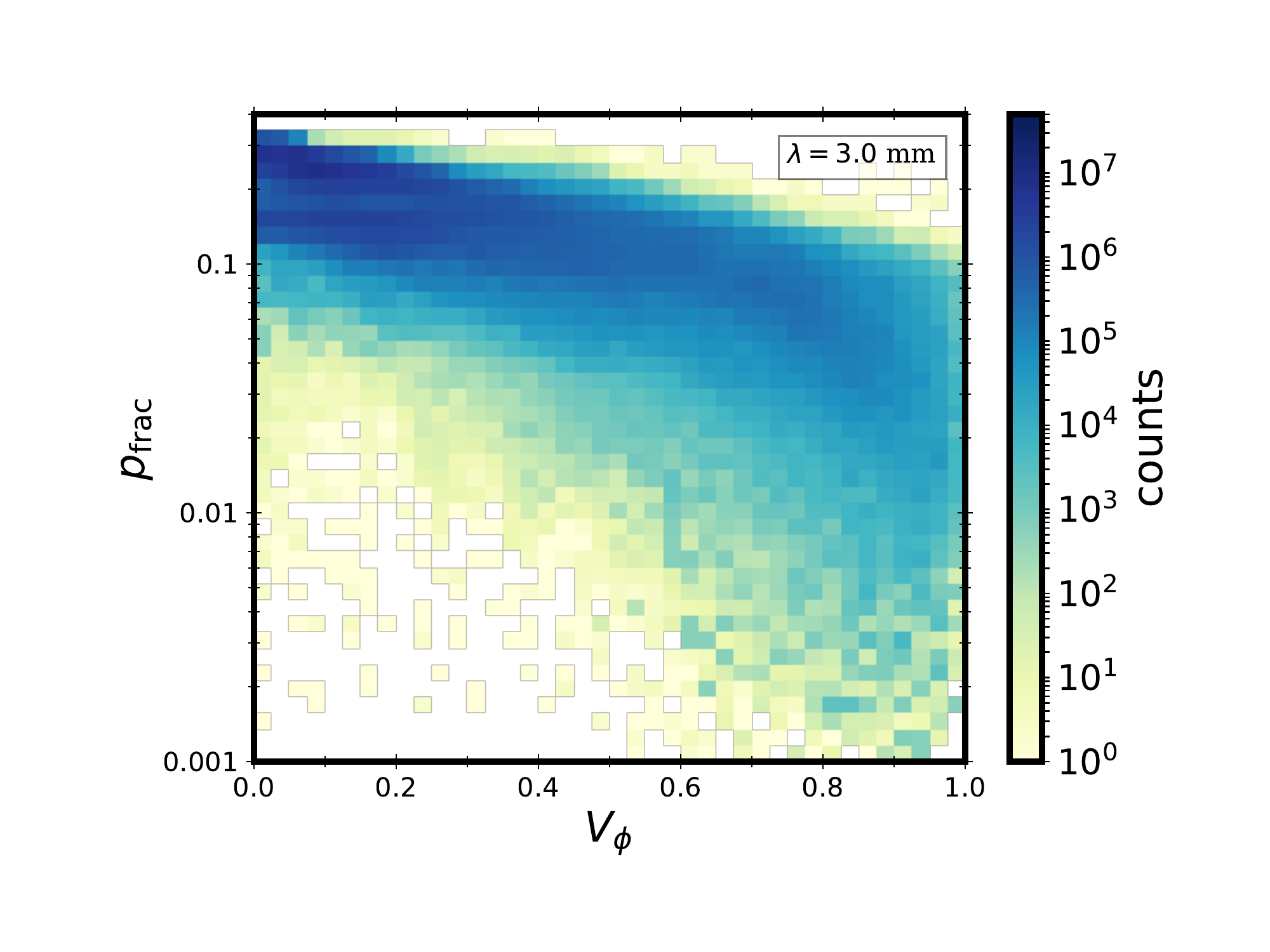}  
  \end{tabular}
  };
    \node [rotate=0, scale=0.8] at (2.0,  9.8) {\fontfamily{phv}\selectfont B-RAT+II+IDG};
    \node [rotate=0, scale=0.8] at (1.8,  4.7) {\fontfamily{phv}\selectfont PA};
\end{tikzpicture}  
\caption{Two-dimensional histograms of the polarization fraction as a function of the circular variance for all the stacked maps at three wavelengths (from left to right $\lambda=0.8, 1.3$ and $3.0~\mathrm{mm}$). The top row shows the results for the case where the alignment is computed according to the B-RAT theory along with the II and IDG mechanisms (B-RAT+II+IDG), as described in Section~\ref{dustalig}, while the bottom row shows the results for the case where dust grains are assumed to be perfectly aligned (PA) with the magnetic field.}
\label{hist2d_pfrac_V}
\end{figure*}

\subsection{Depolarization and magnetic field organization}
Finally, to better understand the influence of the magnetic field topology on the polarization fraction, we compared the distribution of the polarization fraction ($p_\mathrm{frac}$) as a function of the circular variance ($V_\phi$) obtained for our synthetic maps to a test case.  
In the case of our synthetic maps, the alignment was computed according to the B-RAT theory, including the imperfect internal alignment and the imperfect Davis-Greenstein mechanisms, as described in Sect.~\ref{dustalig} (top row of Fig.~\ref{hist2d_pfrac_V}, labeled as B-RAT+II+IDG), while for the test case the alignment is imposed in such a manner that all the dust grains susceptible to align are perfectly aligned with the $B$-field lines (bottom row of Fig.~\ref{hist2d_pfrac_V}, labeled as PA). Fig.~\ref{hist2d_pfrac_V} shows 2D histograms of $p_\mathrm{frac}$ versus $V_\phi$ by stacking the results of all the simulations, time-steps, and projections at three wavelengths. This figure shows a clear anti-correlation between these two quantities, indicating that the degree of polarization decreases with an increasing level of disorganization of the magnetic field along the line of sight, which is consistent with the anticorrelation between the polarization fraction and the dispersion of the position angles found in ALMA observations of Class 0 objects \citep{LeGouellec2020}. \\ 
The fact that both cases display the same behavior supports the idea that a geometrically depolarized signal due to the nonorganized component of the magnetic field could be greatly responsible for the lower levels of polarization observed in denser regions down to the scales of protostellar envelopes, which is  in agreement with observations at larger scales \citep{Doi2020, PlanckXIX2015, PlanckXX2015}. However, while at large scales the turbulence of the gas may be the main physical cause of the magnetic field disorganization, inside protostellar cores it seems the gravitational field and the launching of protostellar outflows are more likely responsible, as models with turbulence do not show significantly more disorganized $B$-field, nor poorer fidelity, than models without, as shown in Table~\ref{table:turb}. \\

 \subsection{Significance of other polarization-producing processes in protostellar envelopes}
 
Our analysis relies on the hypothesis that dust grains imperfectly align with the local magnetic field  {in a way that reflects the }local radiation conditions (B-RATs) and grain properties. 
Dedicated studies of other physical effects producing polarization of the dust thermal emission, along with better dust models, are necessary to shed light on the precise conditions in which polarization seen at mm wavelengths could trace other processes. However, here we reinforce our hypothesis by recalling a few of them, and give simple qualitative arguments as to why they are not expected to be dominating in typical conditions reigning in protostellar envelopes. 

While at the protostar disk scales the alignment of dust grains with local magnetic field lines can become more difficult to realize because of the higher rate of collisions \citep{Tazaki2017}, the gaseous damping timescale is inversely proportional to gas density and gas temperatures. Hence, such loss of efficiency is not expected in protostellar envelopes where densities are lower than $10^6~\mathrm{cm^{-3}}$ and temperatures are quite low ($T<20~\mathrm{K}$). 
It has also been suggested that in conditions of high opacity and anisotropic radiation field, usually met in disks, the dust self-scattering of the radiation field from large grains ($\sim 100~\mathrm{\mu m}$) could dominate the polarized signal at mm wavelengths (\citealt{Kataoka2015, Kataoka2017}, see also \citealt{Kirchschlager2020, Brunngraber2020}). Despite the indications of large grains at the protostellar scales probed in our analysis, not only are the conditions not met to produce significant polarization from self-scattering, but the levels of observed polarization in embedded protostars (a few percent to a few tens of percent) cannot be produced with self-scattering polarization.

A highly anisotropic radiation field could reduce the radiative precession timescale to values shorter than the Larmor timescale, inducing polarization of dust thermal emission due to the k-RAT mechanism,  which is related to the radiation anisotropy and not to the magnetic field. This mechanism is favored for the large grains (for a discussion, see e.g. the work of \citealt{Pattle2021} in the Orion Source I), and necessitate extremely irradiated conditions at wavelengths similar to the dust grain sizes, which are not typical of embedded low-mass protostars (Le Gouellec et al. in prep). 

Finally, mechanical alignment of dust grains with the local gas flow has been proposed as an extra mechanism to produce polarized dust emission from irregular grains \citep{Hoang2018}, with a rotation axis parallel to the velocity drift direction. While this mechanism could in principle be at work for example around massive protostars where radiative pressure drives the outflow \citep{Mignon2021}, its conditions seem very sensitive to the flow properties (high velocities to enhance the alignment efficiency, but not as high as to disrupt the dust grains) and quite unable to explain the large range of observations showing polarized dust emission in protostars. We also stress that no convincing observational signature of such alignment process has yet been found in protostellar environments \citep{Cortes2021,Aso2021}.

\section{Conclusions}

In this paper, we present a thorough investigation of the fidelity and limitations of using dust polarized emission as a tracer of the magnetic field topologies in the dense regions of protostellar cores forming solar type stars. 
We focus here mostly on quantifying the errors induced by \textit{(i)}~widely varying physical conditions along each individual line of sight used to probe the dust grain emission of the protostellar objects, and \textit{(ii)}~its potentially devastating combination with the averaging effects of complex $B$-field line orientations along those same lines of sight.

To assess the importance of these effects, we performed an analysis of magnetic field properties in 27 realizations of MHD models following the evolution of physical properties in star-forming cores.
Assuming a uniform population of oblate dust grains whose sizes follow the standard MRN distribution \citep{Mathis1977}, we produced synthetic polarized dust emission maps for dust grains aligned with the local $B$-field thanks to B-RATs mechanism. We provide a detailed comparison of these synthetic maps and their relative models. 

We find that, in most cases, \mbox{(sub-)mm} polarized dust emission is a robust tracer of the magnetic field topologies in inner protostellar envelopes and is successful at capturing the details of the magnetic field spatial distribution down to radii $\sim 100$ au. 
Measurements of the line-of-sight-averaged magnetic field line orientation using the polarized dust emission are precise to $<15\deg$ (typical of the error on polarization angles obtained with observations from large mm polarimetric facilities such as ALMA) in about $95 \%$ of the independent lines of sight that pass through protostellar envelopes at all radii. 
When focusing on the smaller envelope radii $<500~\mathrm{au}$, where the magnetic field lines are more likely perturbed from the initially smooth configuration by infall and outflow, $75 \%$ of the lines of sight still give robust results.

Large discrepancies between the integrated $B$-field mean orientation and the orientation reconstructed from the polarized dust emission are mostly observed in lines of sight probing large column densities. Our analysis shows that, at disk scales, physical conditions producing thermal
dust emission with high opacity are mostly responsible for the small fraction of mediocre measurements (inaccuracy of $50\%$ or more on the recovered $B$-field position angle at opacities higher than $1~\mathrm{mag}$). While the use of longer wavelengths allows a slightly better recovery of the true orientation of the magnetic fields, our study suggest that, at those scales, most measurements of magnetic field line orientation will suffer from large ($>30\deg$) error bars.
Discrepancy is also found associated to a small fraction of lines of sight where the magnetic field is highly disorganized, such as outflow cavity walls and accretion shocks: in such lines of sight the concept of integrated $B$-field mean orientation does not make sense, and no good measurement can therefore be done.
We find that the polarization fraction is correlated with the degree of organization of the magnetic field along the line of sight, confirming that the behavior observed at larger scales in less dense gas holds true down to the scales of the envelopes of protostars. Hence, low levels of polarization fraction in the observations could be used as a flag to avoid these measurements affected by large errors.

\begin{acknowledgements}
We thank M. K\"uffmeier for constructive comments to improve the readability of this paper. 
This research has received funding from the European Research Council (ERC) under the European Union's Horizon 2020 research and innovation programme (MagneticYSOS project, Grant Agreement No 679937).
\end{acknowledgements}

%
%

\bibliographystyle{aa} 
\bibliography{biblio_val} 

\appendix

\section{Additional material}

\begin{figure*}
\centering
      \includegraphics[width=0.99\textwidth, trim={2.65cm 1.3cm 2.72cm 4.1cm},clip]{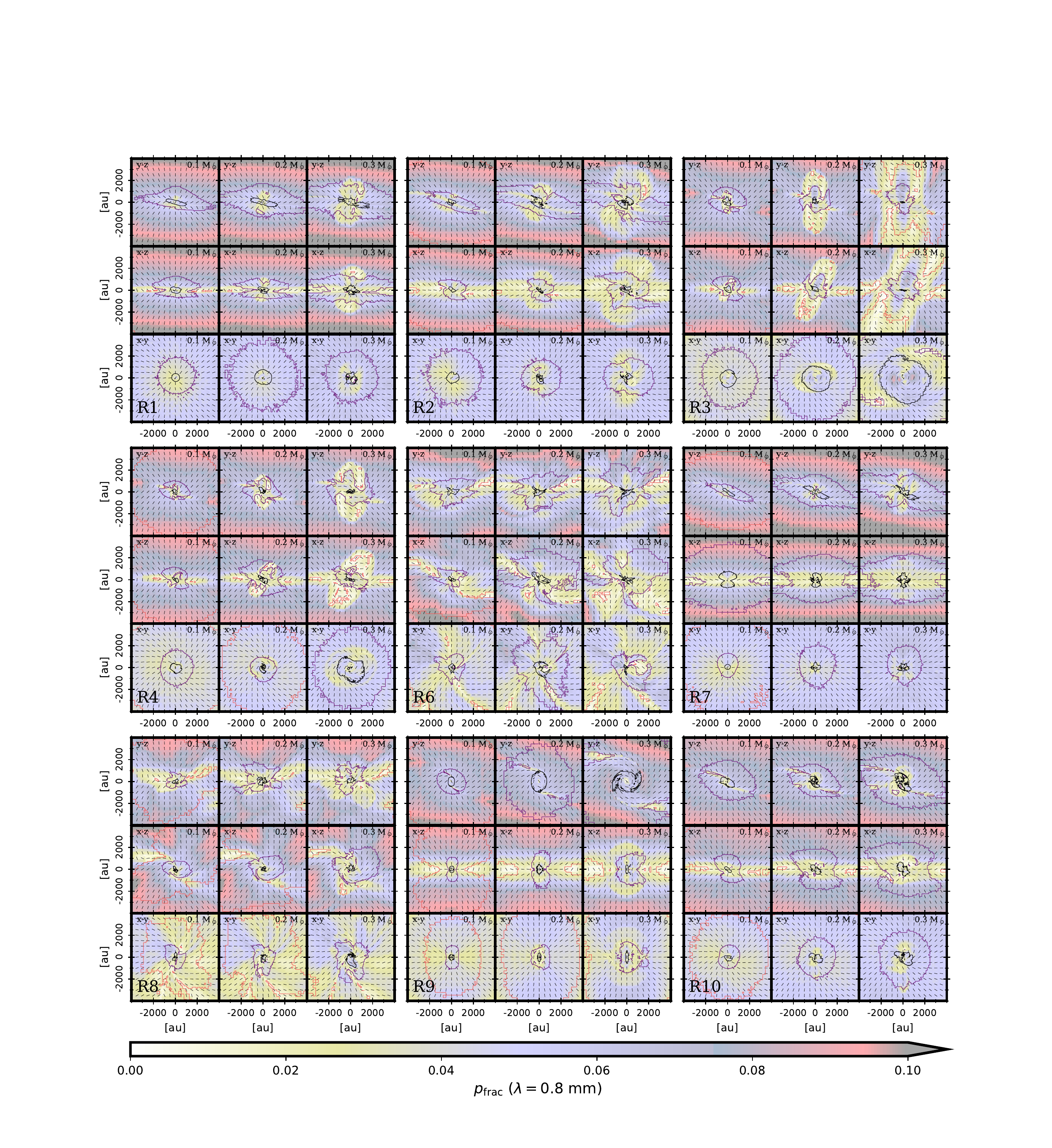}
\caption{Results from the synthetic observation at $\lambda = 0.8~\mathrm{mm}$ showing the polarization fraction (background image), the polarized intensity contours, and the inferred magnetic field orientation vectors. The contours correspond to the linearly polarized intensity normalized to the peak value at levels $10^{-2}$ (black), $10^{-3}$ (purple), $10^{-4}$ (red), and $10^{-5}$ (yellow). Higher resolution figures for individual simulations are available online.}
\label{pfrac_maps_1}
\end{figure*}

\begin{figure*}
\centering
      \includegraphics[width=0.99\textwidth, trim={2.65cm 1.3cm 2.72cm 4.1cm},clip]{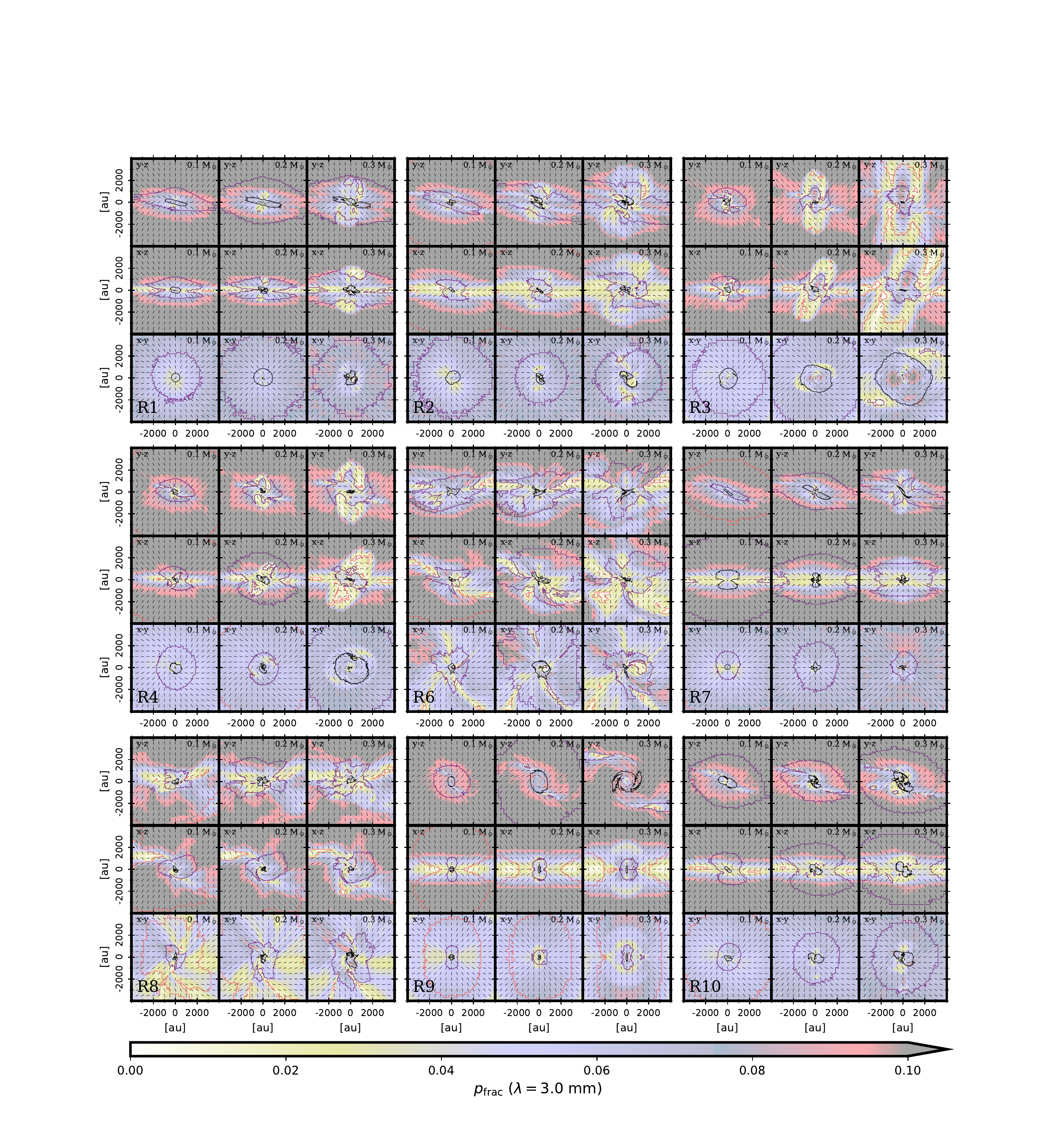}
\caption{Results from the synthetic observation at $\lambda = 3.0~\mathrm{mm}$ showing the polarization fraction (background image), the polarized intensity contours, and the inferred magnetic field orientation vectors. The contours correspond to the linearly polarized intensity normalized to the peak value at levels $10^{-2}$ (black), $10^{-3}$ (purple), $10^{-4}$ (red), and $10^{-5}$ (yellow). Higher resolution figures for individual simulations are available online.}
\label{pfrac_maps_3}
\end{figure*}

\begin{figure*}
\centering
\begin{tikzpicture}
\node[above right] (img) at (0,0) {
  \begin{tabular}{@{}l@{}}
  \includegraphics[width=0.9\textwidth, trim={1.2cm 1.65cm 2.2cm 1.7cm},clip]{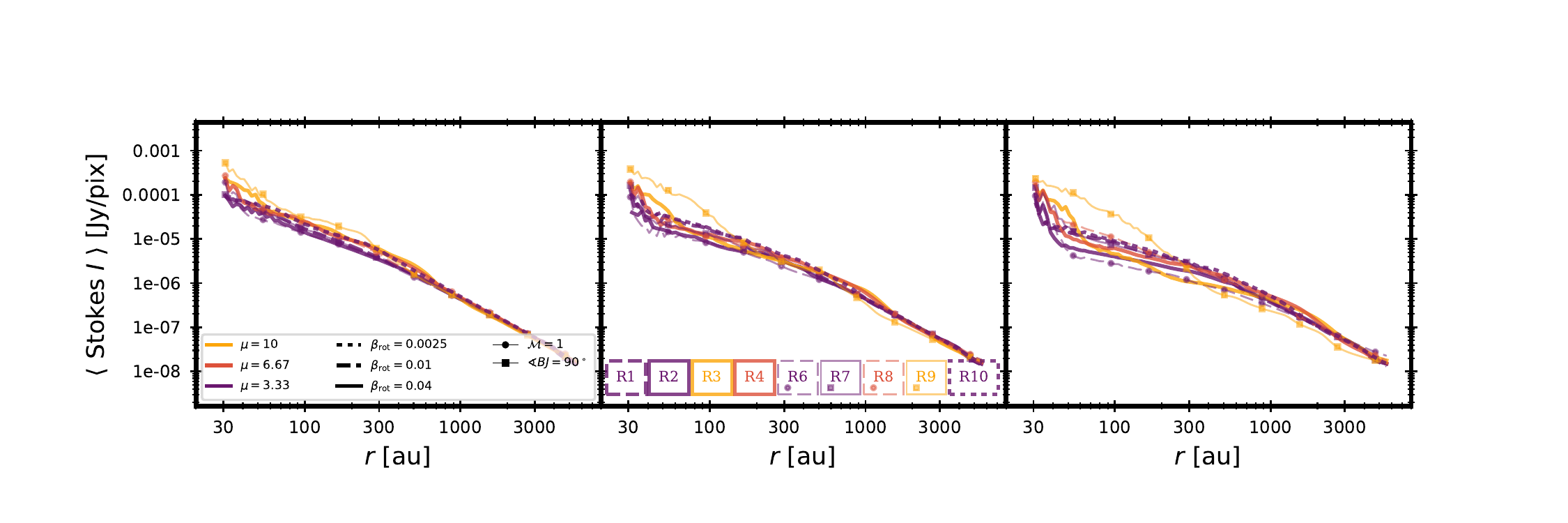}\\
  \includegraphics[width=0.9\textwidth, trim={1.2cm 1.65cm 2.2cm 1.7cm},clip]{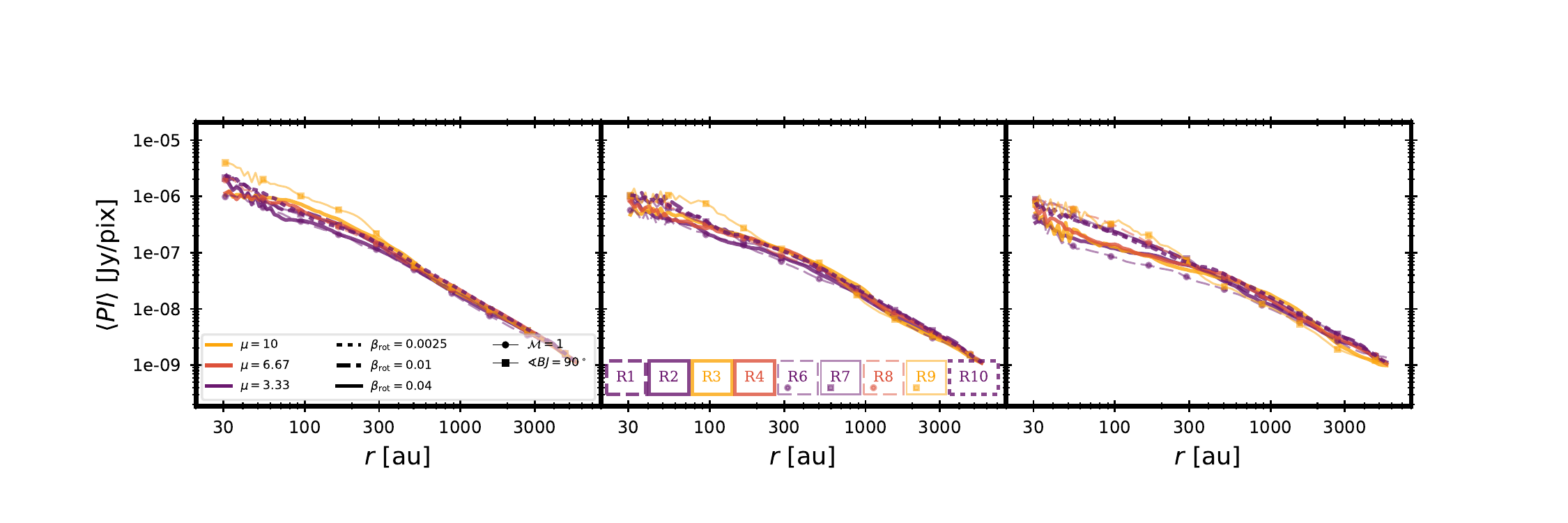}\\
  \includegraphics[width=0.9\textwidth, trim={1.2cm 0.8cm 2.2cm 1.7cm},clip]{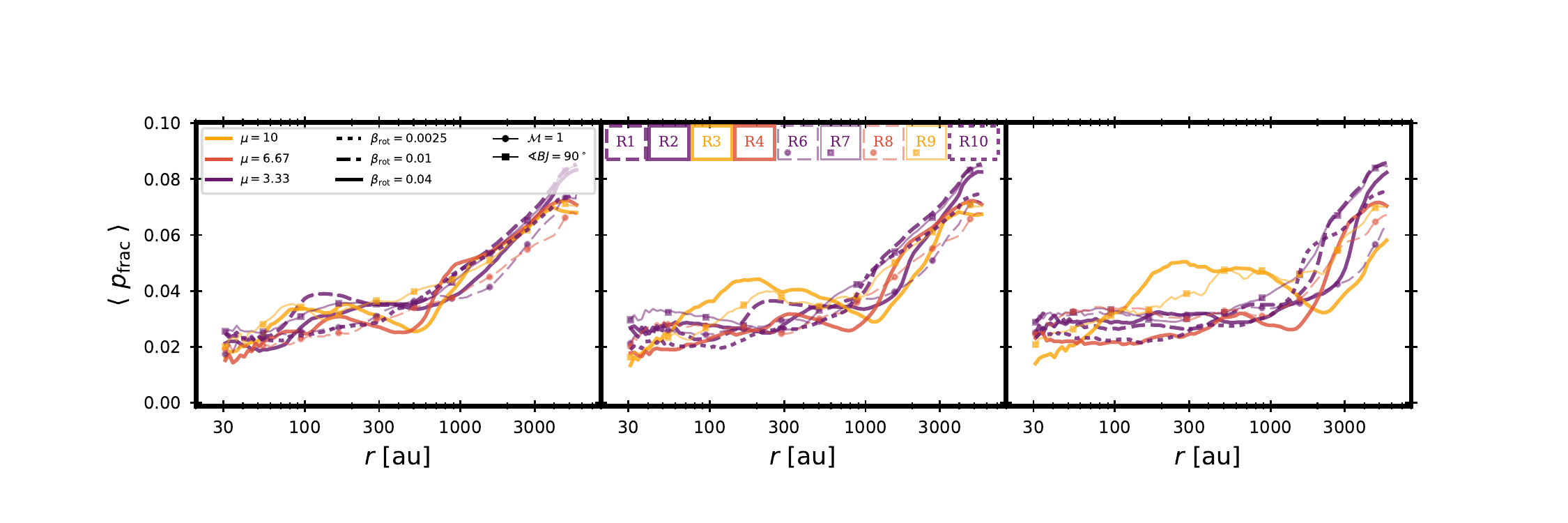}   
  \end{tabular}};
\node [rotate=0, scale=0.8] at (5.4,  11.5) {\fontfamily{phv}\selectfont $0.1~\mathrm{M_\odot}$};
\node [rotate=0, scale=0.8] at (10.1, 11.5) {\fontfamily{phv}\selectfont $0.2~\mathrm{M_\odot}$};
\node [rotate=0, scale=0.8] at (15.0, 11.5) {\fontfamily{phv}\selectfont $0.3~\mathrm{M_\odot}$};        
\end{tikzpicture}  
\caption{Radial profiles of the synthetic observation at $\lambda=0.8~\mathrm{mm}$. From top to bottom: Total dust emission (Stokes $I$), total linearly polarized dust emission ($PI$), and polarization fraction ($p_\mathrm{frac}$).}
\label{radial_Stokes_RAT_1}
\end{figure*}

\begin{figure*}
\centering
\begin{tikzpicture}
\node[above right] (img) at (0,0) {
  \begin{tabular}{@{}l@{}}
  \includegraphics[width=0.9\textwidth, trim={1.2cm 1.65cm 2.2cm 1.7cm},clip]{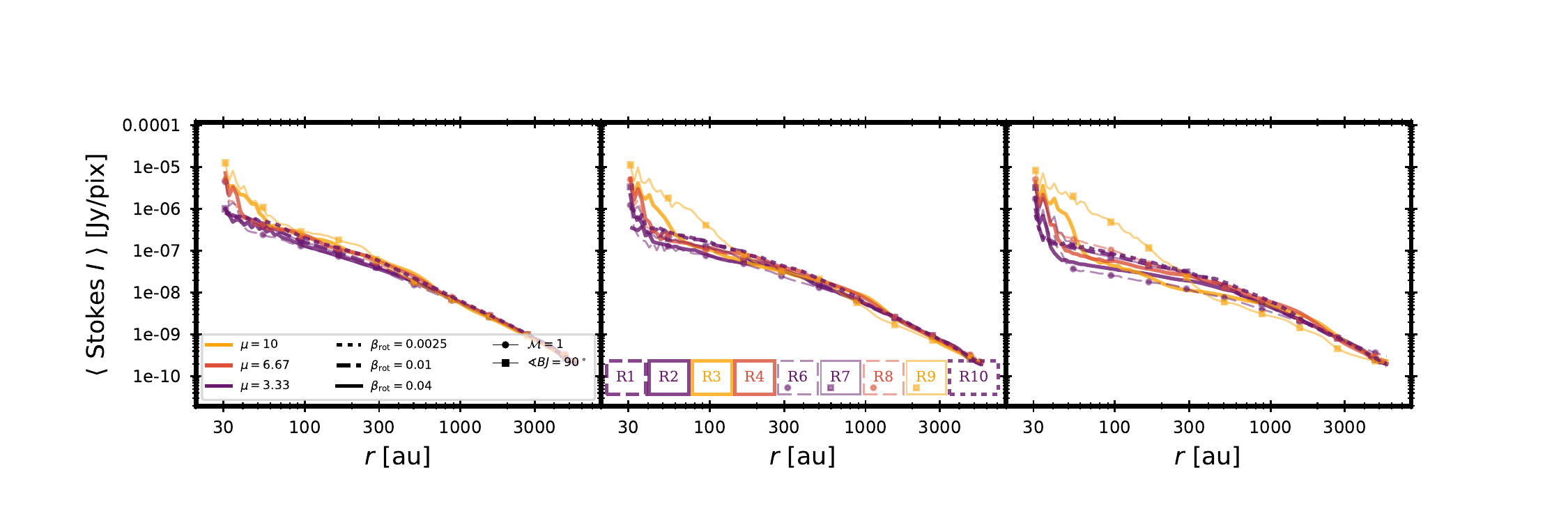}\\
  \includegraphics[width=0.9\textwidth, trim={1.2cm 1.65cm 2.2cm 1.7cm},clip]{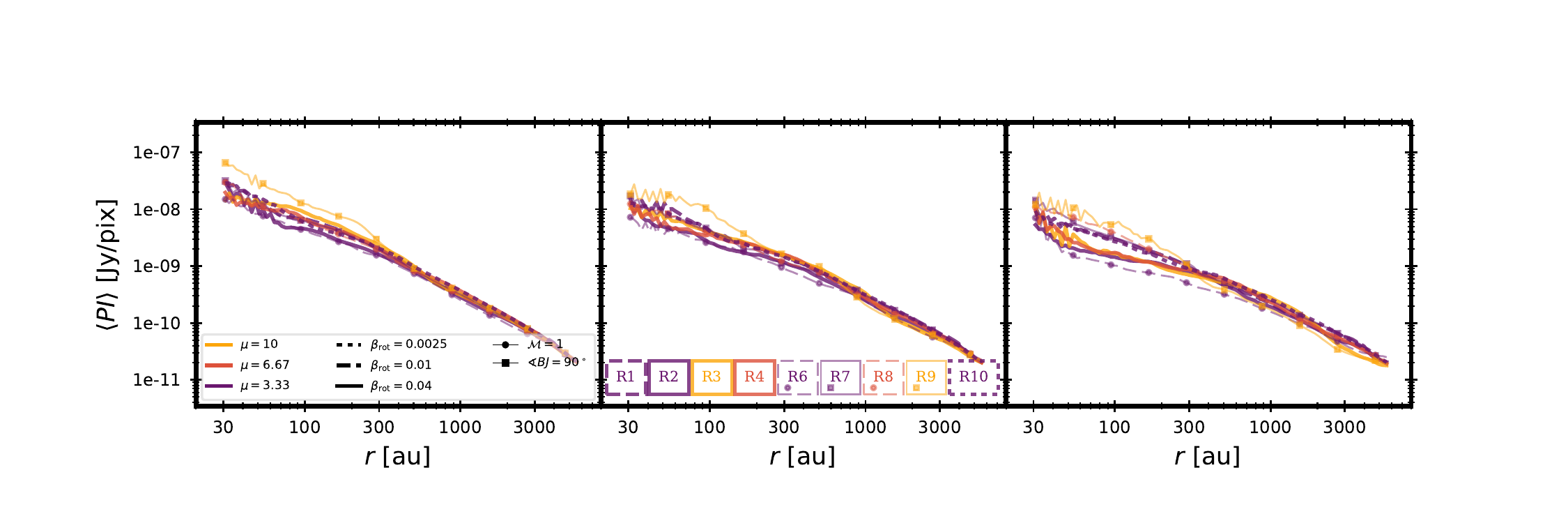}\\
  \includegraphics[width=0.9\textwidth, trim={1.2cm 0.8cm 2.2cm 1.7cm},clip]{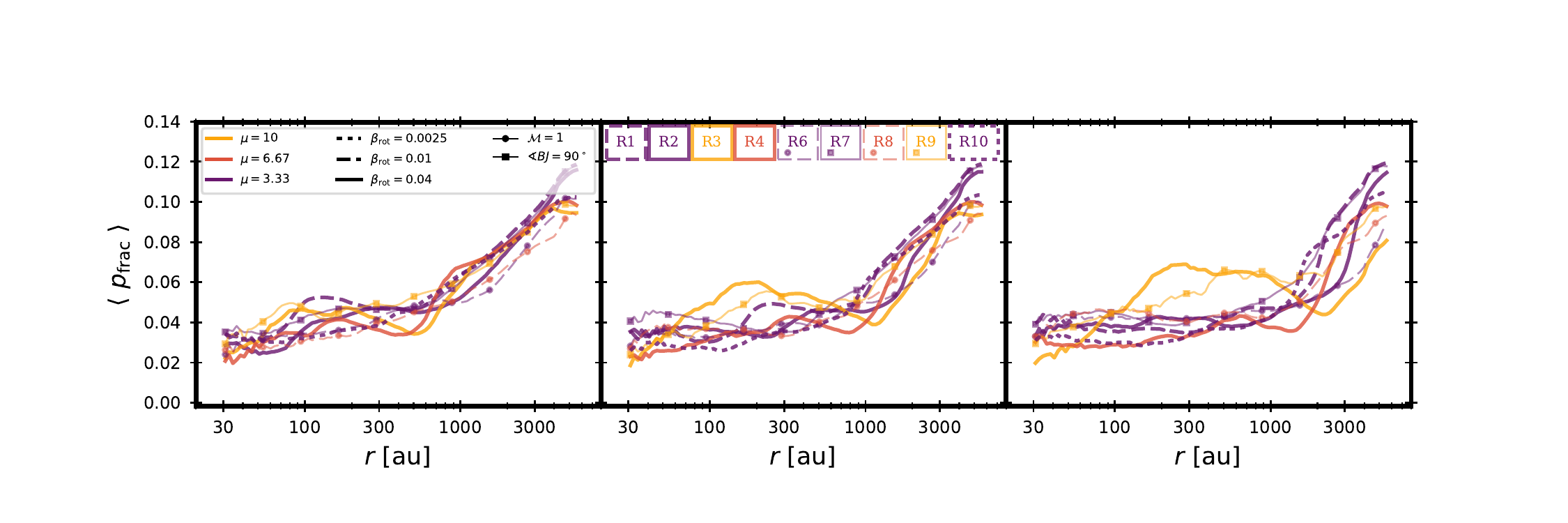}   
  \end{tabular}};
\node [rotate=0, scale=0.8] at (5.4,  11.5) {\fontfamily{phv}\selectfont $0.1~\mathrm{M_\odot}$};
\node [rotate=0, scale=0.8] at (10.1, 11.5) {\fontfamily{phv}\selectfont $0.2~\mathrm{M_\odot}$};
\node [rotate=0, scale=0.8] at (15.0, 11.5) {\fontfamily{phv}\selectfont $0.3~\mathrm{M_\odot}$};        
\end{tikzpicture}  
\caption{Radial profiles of the synthetic observation at $\lambda=3.0~\mathrm{mm}$. From top to bottom: Total dust emission (Stokes $I$), total linearly polarized dust emission ($PI$), and polarization fraction ($p_\mathrm{frac}$).}
\label{radial_Stokes_RAT_3}
\end{figure*}

\begin{figure*}
\centering
  \includegraphics[width=0.99\textwidth, trim={2.65cm 1.3cm 2.72cm 4.1cm},clip]{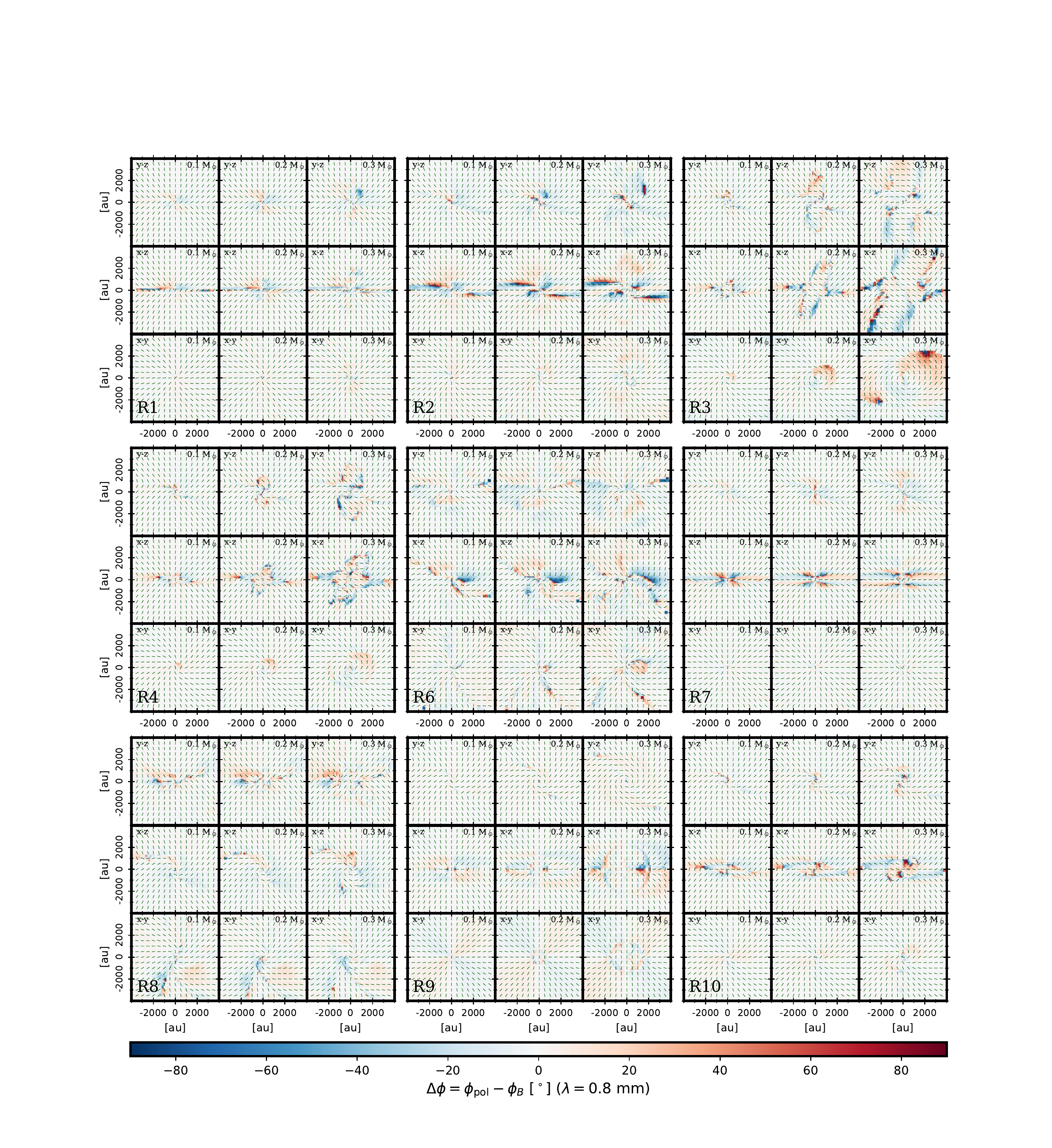}
\caption{Difference maps $\Delta\phi$ at $\lambda=0.8~\mathrm{mm}$. The simulation identifier is shown in the bottom left corner of each block, while the projection and evolutionary step (given by the sink mass) are indicated at the top of each individual map.}
\label{dphi_1}
\end{figure*}

\begin{figure*}
\centering
  \includegraphics[width=0.99\textwidth, trim={2.65cm 1.3cm 2.72cm 4.1cm},clip]{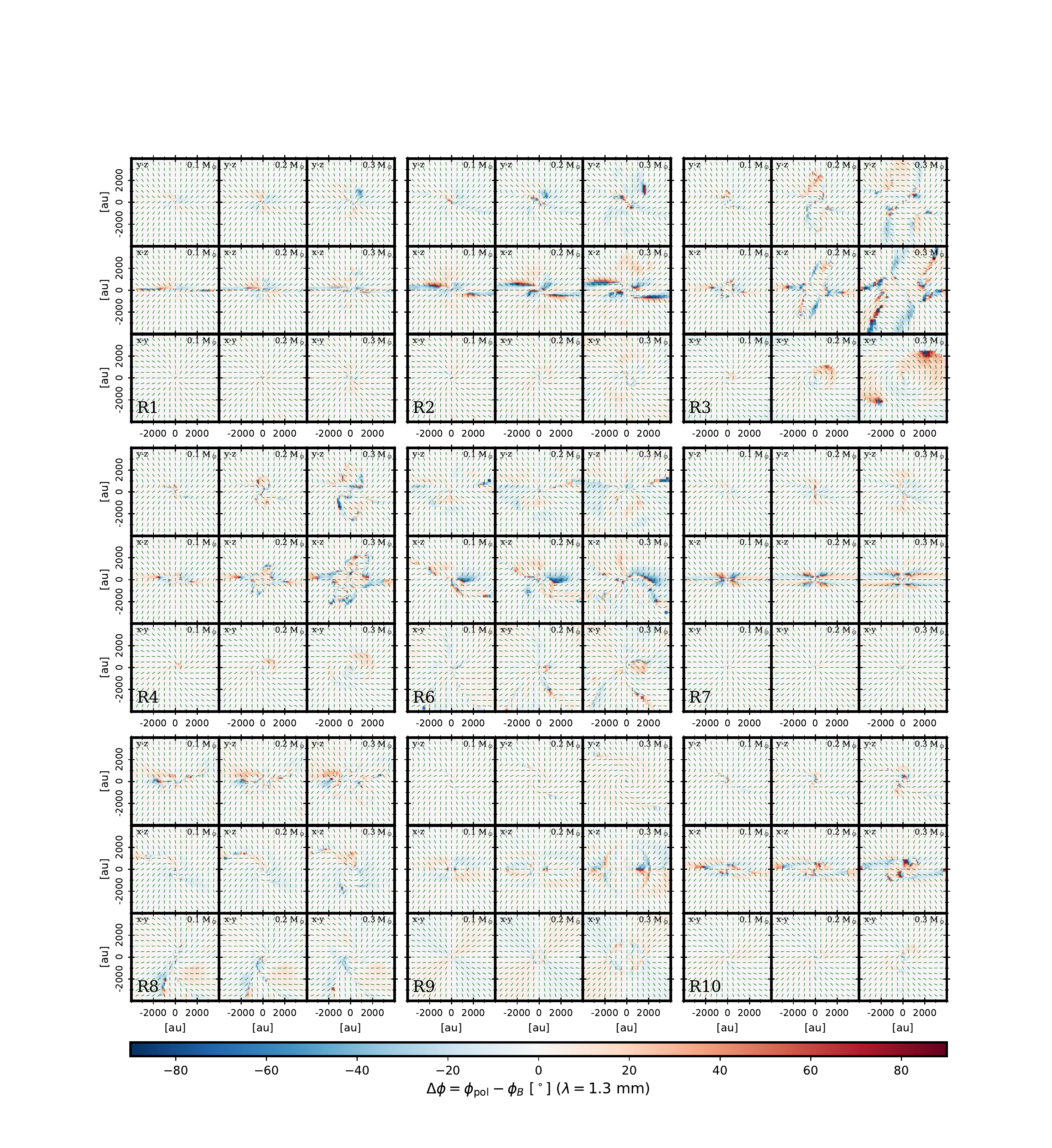}
\caption{Difference maps $\Delta\phi$ at $\lambda=1.3~\mathrm{mm}$. The structure of the figure is the same as for Fig.~\ref{dphi_1}.}
\label{dphi_2}
\end{figure*}

\begin{figure*}
\centering
  \includegraphics[width=0.99\textwidth, trim={2.65cm 1.3cm 2.72cm 4.1cm},clip]{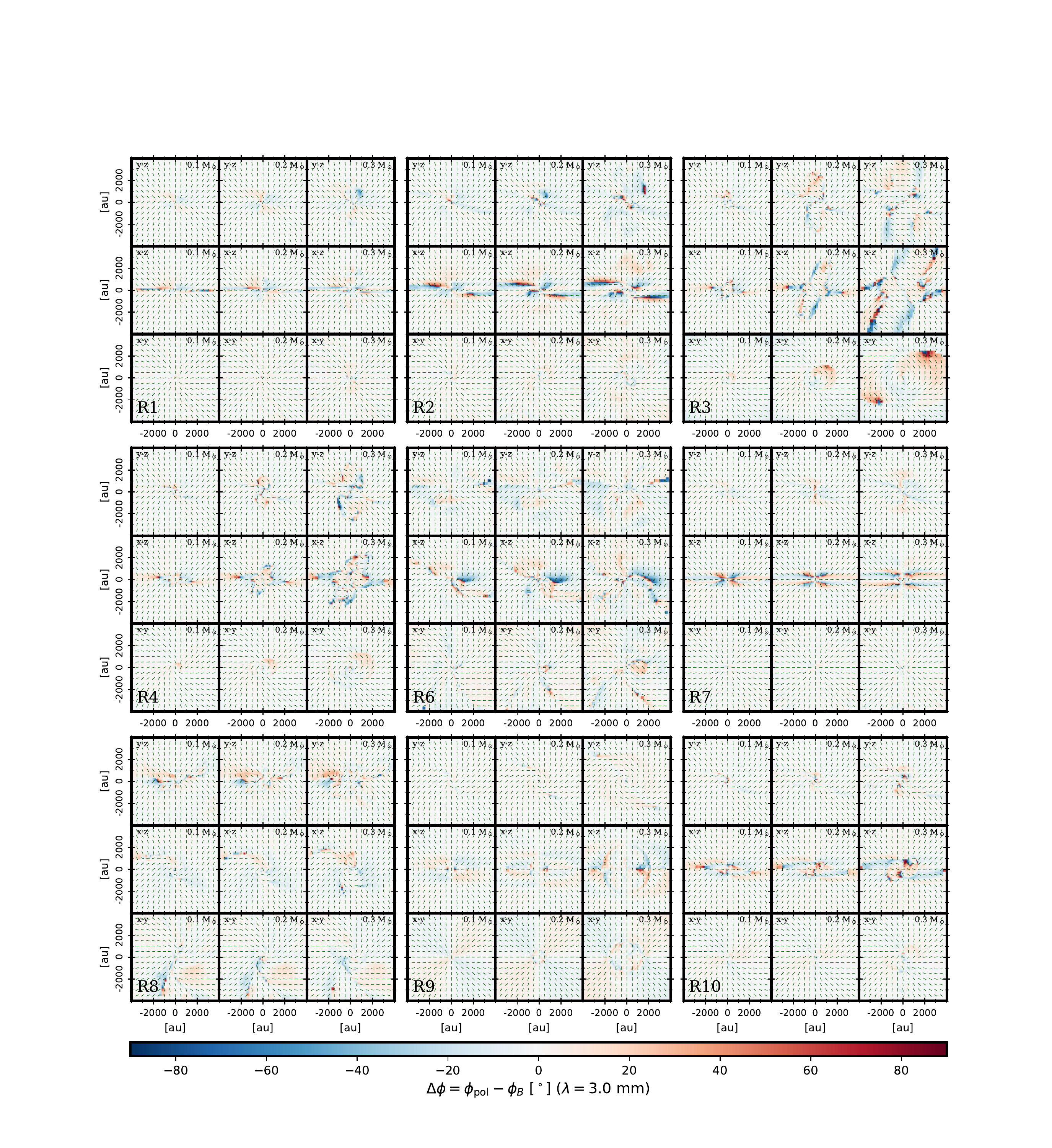}
\caption{$\Delta\phi$ at $\lambda=3.0~\mathrm{mm}$. The structure of the figure is the same as for Figs.~\ref{dphi_1} and \ref{dphi_2}.}
\label{dphi_3}
\end{figure*}

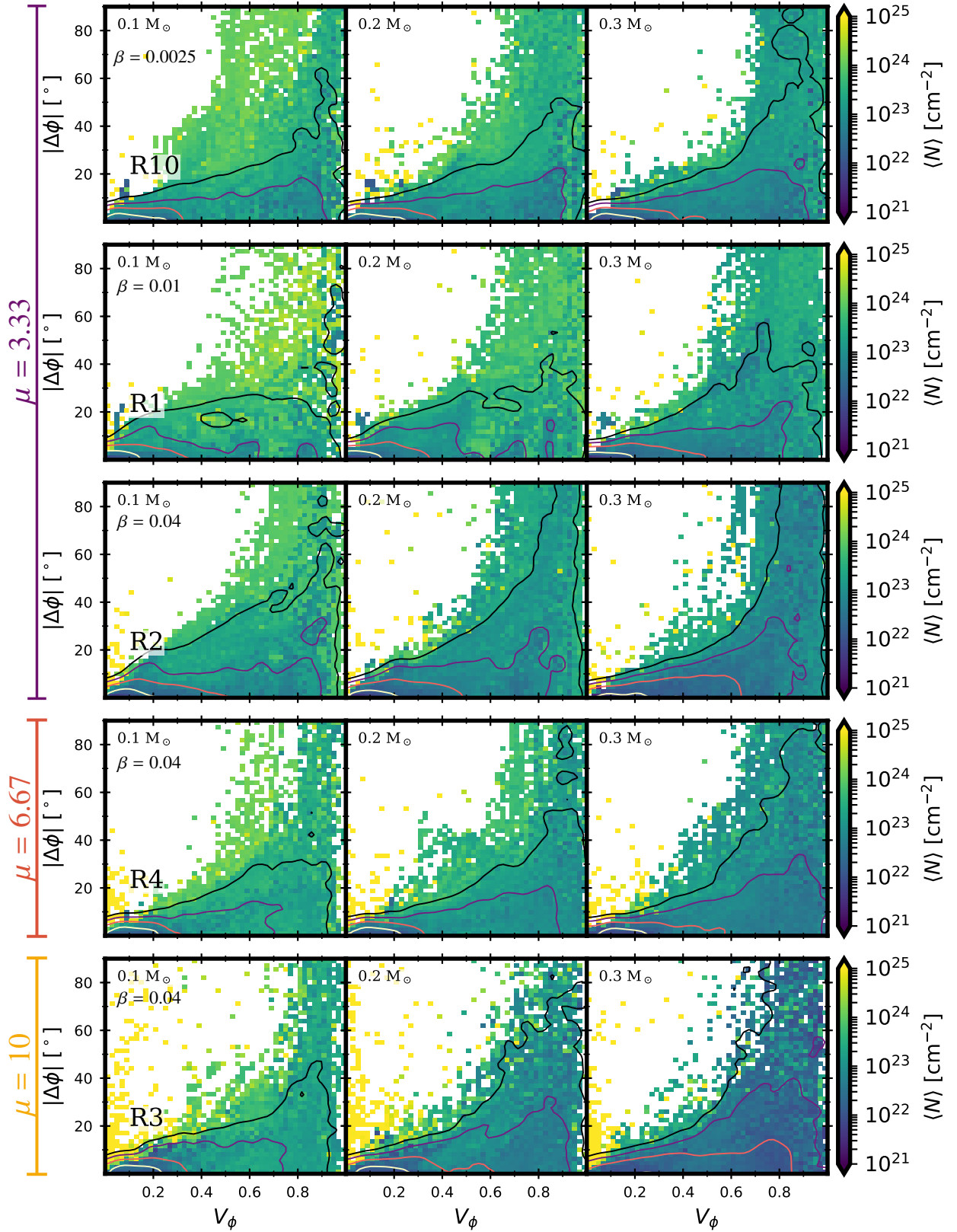
\begin{figure*}
\centering
\begin{tikzpicture}
\node[above right] (img) at (0,0) {
  \begin{tabular}{@{}l@{}}
  \includegraphics[width=0.85\textwidth, trim={0.95cm 4.1cm 0.0cm 4.15cm},clip]{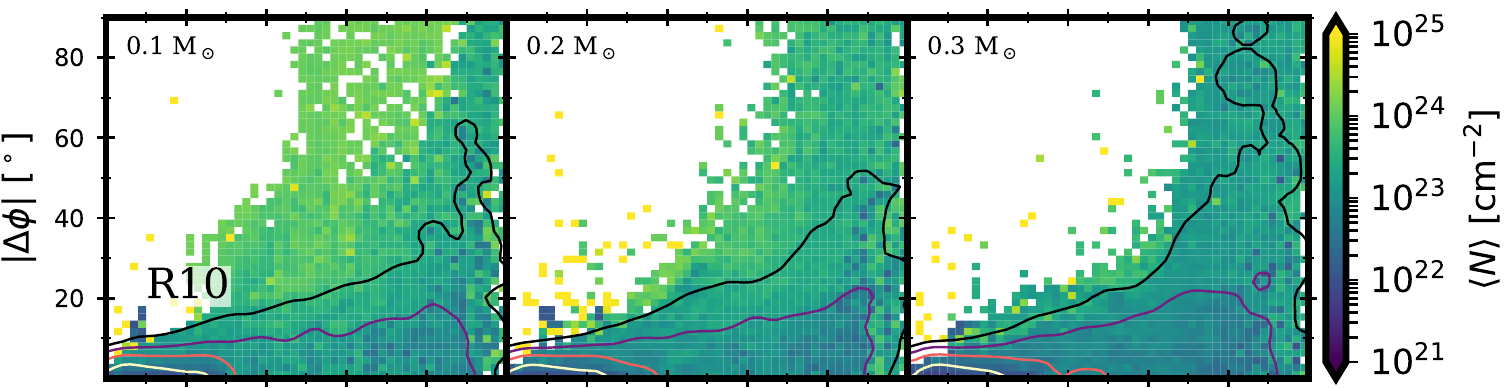}\\
  \includegraphics[width=0.85\textwidth, trim={0.95cm 4.1cm 0.0cm 4.15cm},clip]{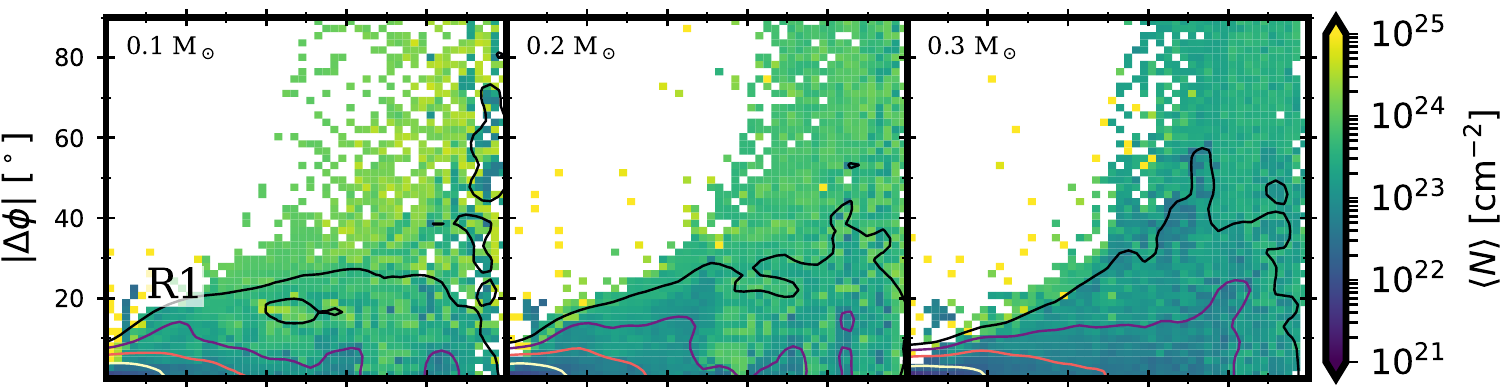}\\
  \includegraphics[width=0.85\textwidth, trim={0.95cm 4.1cm 0.0cm 4.15cm},clip]{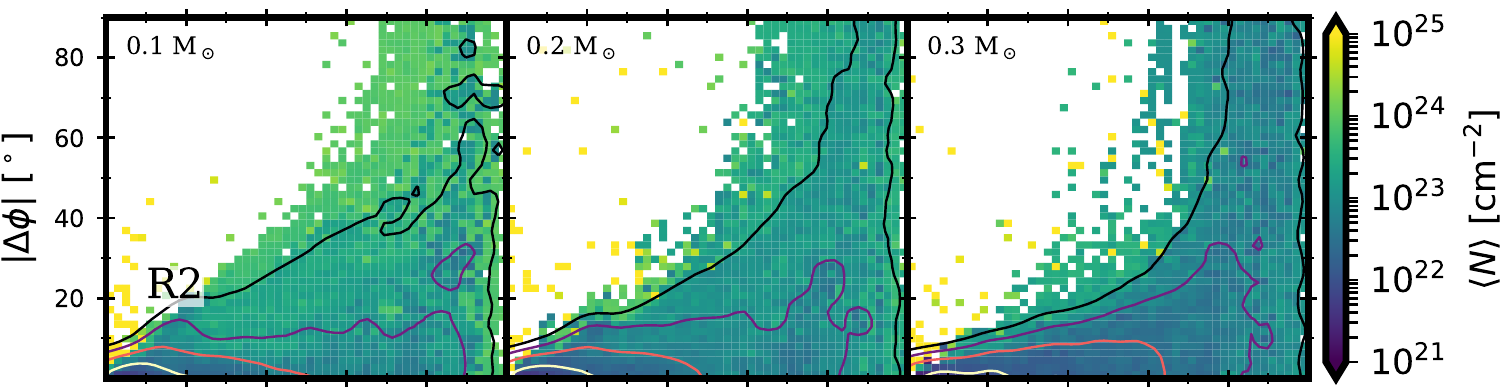}\\
  \includegraphics[width=0.85\textwidth, trim={0.95cm 4.1cm 0.0cm 4.15cm},clip]{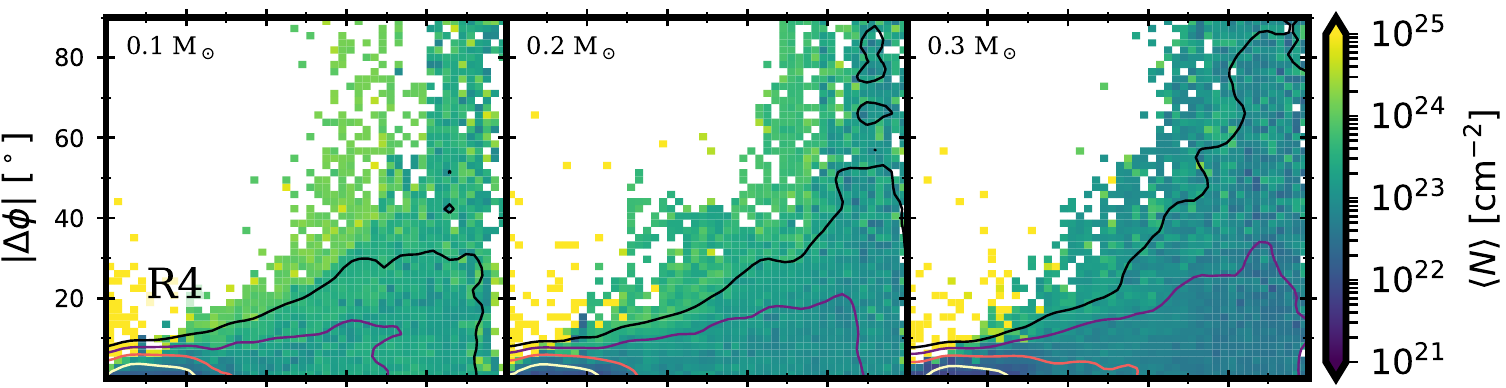}\\
  \includegraphics[width=0.85\textwidth, trim={0.95cm 3.00cm 0.0cm 4.15cm},clip]{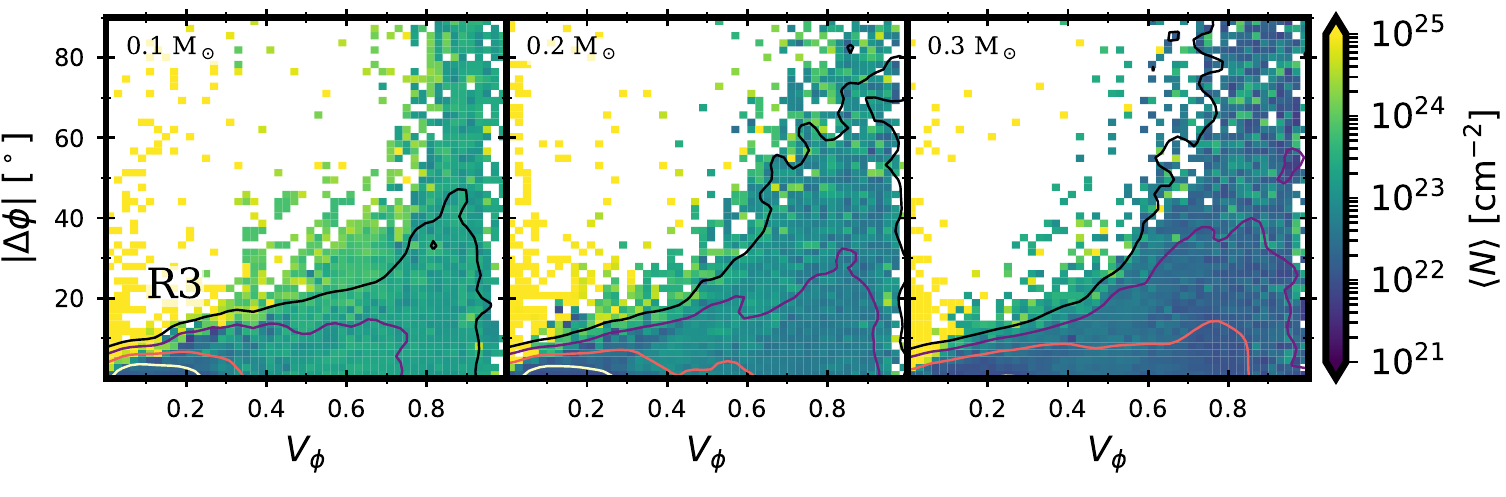}   
  \end{tabular}};
  \draw [|-|, line width=1.3pt, color=col1] (0.05,9.721) -- (0.05,21.827);
  \draw [|-|, line width=1.5pt, color=col2] (0.05,5.576) -- (0.05,9.397);
  \draw [|-|, line width=1.5pt, color=col3] (0.05,1.433) -- (0.05,5.252);
  \node [rotate=90, scale=1.5, color=col1] at (-0.2,15.8) {\fontfamily{phv}\selectfont $\mu = 3.33$};
  \node [rotate=90, scale=1.5, color=col2] at (-0.2,7.5) {\fontfamily{phv}\selectfont $\mu = 6.67$};
  \node [rotate=90, scale=1.5, color=col3] at (-0.2,  3.2) {\fontfamily{phv}\selectfont $\mu = 10$};
  \node [rotate=0, scale=0.9] at (2.1,20.90) {\fontfamily{phv}\selectfont $\beta = 0.0025$};
  \node [rotate=0, scale=0.9] at (2.0,16.90) {\fontfamily{phv}\selectfont $\beta = 0.01$};
  \node [rotate=0, scale=0.9] at (2.0,12.80) {\fontfamily{phv}\selectfont $\beta = 0.04$};
  \node [rotate=0, scale=0.9] at (2.0,  8.60) {\fontfamily{phv}\selectfont $\beta = 0.04$};
  \node [rotate=0, scale=0.9] at (2.0,  4.50) {\fontfamily{phv}\selectfont $\beta = 0.04$};
\end{tikzpicture}
\caption{Distribution of the discrepancy $|\Delta\phi|$ between the $B$ angle inferred from the synthetic polarized dust emission at $\lambda=0.8~\mathrm{mm}$ and the
mean orientation of the $B$ lines in the simulation as a function of the circular variance $V_\phi$ for the simulations with standard conditions. None of these simulations includes any initial turbulence ($\mathcal{M} = 0$) and all have an initial inclination angle between the initial rotation axis and the magnetic field $\theta$ of $30^\circ$. The contour lines show the smoothed 2D histogram contour levels at $10^5$ (yellow), $10^4$ (red), $10^3$  (purple), and $10^2$ (black) counts. The color coded background corresponds to the mean column density.}
\label{DphiVstdcond1}
\end{figure*}

\begin{figure*}
\centering
\begin{tikzpicture}
\node[above right] (img) at (0,0) {
  \begin{tabular}{@{}l@{}}
  \includegraphics[width=0.85\textwidth, trim={0.95cm 4.1cm 0.0cm 4.15cm},clip]{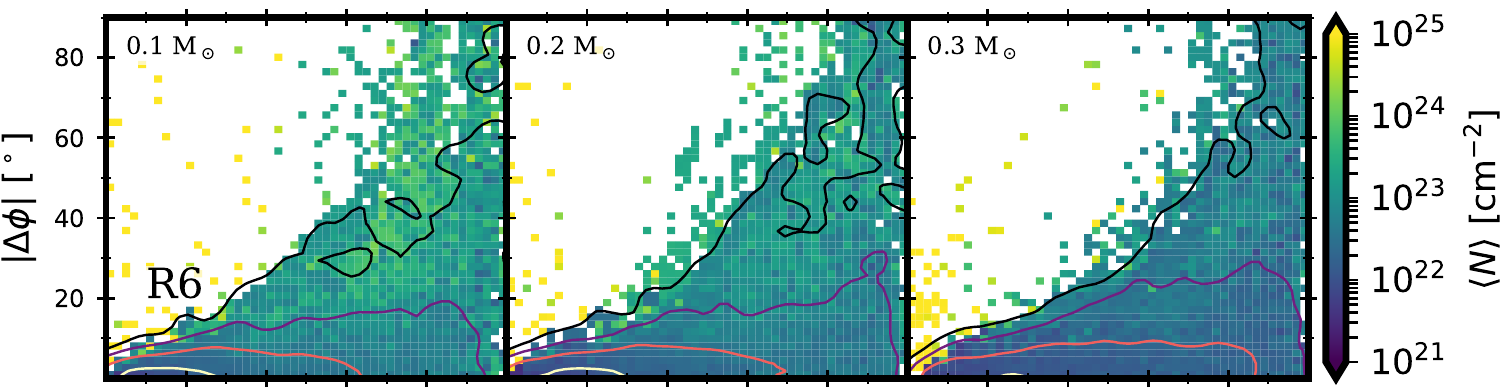}\\
  \includegraphics[width=0.85\textwidth, trim={0.95cm 4.1cm 0.0cm 4.15cm},clip]{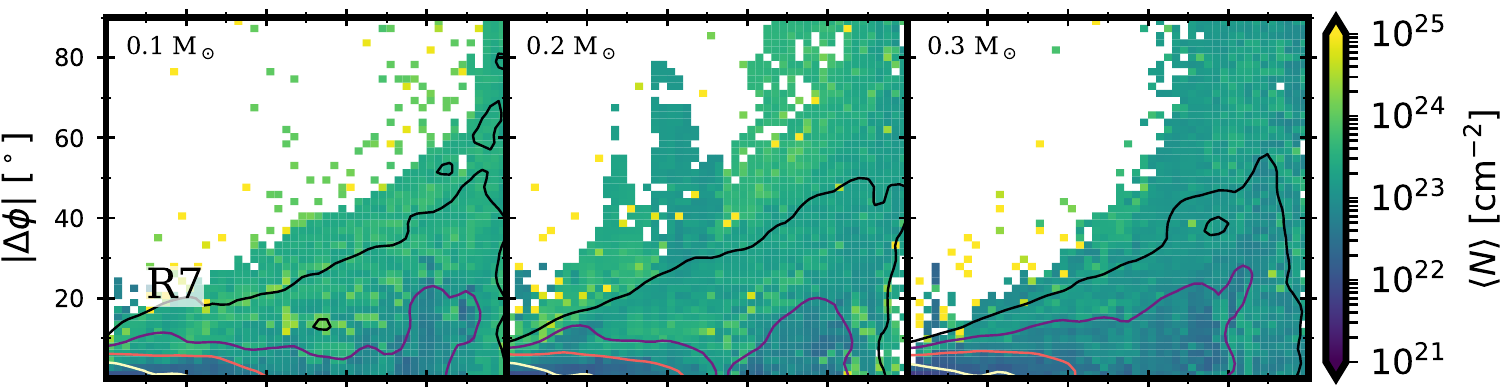}\\
  \includegraphics[width=0.85\textwidth, trim={0.95cm 4.1cm 0.0cm 4.15cm},clip]{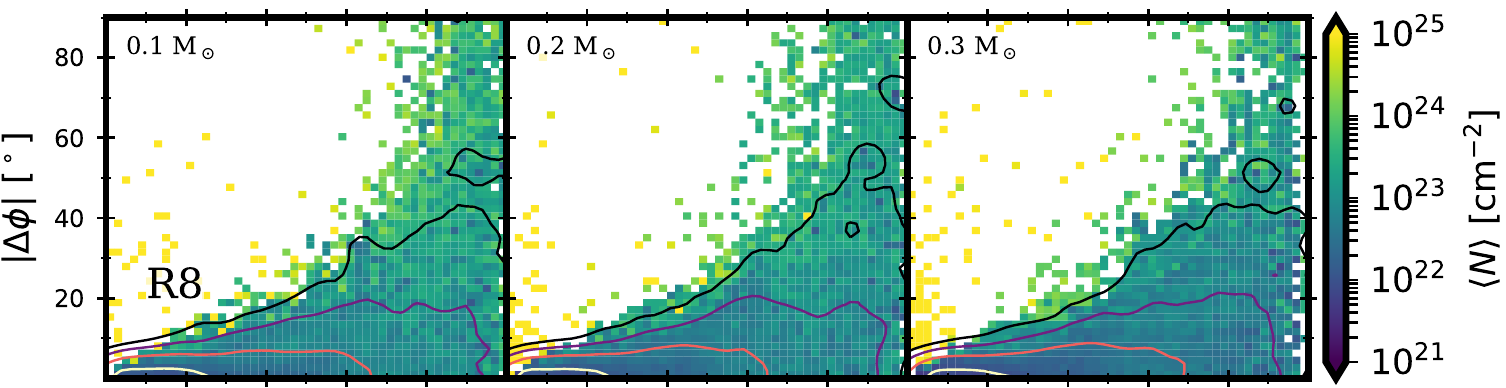}\\
  \includegraphics[width=0.85\textwidth, trim={0.95cm 3.00cm 0.0cm 4.15cm},clip]{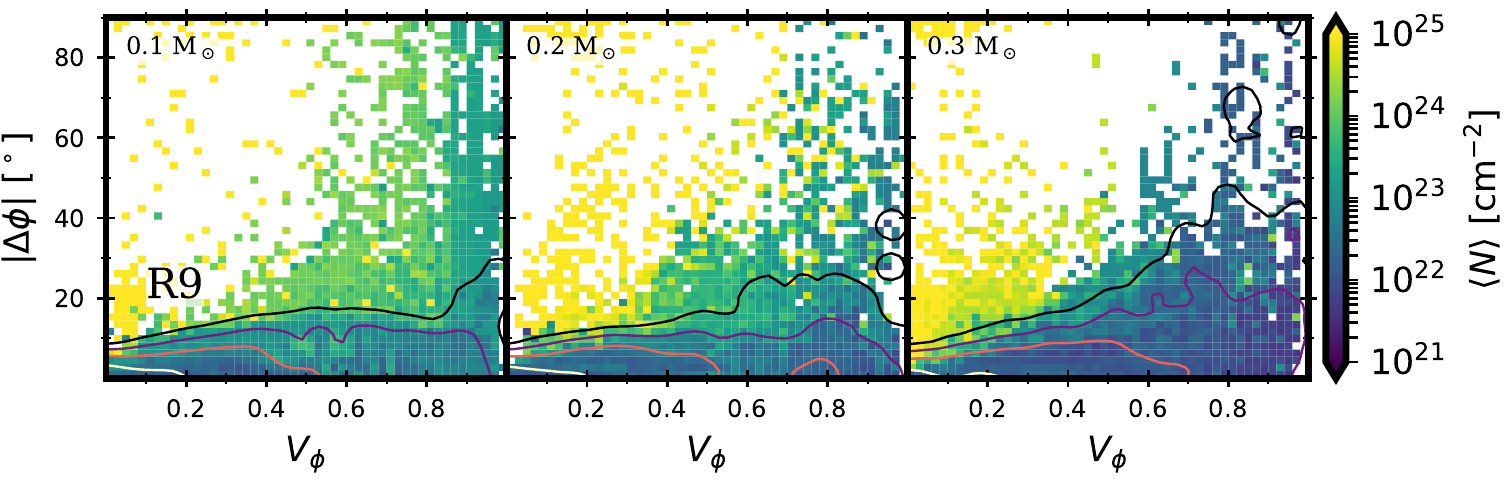}   
  \end{tabular}};
  \draw [|-|, line width=1.3pt, color=col1] (0.05,9.721) -- (0.05,17.684);
  \draw [|-|, line width=1.5pt, color=col2] (0.05,5.576) -- (0.05,9.397);
  \draw [|-|, line width=1.5pt, color=col3] (0.05,1.433) -- (0.05,5.252);
  \node [rotate=90, scale=1.5, color=col1] at (-0.2,14.0) {\fontfamily{phv}\selectfont $\mu = 3.33$};
  \node [rotate=90, scale=1.5, color=col2] at (-0.2,7.5) {\fontfamily{phv}\selectfont $\mu = 6.67$};
  \node [rotate=90, scale=1.5, color=col3] at (-0.2,  3.2) {\fontfamily{phv}\selectfont $\mu = 10$};
  \node [rotate=0, scale=0.9] at (2.5,16.90) {\fontfamily{phv}\selectfont $\beta = 0.01, \mathcal{M} = 1$};
  \node [rotate=0, scale=0.9] at (2.5,12.80) {\fontfamily{phv}\selectfont $\beta = 0.04, \theta = 90^\circ$};
  \node [rotate=0, scale=0.9] at (2.5,  8.60) {\fontfamily{phv}\selectfont $\beta = 0.01, \mathcal{M} = 1$};
  \node [rotate=0, scale=0.9] at (2.5,  4.450) {\fontfamily{phv}\selectfont $\beta = 0.04, \theta = 90$};  
\end{tikzpicture}
\caption{Distribution of $|\Delta \phi|$ at $\lambda=0.8~\mathrm{mm}$ as a function of the circular variance $V_\phi$ for the simulations with nonstandard conditions. The contours and the background image are the same as in Fig.~\ref{DphiVstdcond1}. Simulations R6 and R8 include turbulence ($\mathcal{M}=1$), while the simulations R7 and R9 have an initial angle between the rotation axis and the magnetic field axis of $\theta=90^\circ$.}
\label{DphiVnonstd1}
\end{figure*}

\begin{figure*}
\centering
\begin{tikzpicture}
\node[above right] (img) at (0,0) {
  \begin{tabular}{@{}l@{}}
  \includegraphics[width=0.85\textwidth, trim={0.95cm 4.1cm 0.0cm 4.15cm},clip]{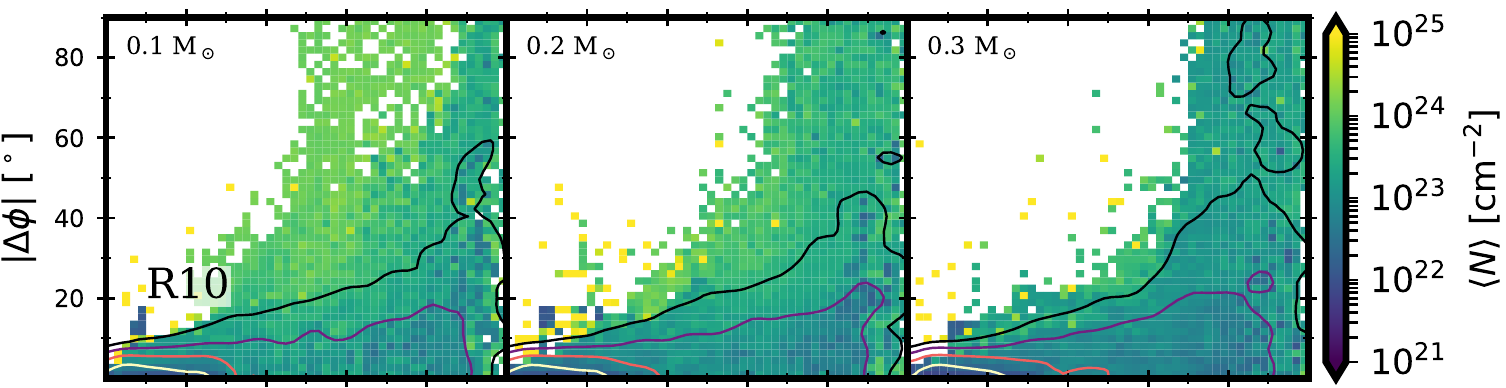}\\
  \includegraphics[width=0.85\textwidth, trim={0.95cm 4.1cm 0.0cm 4.15cm},clip]{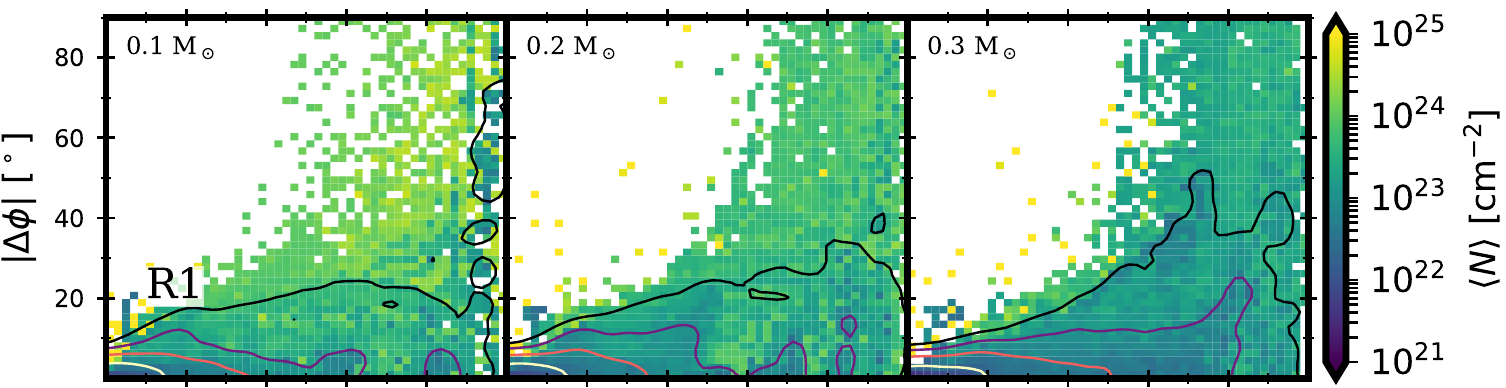}\\
  \includegraphics[width=0.85\textwidth, trim={0.95cm 4.1cm 0.0cm 4.15cm},clip]{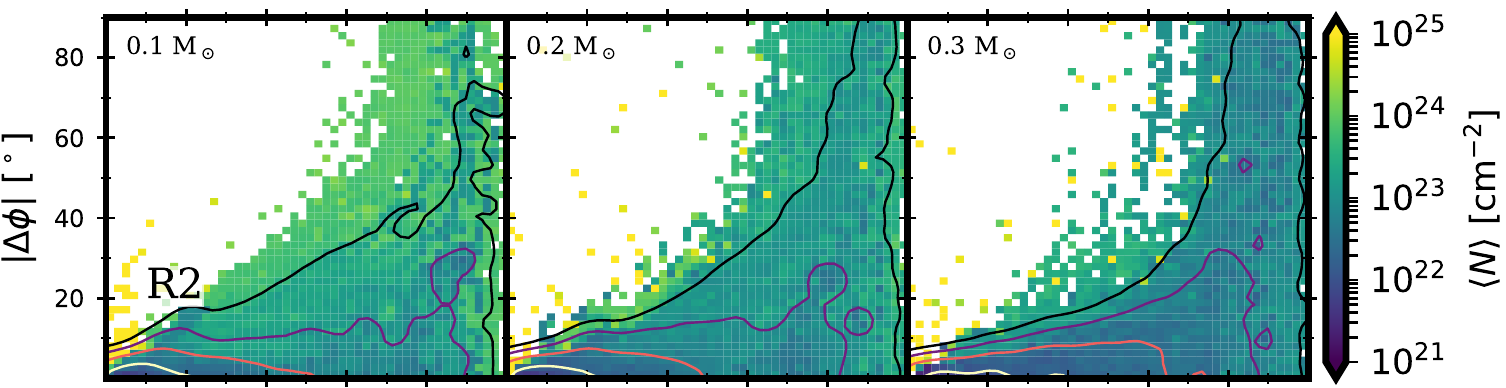}\\
  \includegraphics[width=0.85\textwidth, trim={0.95cm 4.1cm 0.0cm 4.15cm},clip]{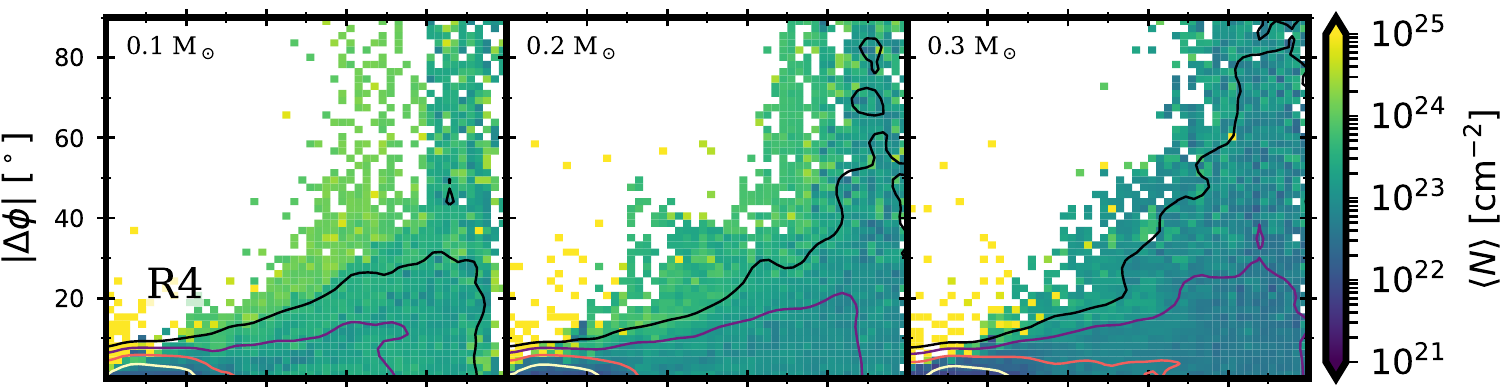}\\
  \includegraphics[width=0.85\textwidth, trim={0.95cm 3.00cm 0.0cm 4.15cm},clip]{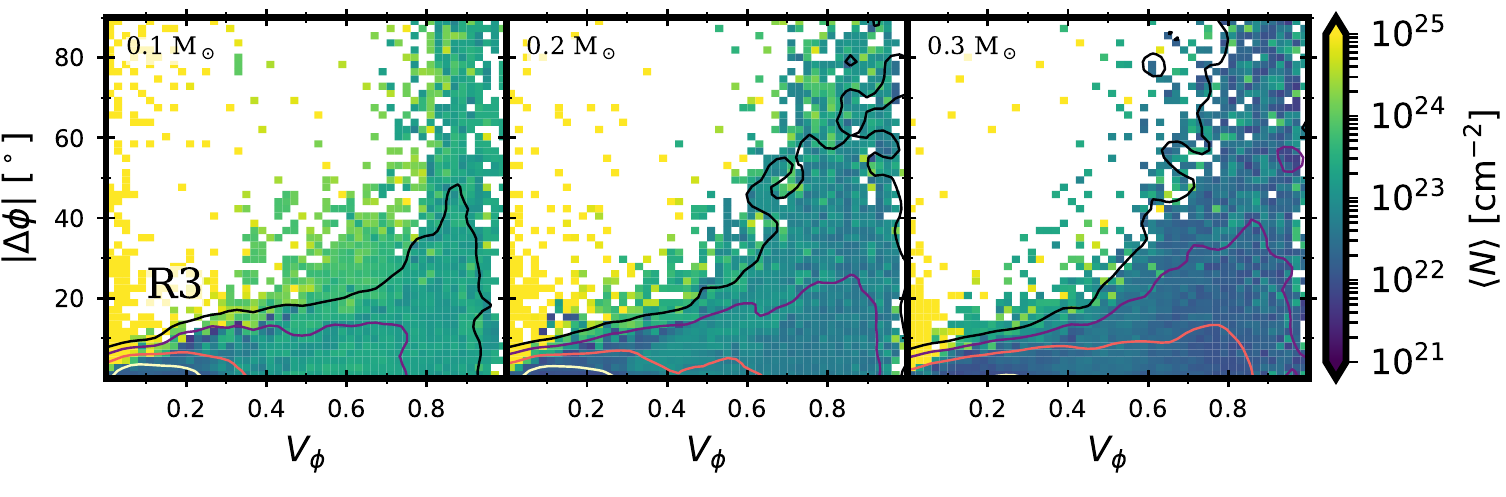}   
  \end{tabular}};
  \draw [|-|, line width=1.3pt, color=col1] (0.05,9.721) -- (0.05,21.827);
  \draw [|-|, line width=1.5pt, color=col2] (0.05,5.576) -- (0.05,9.397);
  \draw [|-|, line width=1.5pt, color=col3] (0.05,1.433) -- (0.05,5.252);
  \node [rotate=90, scale=1.5, color=col1] at (-0.2,15.8) {\fontfamily{phv}\selectfont $\mu = 3.33$};
  \node [rotate=90, scale=1.5, color=col2] at (-0.2,7.5) {\fontfamily{phv}\selectfont $\mu = 6.67$};
  \node [rotate=90, scale=1.5, color=col3] at (-0.2,  3.2) {\fontfamily{phv}\selectfont $\mu = 10$};
  \node [rotate=0, scale=0.9] at (2.1,20.90) {\fontfamily{phv}\selectfont $\beta = 0.0025$};
  \node [rotate=0, scale=0.9] at (2.0,16.90) {\fontfamily{phv}\selectfont $\beta = 0.01$};
  \node [rotate=0, scale=0.9] at (2.0,12.80) {\fontfamily{phv}\selectfont $\beta = 0.04$};
  \node [rotate=0, scale=0.9] at (2.0,  8.60) {\fontfamily{phv}\selectfont $\beta = 0.04$};
  \node [rotate=0, scale=0.9] at (2.0,  4.50) {\fontfamily{phv}\selectfont $\beta = 0.04$};
\end{tikzpicture}
\caption{Distribution of the discrepancy between the $B$ angle inferred from the synthetic polarized dust emission at $\lambda=3.0~\mathrm{mm}$ and the
mean orientation of the $B$ lines in the simulation as a function of the circular variance $V_\phi$ for the simulations with standard conditions. All these simulations do not include any initial turbulence ($\mathcal{M} = 0$) and have an initial inclination angle between the initial rotation axis and the magnetic field $\theta$ of $30^\circ$. The contour lines show the smoothed 2D histogram contour levels at $10^5$ (yellow), $10^4$ (red), $10^3$  (purple), and $10^2$ (black) counts. The color coded background corresponds to the mean column density.}
\label{DphiVstdcond3}
\end{figure*}

\begin{figure*}
\centering
\begin{tikzpicture}
\node[above right] (img) at (0,0) {
  \begin{tabular}{@{}l@{}}
  \includegraphics[width=0.85\textwidth, trim={0.95cm 4.1cm 0.0cm 4.15cm},clip]{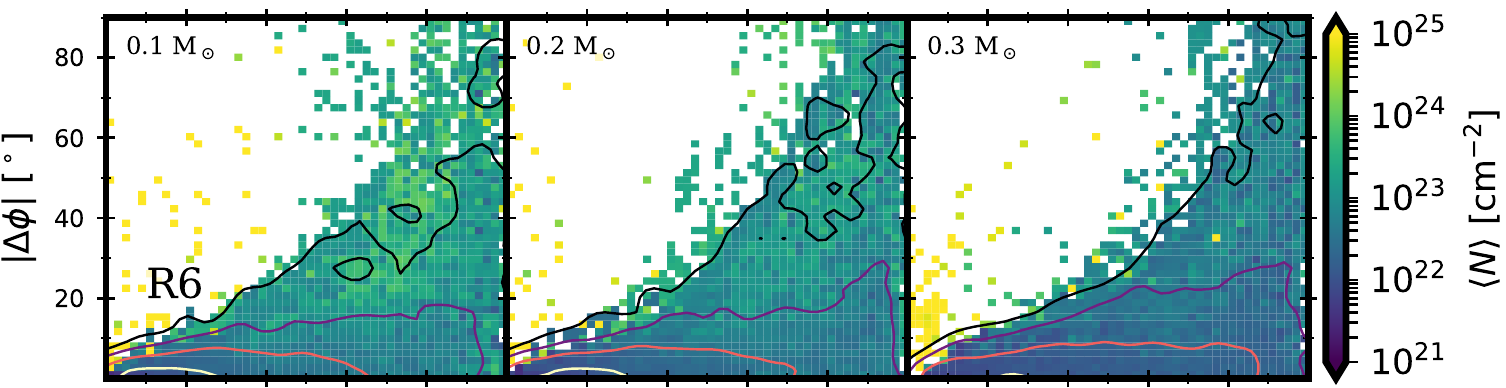}\\
  \includegraphics[width=0.85\textwidth, trim={0.95cm 4.1cm 0.0cm 4.15cm},clip]{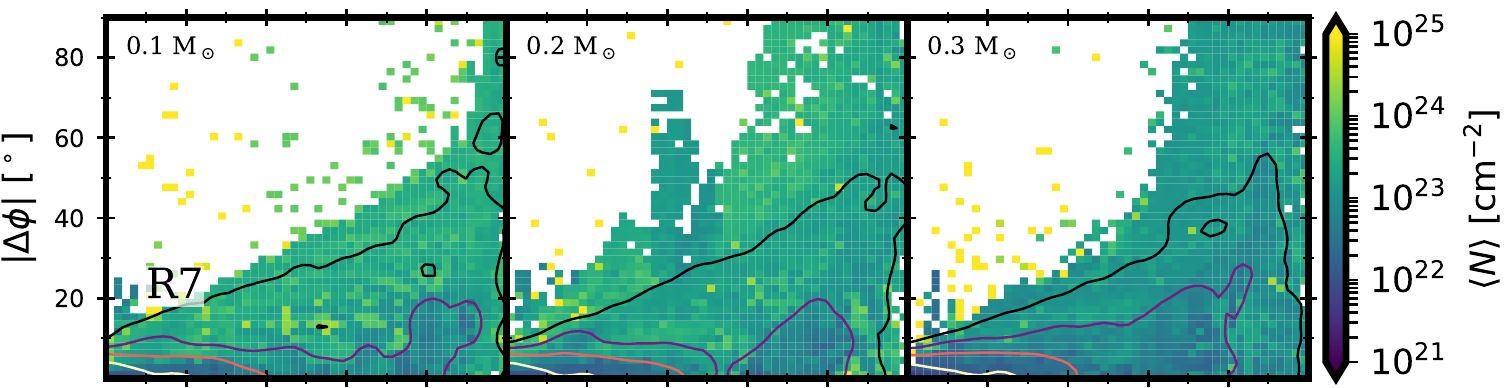}\\
  \includegraphics[width=0.85\textwidth, trim={0.95cm 4.1cm 0.0cm 4.15cm},clip]{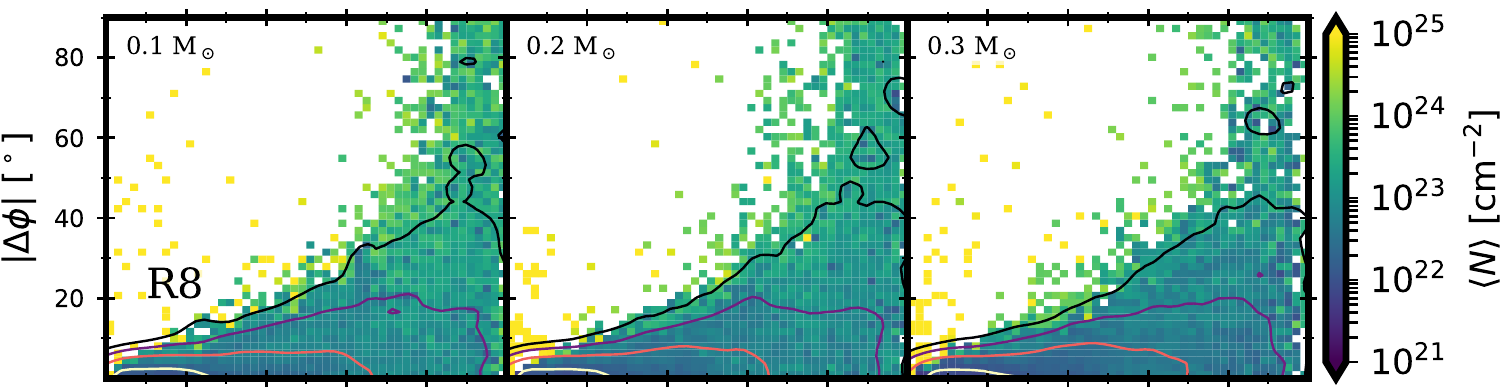}\\
  \includegraphics[width=0.85\textwidth, trim={0.95cm 3.00cm 0.0cm 4.15cm},clip]{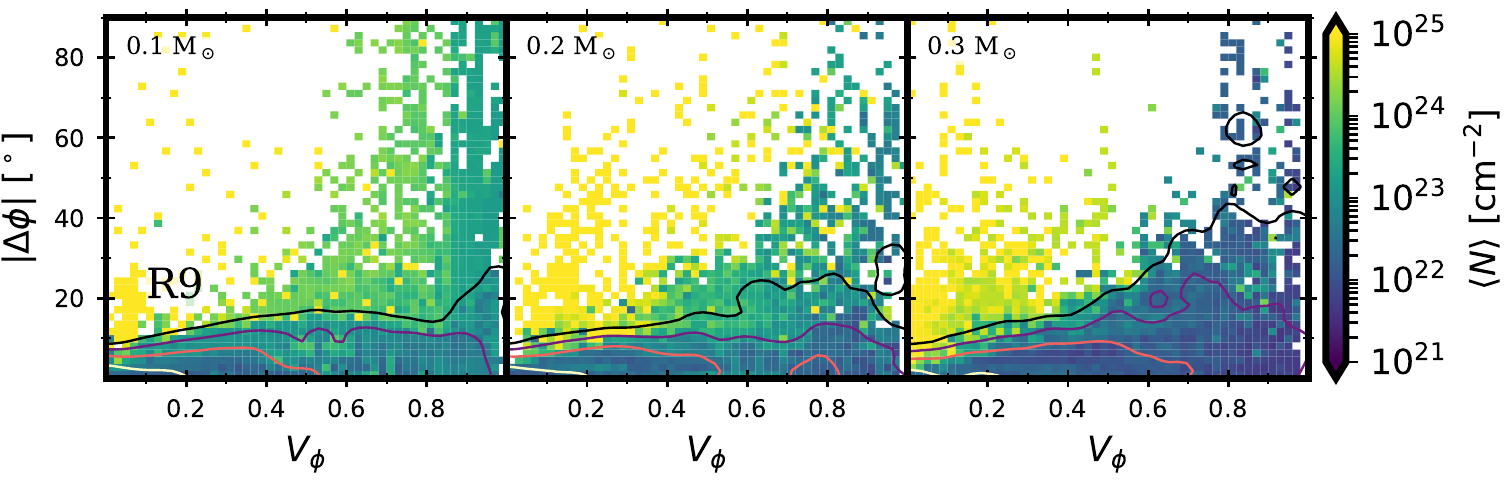}   
  \end{tabular}};
  \draw [|-|, line width=1.3pt, color=col1] (0.05,9.721) -- (0.05,17.684);
  \draw [|-|, line width=1.5pt, color=col2] (0.05,5.576) -- (0.05,9.397);
  \draw [|-|, line width=1.5pt, color=col3] (0.05,1.433) -- (0.05,5.252);
  \node [rotate=90, scale=1.5, color=col1] at (-0.2,14.0) {\fontfamily{phv}\selectfont $\mu = 3.33$};
  \node [rotate=90, scale=1.5, color=col2] at (-0.2,7.5) {\fontfamily{phv}\selectfont $\mu = 6.67$};
  \node [rotate=90, scale=1.5, color=col3] at (-0.2,  3.2) {\fontfamily{phv}\selectfont $\mu = 10$};
  \node [rotate=0, scale=0.9] at (2.5,16.90) {\fontfamily{phv}\selectfont $\beta = 0.01, \mathcal{M} = 1$};
  \node [rotate=0, scale=0.9] at (2.5,12.80) {\fontfamily{phv}\selectfont $\beta = 0.04, \theta = 90^\circ$};
  \node [rotate=0, scale=0.9] at (2.5,  8.60) {\fontfamily{phv}\selectfont $\beta = 0.01, \mathcal{M} = 1$};
  \node [rotate=0, scale=0.9] at (2.5,  4.450) {\fontfamily{phv}\selectfont $\beta = 0.04, \theta = 90$};  
\end{tikzpicture}
\caption{Distribution of $|\Delta \phi|$ at $\lambda=3.0~\mathrm{mm}$ as a function of the circular variance $V_\phi$ for the simulations with non standard conditions. The contours and the background image are the same as in Fig.~\ref{DphiVstdcond3}. Simulations R6 and R8 include turbulence ($\mathcal{M}=1$), while the simulations R7 and R9 have an initial angle between the rotation axis and the magnetic field axis of $\theta=90^\circ$.}
\label{DphiVnonstd3}
\end{figure*}

In this section, we provide complementary material on synthetic observations and their comparisons to properties measured from the models. Figures~\ref{pfrac_maps_1} and \ref{pfrac_maps_3} show the maps from the synthetic observations of the polarized dust emission of the 27 models, at $\lambda=0.8$ and $3.0~\mathrm{mm}$. Figures~\ref{radial_Stokes_RAT_1} and \ref{radial_Stokes_RAT_3} show the azimuthally averaged profiles of the total dust emission flux density, linearly polarized flux density, and the polarization fraction (averaged for the three projections) at $\lambda=0.8$ and $3.0~\mathrm{mm}$, respectively.\\ Figures~\ref{dphi_1}, \ref{dphi_2}, and \ref{dphi_3} are maps of the difference in measured mean $B$-field orientation between the model value and the value inferred from the polarized dust emission at $\lambda=0.8$, $1.3,$ and $3.0~\mathrm{mm}$. We find mostly small differences (light colors or white), and highlight in dark colors the regions where the $B$-field orientation inferred from the polarized dust emission is very different from the mean magnetic field orientation.\\
Figures~\ref{DphiVstdcond1} and \ref{DphiVnonstd1} (Figures~\ref{DphiVstdcond3} and \ref{DphiVnonstd3}) show the distribution of the discrepancy between the inferred orientation of the magnetic field and the true mean orientation at $\lambda=0.8~\mathrm{mm}$ ($\lambda=3.0~\mathrm{mm}$) for the simulations with standard conditions and the ones starting from less prototypical conditions (either $\mathcal{M}=1$ or $\theta=90^\circ$), respectively.\\
Table~\ref{table:Ap} gives the detailed recovery rates at $\lambda=0.8$ for all the simulations (ID), projections ($los$), and evolutionary stages ($m_\mathrm{sink}$) at different scales. The fourth column shows the recovery rate over the whole map, while columns 5, 6, and 7 correspond to the recovery rate in concentric rings of radial distances comprised in the interval given in the header of the table.\\
We verified the influence of the presence of turbulence in Table~\ref{table:turb}. This table shows the mean recovery rates for simulations R1 and R6. These two otherwise identical simulations ($\mu=3.33$, $\beta_\mathrm{rot}=0.01$ and $\theta=30^\circ$), differing only in the degree of turbulence (R6 has $\mathcal{M}=1$), show that the degree of turbulence does not have a strong impact on the recovery rate.\\
Finally, Tables~\ref{table:ap.1.3} and \ref{table:ap.3.0} show the mean recovery rates separated into two cases (moderate and low magnetization degree) similar to Table~\ref{table:3}, but at $\lambda=1.3$, and $3.0~\mathrm{mm}$, respectively.

\begin{table*}[h]
\caption{Fidelity of the $B$-field mapping from polarized dust emission at $\lambda=0.8~\mathrm{mm}$ at different scales.}              
\label{table:Ap}      
\centering                                      
\begin{tabular}{c c c r r r r | r c c r r r r}          
\hline\hline                        
 \noalign{\smallskip}
 ID & $m_\mathrm{sink}$ & $los$ & \multicolumn{4}{c}{\textit{Recovery rate at different scales $[\%]$  }} & ID & $m_\mathrm{sink}$ & $los$ & \multicolumn{4}{c}{\textit{Recovery rate at different scales $[\%]$  }}\\
    &\tiny{$[\mathrm{M_\odot}]$} & & \tiny{All scales}  & \tiny{$30-100$ } & \tiny{$100-500$}  & \tiny{$500-1500$} &   &\tiny{$[\mathrm{M_\odot}]$} & & \tiny{All scales}  & \tiny{$30-100$} & \tiny{$100-500$}  & \tiny{$500-1500$}\\    
    \tiny{}&\tiny{} & \tiny{}& \tiny{} & \tiny{au}& \tiny{au} & \tiny{au} & & & & \tiny{} & \tiny{au}& \tiny{au} & \tiny{au}\\
\hline\hline 
 \noalign{\smallskip}
R1  & 0.1   & $x$ &      99.1   &        82.2 &  96.8   &        99.0   &   R7  & 0.1   & $x$ &    99.3   &        88.9 &  96.6   &        99.5\\
    &       & $y$ &      96.8   &        60.9 &  76.3   &        90.5   &       &       & $y$ &    92.1   &        55.1 &  33.6   &        71.1\\
    &       & $z$ &      >99.9  &        68.7 &  >99.9  &        100.0  &       &       & $z$ &    >99.9   &       95.1 & >99.9   &        100.0\\
    & 0.2   & $x$ &      97.7   &        80.5 &  72.2   &        95.5   &       & 0.2   & $x$ &    97.8   &        76.2 &  70.4   &        98.6\\
    &       & $y$ &      94.4   &        66.3 &  70.3   &        82.3   &       &       & $y$ &    91.9   &        93.4 &  46.1   &        66.1\\
    &       & $z$ &      >99.9  &        73.4 &  99.0   &        100.0  &       &       & $z$ &    >99.9  &        92.7 &  99.5   &        100.0\\
    & 0.3   & $x$ &      97.2   &        72.1 &  90.4   &        87.0   &       & 0.3   & $x$ &    95.8   &        76.0 &  83.6   &        92.4\\
    &       & $y$ &      93.3   &        61.5 &  76.0   &        84.4   &       &       & $y$ &    89.4   &        76.6 &  66.8   &        63.1\\
    &       & $z$ &      99.5   &        72.5 &  90.3   &        99.5   &       &       & $z$ &    99.8   &        81.6 &  96.3   &       >99.9\\
\hline
 \noalign{\smallskip}
R2  & 0.1   & $x$ &      97.7   &        51.0 &  75.2   &        97.2   &   R8  & 0.1   & $x$ &    94.8   &        80.4 &  70.9   &        88.0\\
    &       & $y$ &      91.0   &        50.3 &  57.0   &        76.2   &       &       & $y$ &    97.4   &        70.8 &  96.6   &        96.9\\
    &       & $z$ &      99.8   &        75.7 &  99.4   &        >99.9  &       &       & $z$ &    93.2   &        75.3 &  90.8   &        96.6\\
    & 0.2   & $x$ &      96.5   &        67.2 &  65.7   &        86.9   &       & 0.2   & $x$ &    93.5   &        72.7 &  79.9   &        86.5\\ 
    &       & $y$ &      88.1   &        79.3 &  48.4   &        68.5   &       &       & $y$ &    96.9   &        78.3 &  88.5   &        97.8\\
    &       & $z$ &      99.7   &        82.8 &  96.2   &        99.1   &       &       & $z$ &    93.9   &        92.0 &  83.7   &        95.6\\
    & 0.3   & $x$ &      92.6   &        69.5 &  73.8   &        85.1   &       & 0.3   & $x$ &    92.5   &        86.8 &  95.6   &        87.3\\
    &       & $y$ &      83.7   &        81.7 &  79.7   &        69.6   &       &       & $y$ &    94.7   &        85.7 &  87.7   &        88.7\\
    &       & $z$ &      98.4   &        67.4 &  87.1   &        97.8   &       &       & $z$ &    95.5   &        85.1 &  92.2   &        92.7\\
\hline  
 \noalign{\smallskip}
R3  & 0.1   & $x$ &      98.7   &        89.4 &  87.8   &        94.3   &   R9  & 0.1   & $x$ &    99.8   &        91.8 &  99.7   &        99.5\\
    &       & $y$ &      96.4   &        93.0 &  89.7   &        86.2   &       &       & $y$ &    94.9   &        97.2 &  71.4   &        98.0\\
    &       & $z$ &      98.2   &        80.2 &  95.2   &        99.2   &       &       & $z$ &    99.8   &        80.7 &  95.5   &        100.0\\
    & 0.2   & $x$ &      94.5   &        86.5 &  92.8   &        87.4   &       & 0.2   & $x$ &    99.5   &        63.7 &  99.7   &        98.7\\
    &       & $y$ &      90.5   &        92.9 &  97.6   &        78.7   &       &       & $y$ &    97.1   &        91.1 &  91.9   &        95.1\\
    &       & $z$ &      95.4   &        77.1 &  91.1   &        86.9   &       &       & $z$ &    99.3   &        79.7 &  94.9   &        97.2\\
    & 0.3   & $x$ &      88.7   &        68.9 &  96.0   &        89.1   &       & 0.3   & $x$ &    99.0   &        67.5 &  98.5   &        >99.9\\
    &       & $y$ &      79.7   &        84.3 &  97.9   &        94.6   &       &       & $y$ &    84.4   &        91.4 &  90.5   &        90.0\\
    &       & $z$ &      81.1   &        67.3 &  87.6   &        91.4   &       &       & $z$ &    96.3   &        90.4 &  89.4   &        87.1\\
\hline
 \noalign{\smallskip}
R4  & 0.1   & $x$ &      98.9   &        81.7 &  73.9   &        96.4   &   R10 & 0.1   & $x$ &    99.3   &        54.7 &  84.5   &        99.2\\
    &       & $y$ &      96.1   &        86.8 &  78.1   &        90.6   &       &       & $y$ &    93.1   &        82.1 &  62.1   &        86.3\\
    &       & $z$ &      99.6   &        77.2 &  86.9   &        99.3   &       &       & $z$ &    99.9   &        78.0 &  98.5   &        99.9\\
    & 0.2   & $x$ &      96.9   &        69.5 &  80.1   &        86.0   &       & 0.2   & $x$ &    98.9   &        69.6 &  80.6   &        97.3\\
    &       & $y$ &      92.7   &        73.4 &  87.1   &        75.6   &       &       & $y$ &    91.6   &        79.0 &  46.2   &        83.7\\
    &       & $z$ &      98.8   &        74.9 &  97.3   &        99.7   &       &       & $z$ &    99.7   &        59.4 &  94.5   &        100.0\\
    & 0.3   & $x$ &      93.0   &        64.6 &  70.4   &        84.0   &       & 0.3   & $x$ &    97.1   &        61.4 &  65.5   &        90.3\\
    &       & $y$ &      85.8   &        47.8 &  73.0   &        73.7   &       &       & $y$ &    89.1   &        90.4 &  44.4   &        65.2\\
    &       & $z$ &      97.1   &        62.5 &  88.6   &        96.0   &       &       & $z$ &    99.0   &        49.8 &  84.7   &        98.6\\
\hline  
 \noalign{\smallskip}
R6  & 0.1   & $x$ &      95.5   &        81.9 &  91.5   &        96.0 & & & & & & &\\
    &       & $y$ &      92.5   &        49.5 &  74.2   &        80.0 & & & & & & &\\
    &       & $z$ &      96.8   &        76.7 &  89.3   &        98.4 & & & & & & &\\
    & 0.2   & $x$ &      93.6   &        74.8 &  89.9   &        97.5 & & & & & & & \\
    &       & $y$ &      87.1   &        50.5 &  66.3   &        68.3 & & & & & & &\\
    &       & $z$ &      96.8   &        68.9 &  91.7   &        94.1 & & & & & & &\\
    & 0.3   & $x$ &      89.5   &        89.5 &  82.0   &        93.2 & & & & & & &\\
    &       & $y$ &      83.4   &        73.1 &  67.7   &        78.7 & & & & & & &\\
    &       & $z$ &      91.9   &        71.5 &  85.1   &        84.0 & & & & & & &\\
\hline  

\end{tabular}
\tablefoot{
Here we present the recovery rates obtained for each simulation (ID), evolutionary stage ($m_\mathrm{sink}$), and projection ($los$) using the synthetic observations at $\lambda=0.8~\mathrm{mm}$. The fourth column shows the recovery rate over the whole map, while the rest of the scales correspond to the lines of sight with radial distances to the central source comprised in the interval indicated in the header. A synthesis of this table is given in Table~\ref{table:3}.  
}
\end{table*}


\begin{table}[h]
\caption{Influence of the turbulence on the mean fidelity of $B$-field mapping from polarized dust emission at $\lambda=0.8~\mathrm{mm}$ in protostellar envelopes at different scales.}              
\label{table:turb}      
\centering                                      
\begin{tabular}{c c c c c }          
\hline\hline                        
 \noalign{\smallskip}
  $m_\mathrm{sink}$ & \multicolumn{4}{c}{\textit{$B$-field recovery rates and $\sigma$ $[\%]$   }}\\
 \tiny{$[\mathrm{M_\odot}]$} & \tiny{All scales}  & \tiny{$30-100$ au} & \tiny{$100-500$ au} & \tiny{$500-1500$ au} \\    
\hline\hline      
 \noalign{\smallskip}
 \noalign{\smallskip}
%
R1 &    97.5~ \tiny{2.4} &       70.9~~ \tiny{7.5} &     85.7~ \tiny{12.0} &        93.1~~ \tiny{7.2} \\
R6 &    91.9~ \tiny{4.5} &       70.7~ \tiny{13.2} &     82.0~ \tiny{10.1} &        87.8~ \tiny{10.5} \\
\hline
\end{tabular}
\tablefoot{
Mean recovery rates at different scales for an error bar of $15^\circ$. The values have been averaged for all the evolutionary stages and projections for simulations R1 (non-turbulent) and R6 (turbulent) given in Table~\ref{table:Ap}. The values given in a smaller font correspond to the standard deviation.
}
\end{table}


\begin{table}[h]
\caption{Fidelity of $B$-field mapping from polarized dust emission in protostellar envelopes at $\lambda=1.3~\mathrm{mm}$.}              
\label{table:ap.1.3}      
\centering                                      
\begin{tabular}{c c c c c }          
\hline\hline                        
 \noalign{\smallskip}
  $m_\mathrm{sink}$ & \multicolumn{4}{c}{\textit{$B$-field recovery rates and $\sigma$ $[\%]$   }}\\
 \tiny{$[\mathrm{M_\odot}]$} & \tiny{All scales}  & \tiny{$30-100$ au} & \tiny{$100-500$ au} & \tiny{$500-1500$ au} \\    
\hline\hline      
 \noalign{\smallskip}
        \multicolumn{5}{c}{\textit{$\mu = 3.33$}}                       \\ 
\hline                                  
 \noalign{\smallskip}
%
0.1 &    97.1~ \tiny{3.0} &      71.9~ \tiny{14.9} &     83.7~ \tiny{18.4} &        93.4~~ \tiny{9.1} \\
0.2 &    95.9~ \tiny{4.0} &      76.2~ \tiny{11.3} &     77.3~ \tiny{17.9} &        90.2~ \tiny{11.5} \\
0.3 &    93.7~ \tiny{5.1} &      74.6~ \tiny{10.4} &     79.4~ \tiny{12.5} &        86.9~ \tiny{11.5} \\
\hline
 \noalign{\smallskip}
        \multicolumn{5}{c}{\textit{$\mu = 6.67, 10.$}}                  \\ 
\hline                                  
 \noalign{\smallskip}
0.1 &    97.5~ \tiny{2.0} &      85.6~~ \tiny{7.8} &     87.9~ \tiny{9.1} &        95.7~ \tiny{4.4} \\
0.2 &    96.0~ \tiny{2.7} &      82.2~~ \tiny{9.5} &     91.3~ \tiny{6.0} &        90.6~ \tiny{7.1} \\
0.3 &    91.2~ \tiny{5.8} &      77.8~ \tiny{13.7} &     90.7~ \tiny{7.4} &        90.6~ \tiny{6.5} \\
\hline                                             
\end{tabular}
\tablefoot{
Mean recovery rates as in Table~\ref{table:3}, but for maps at $\lambda=1.3~\mathrm{mm}$.
}
\end{table}

\begin{table}[h]
\caption{Fidelity of $B$-field mapping from polarized dust emission in protostellar envelopes at $\lambda=3.0~\mathrm{mm}$.}              
\label{table:ap.3.0}      
\centering                                      
\begin{tabular}{c c c c c }          
\hline\hline                        
 \noalign{\smallskip}
  $m_\mathrm{sink}$ & \multicolumn{4}{c}{\textit{$B$-field recovery rates and $\sigma$ $[\%]$   }}\\
 \tiny{$[\mathrm{M_\odot}]$} & \tiny{All scales}  & \tiny{$30-100$ au} & \tiny{$100-500$ au} & \tiny{$500-1500$ au} \\    
\hline\hline      
 \noalign{\smallskip}
        \multicolumn{5}{c}{\textit{$\mu = 3.33$}}                       \\ 
\hline                                  
 \noalign{\smallskip}
%
0.1 &    97.2~ \tiny{3.0} &      72.9~ \tiny{14.7} &     84.9~ \tiny{18.0} &        93.9~~ \tiny{8.8} \\
0.2 &    96.2~ \tiny{3.8} &      77.4~ \tiny{11.3} &     78.8~ \tiny{17.3} &        90.9~ \tiny{11.0} \\
0.3 &    94.0~ \tiny{5.0} &      75.8~ \tiny{10.5} &     80.5~ \tiny{12.0} &        87.7~ \tiny{11.2} \\
\hline
 \noalign{\smallskip}
        \multicolumn{5}{c}{\textit{$\mu = 6.67, 10.$}}                  \\ 
\hline                                  
 \noalign{\smallskip}
0.1 &    97.6~ \tiny{2.0} &      86.4~~ \tiny{7.9} &     89.1~ \tiny{8.4} &        95.9~ \tiny{4.3} \\
0.2 &    96.2~ \tiny{2.7} &      83.4~ \tiny{10.0} &     92.1~ \tiny{5.6} &        91.3~ \tiny{7.1} \\
0.3 &    91.7~ \tiny{5.6} &      79.3~ \tiny{13.6} &     92.2~ \tiny{6.4} &        91.3~ \tiny{6.4} \\
\hline                                             
\end{tabular}
\tablefoot{
Mean recovery rates as in Table~\ref{table:3}, but for maps at $\lambda=3.0~\mathrm{mm}$.
}
\end{table}

\FloatBarrier
\section{Additional tests for understanding the origin of points with low dispersion and high discrepancy}\label{lowVhighdelta}

\begin{figure*}
\centering
\begin{tabular}{l l l }          
 \noalign{\smallskip}
  \includegraphics[width=0.25\textwidth, trim={1.9cm 1.4cm 1.7cm 1.5cm},clip]{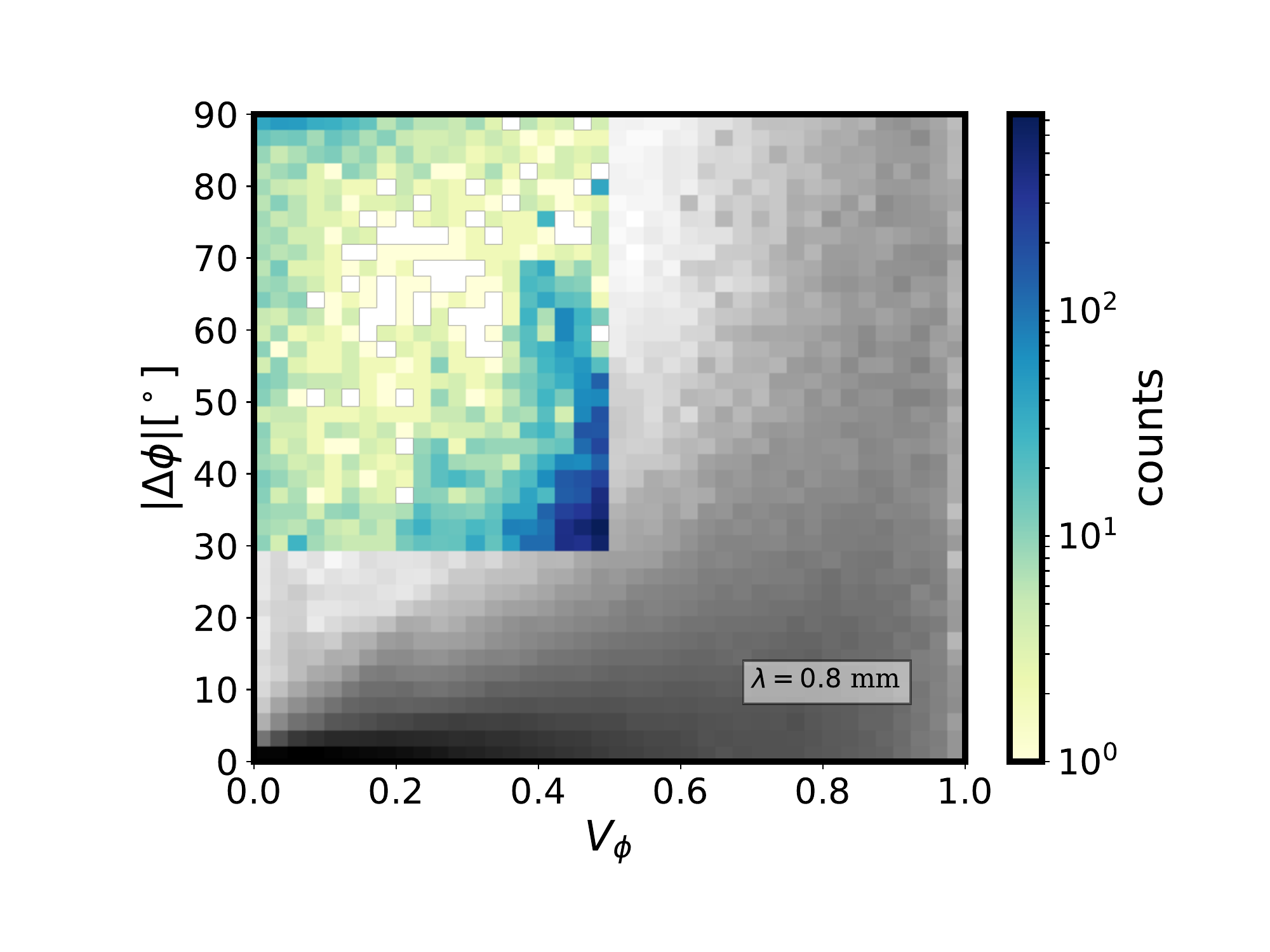}&
  \includegraphics[width=0.25\textwidth, trim={1.9cm 1.4cm 1.7cm 1.5cm},clip]{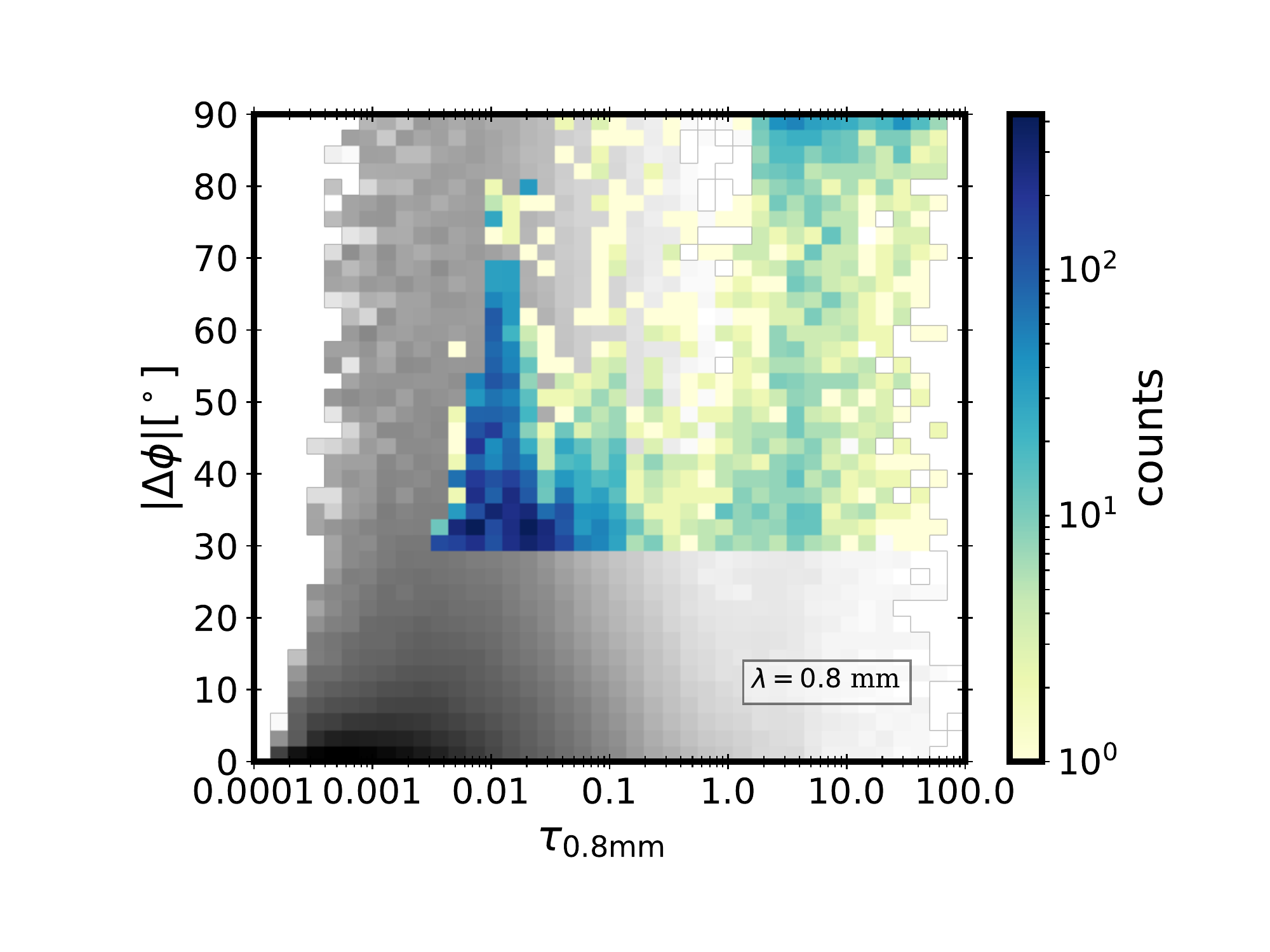}&
  \includegraphics[width=0.25\textwidth, trim={1.9cm 1.4cm 1.7cm 1.5cm},clip]{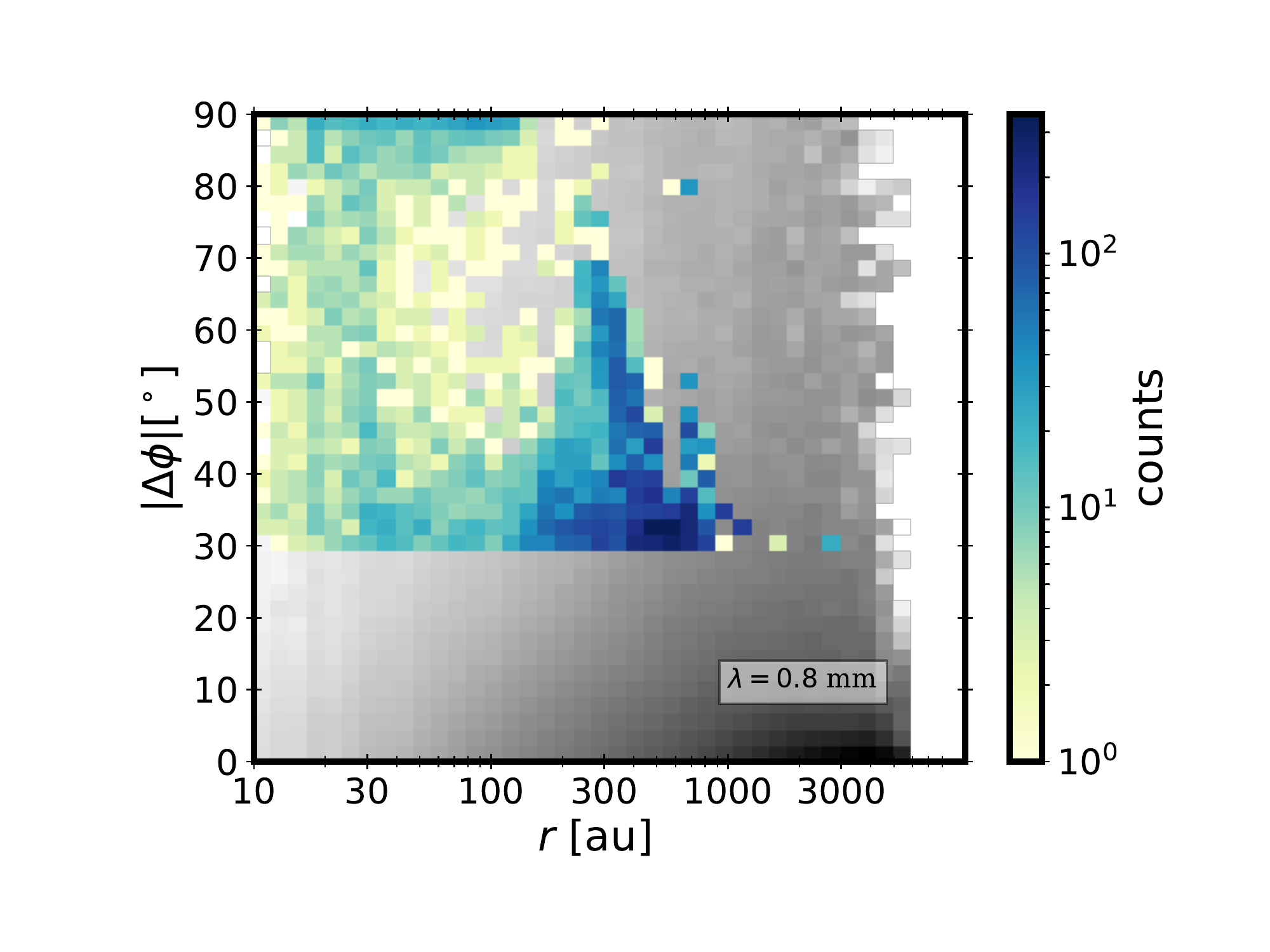}\\
  \includegraphics[width=0.25\textwidth, trim={1.9cm 1.4cm 1.7cm 1.5cm},clip]{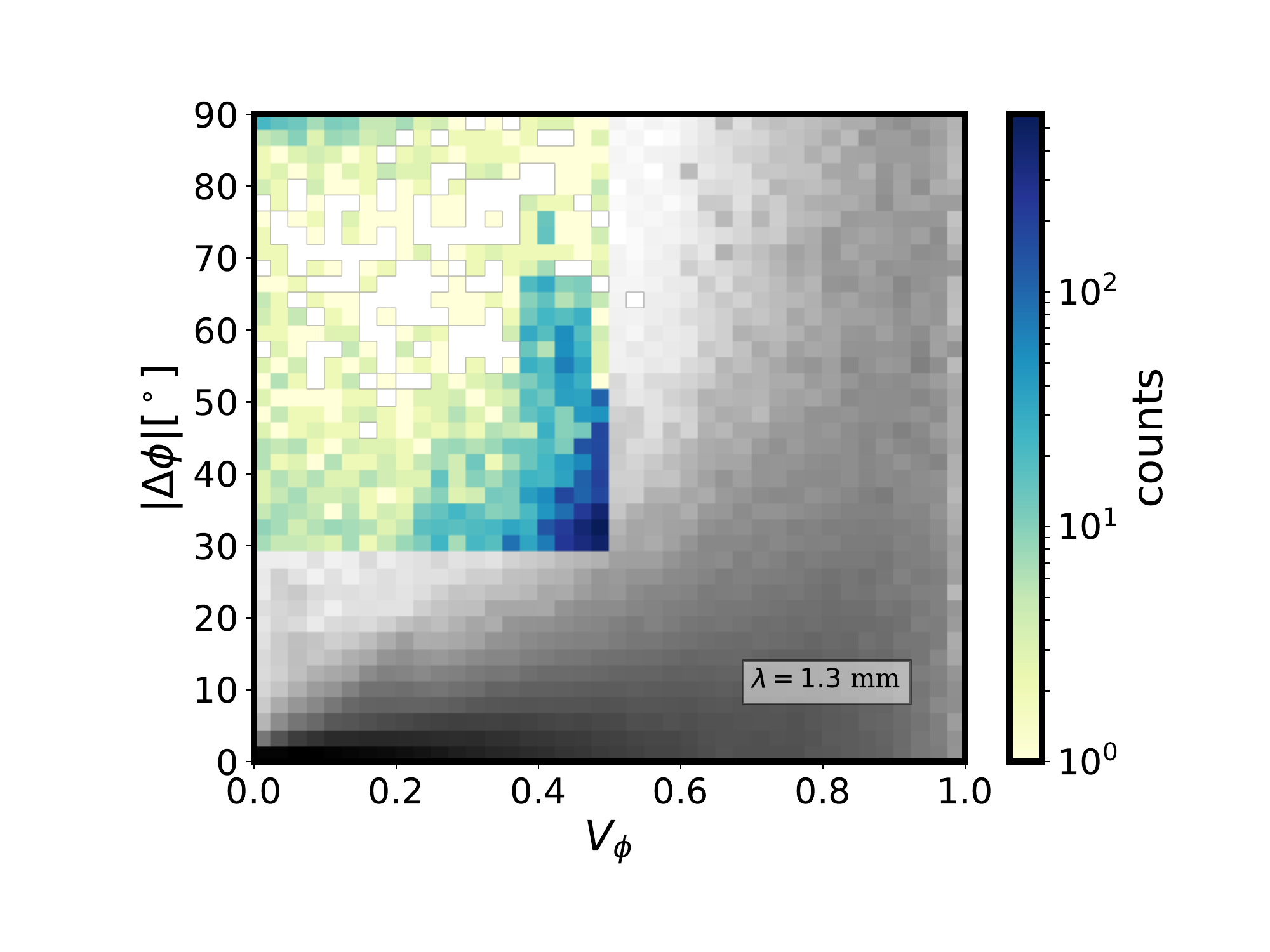}&
  \includegraphics[width=0.25\textwidth, trim={1.9cm 1.4cm 1.7cm 1.5cm},clip]{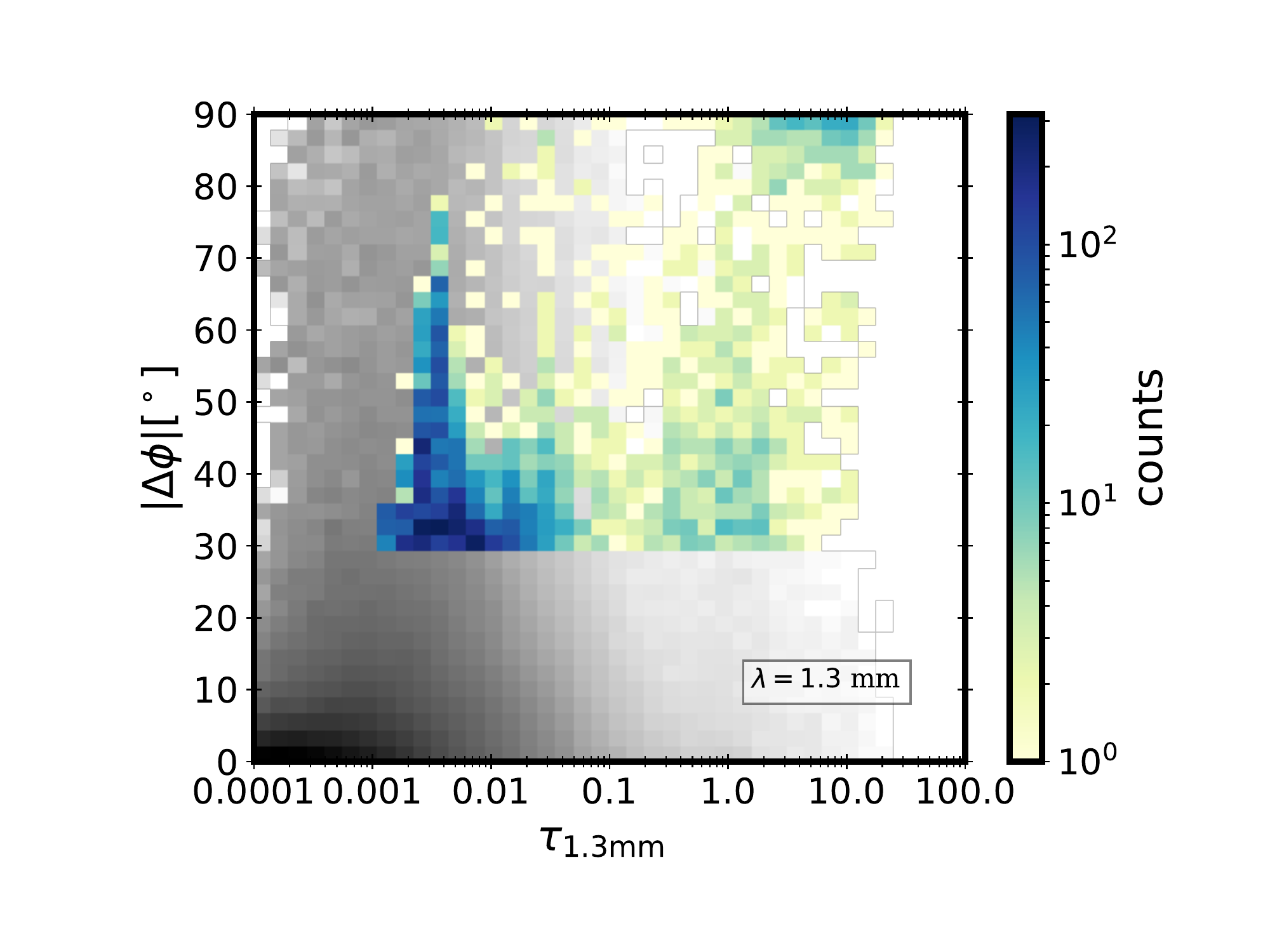}&
  \includegraphics[width=0.25\textwidth, trim={1.9cm 1.4cm 1.7cm 1.5cm},clip]{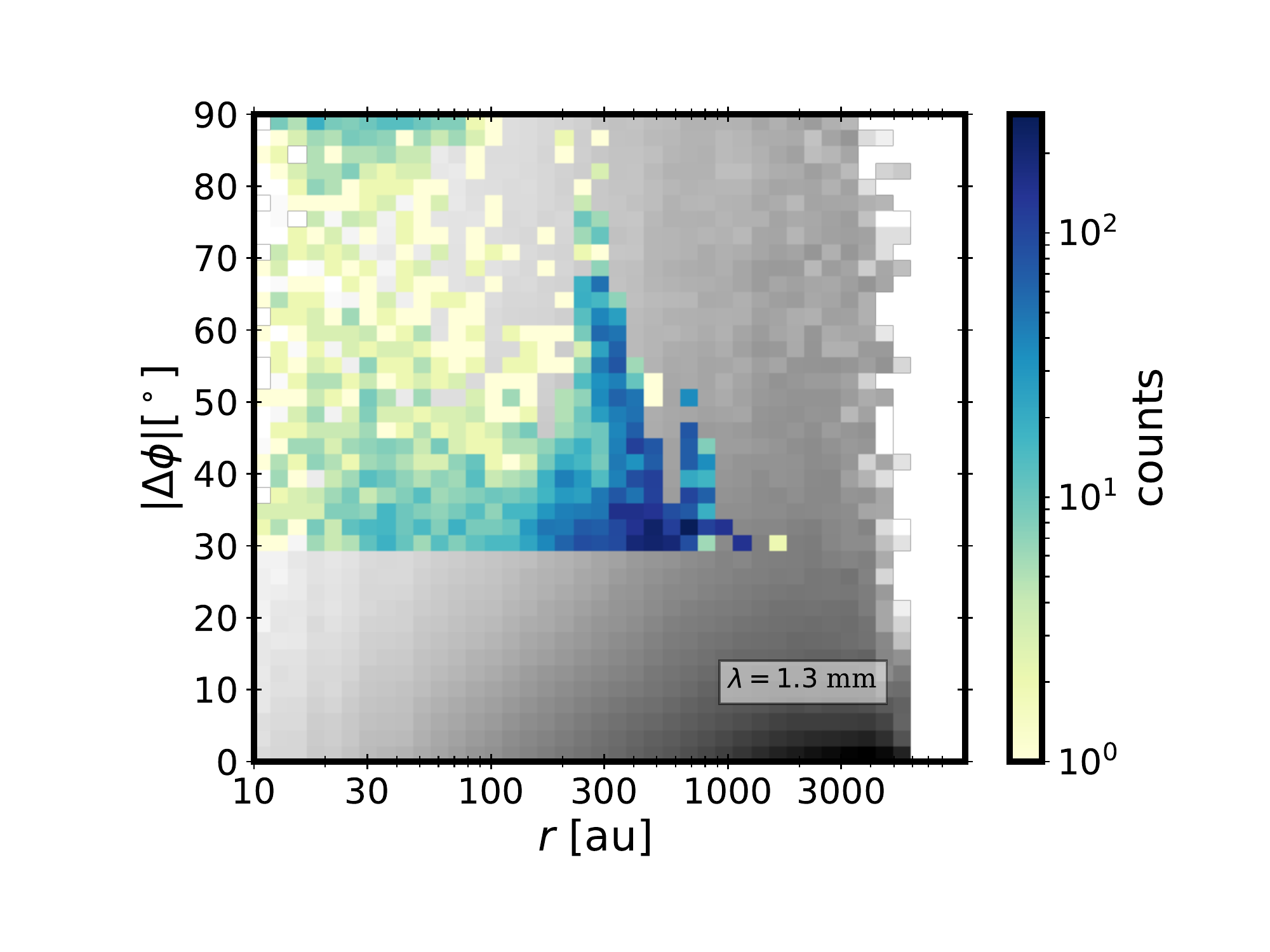}\\
  \includegraphics[width=0.25\textwidth, trim={1.9cm 1.4cm 1.7cm 1.5cm},clip]{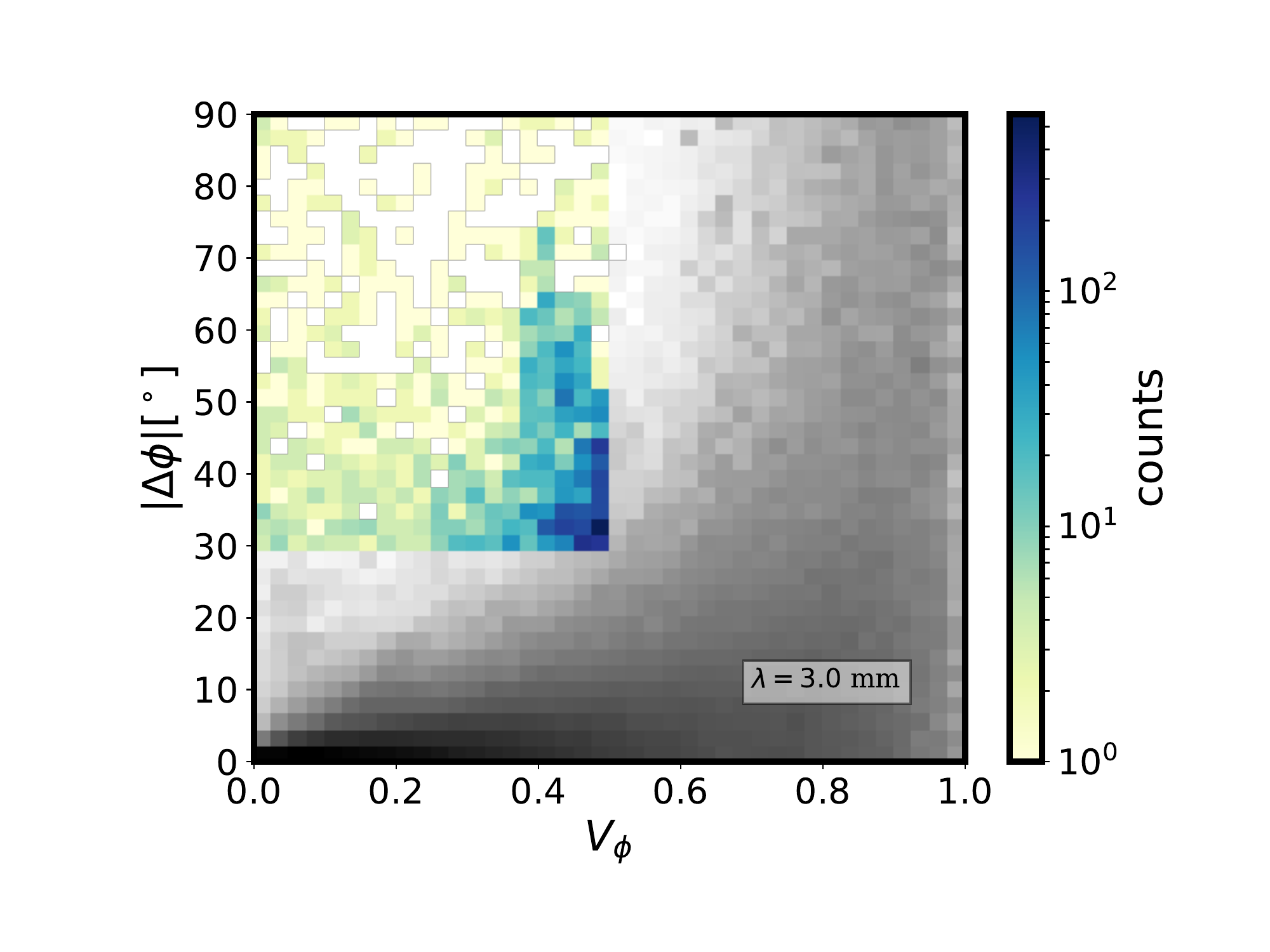}&
  \includegraphics[width=0.25\textwidth, trim={1.9cm 1.4cm 1.7cm 1.5cm},clip]{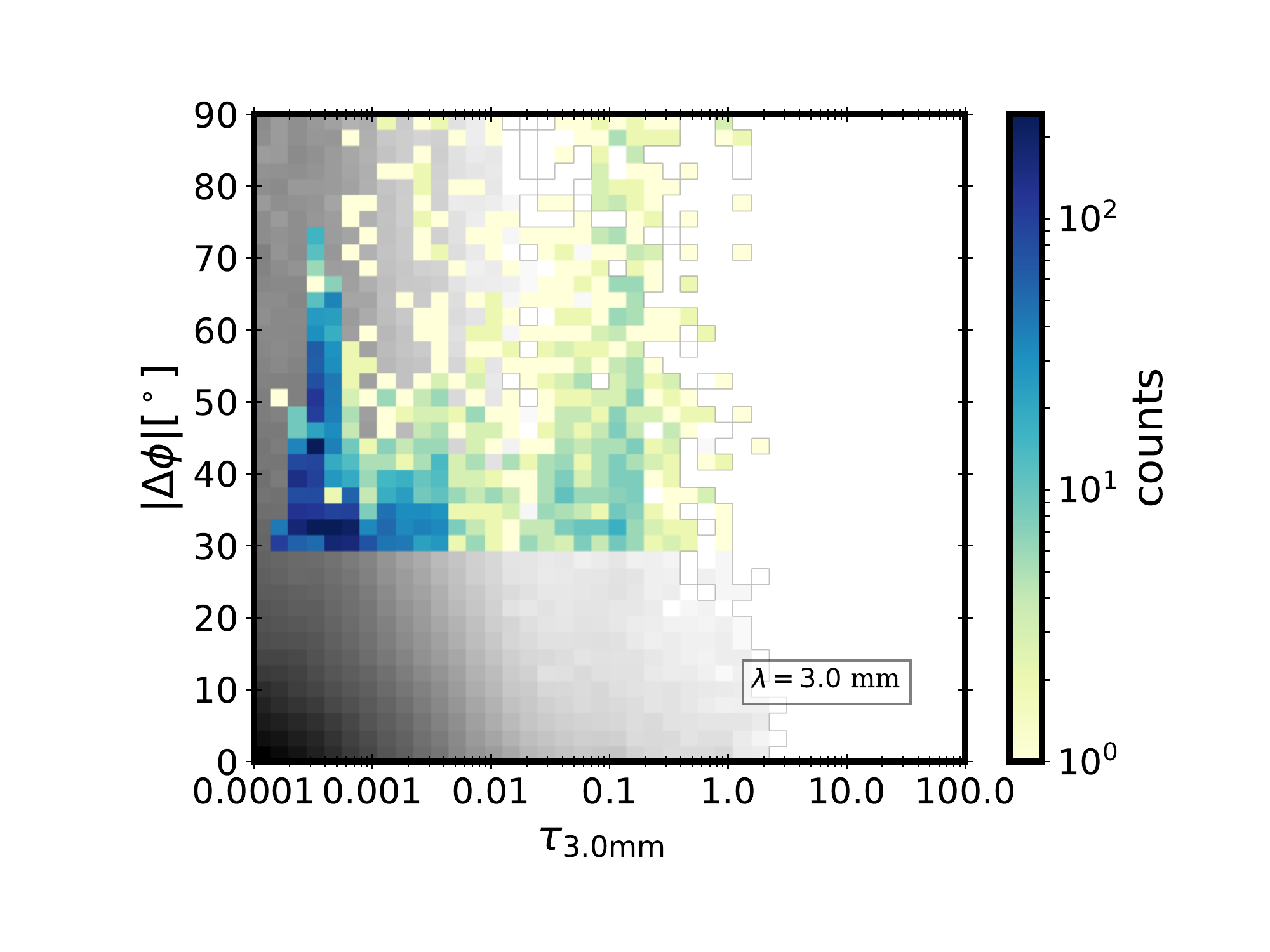}&
  \includegraphics[width=0.25\textwidth, trim={1.9cm 1.4cm 1.7cm 1.5cm},clip]{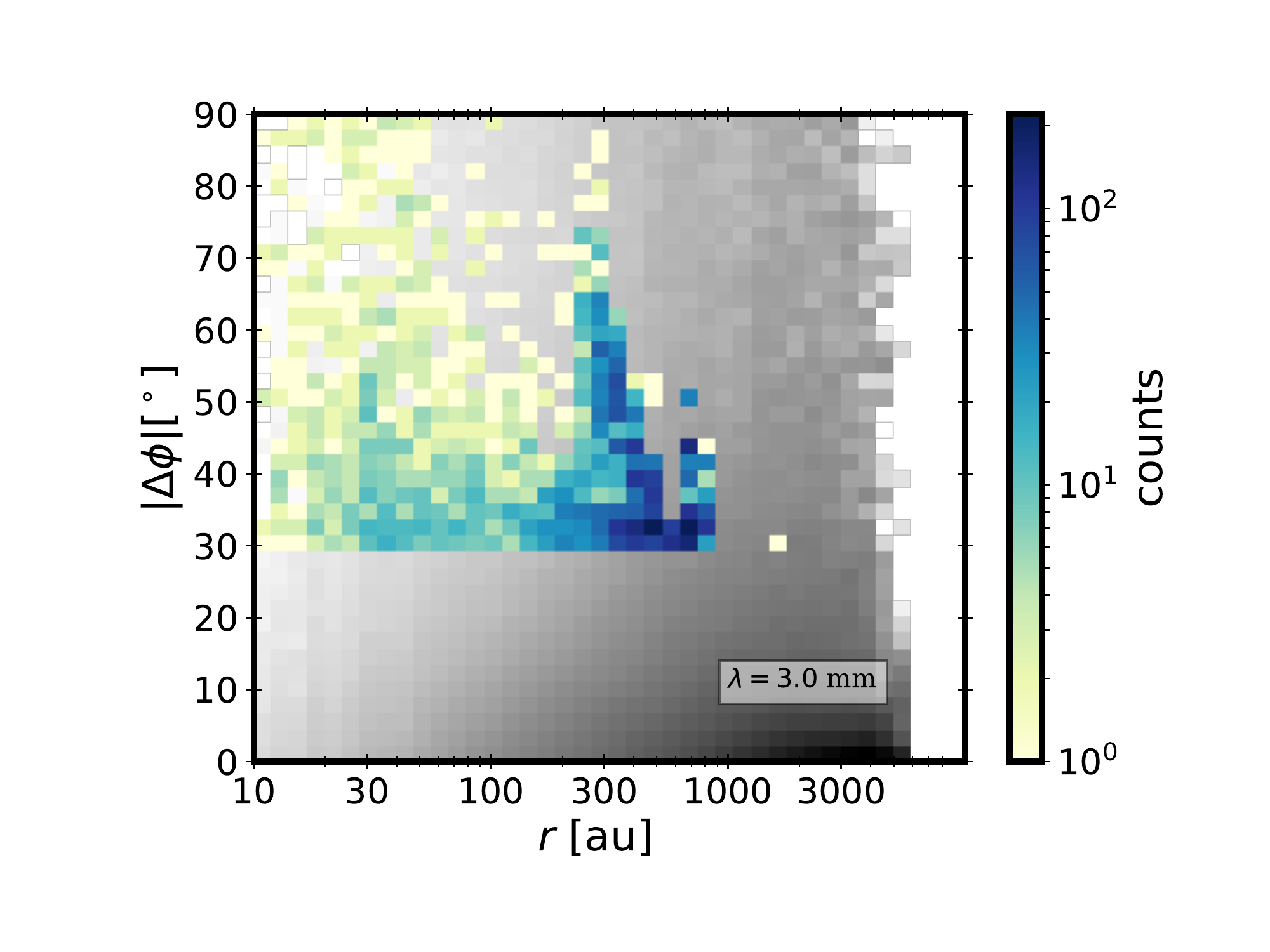}
\end{tabular}  
\caption{Two-dimensional histograms for the RAT case of the stacked maps (all simulations, time-steps, and projections) showing the $\Delta\phi$ as a function of $V_\phi$, vs. optical depth and vs. the radial distance $r$ (from left to right), at three wavelengths $\lambda=0.8, 1.3,  3.0~\mathrm{mm}$ (top to bottom).}
\label{dphi_weird_RAT}
\end{figure*}
\begin{figure*}
\centering
\begin{tabular}{l l l }          
 \noalign{\smallskip}
  \includegraphics[width=0.25\textwidth, trim={1.9cm 1.4cm 1.7cm 1.5cm},clip]{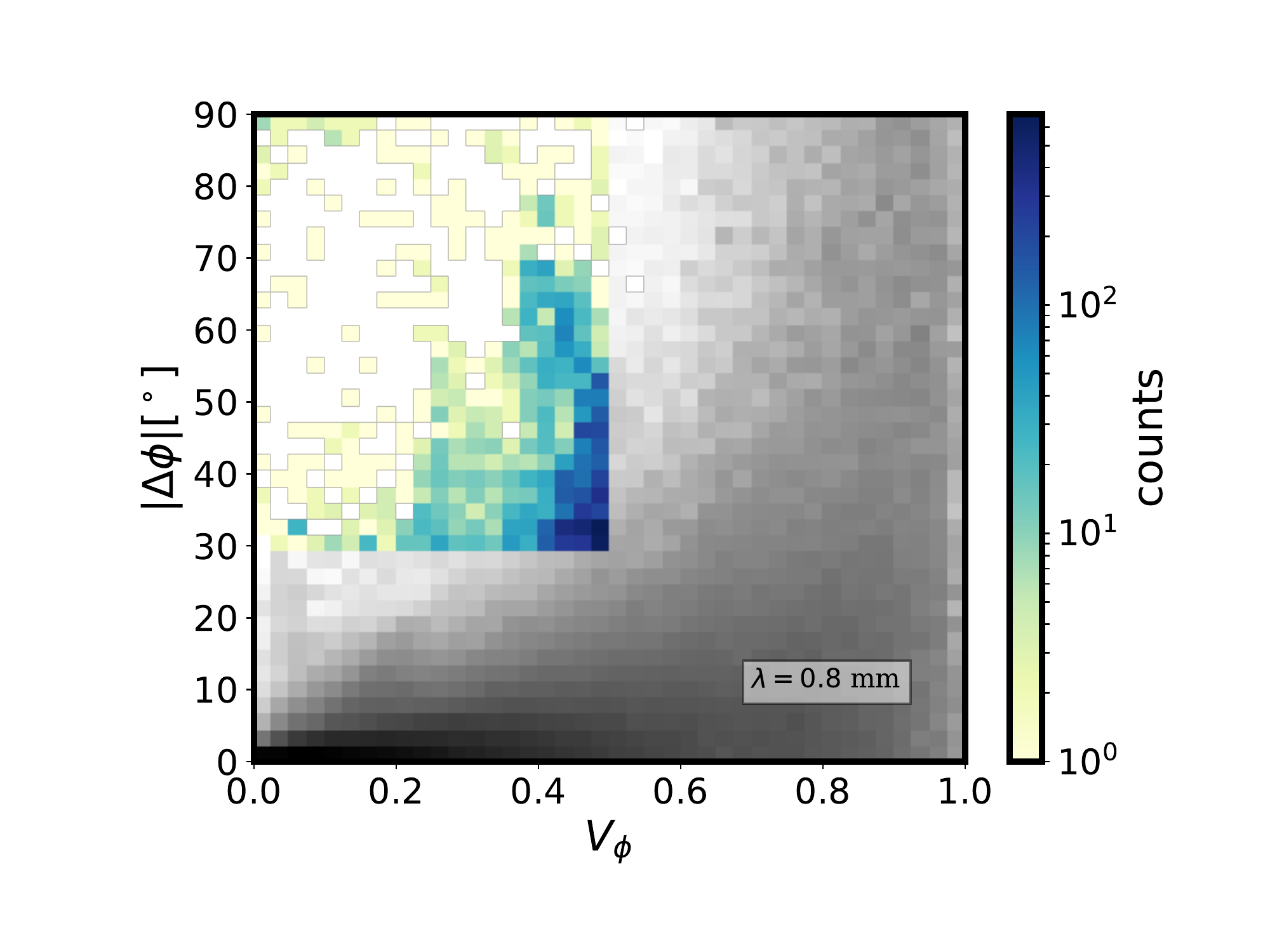}&
  \includegraphics[width=0.25\textwidth, trim={1.9cm 1.4cm 1.7cm 1.5cm},clip]{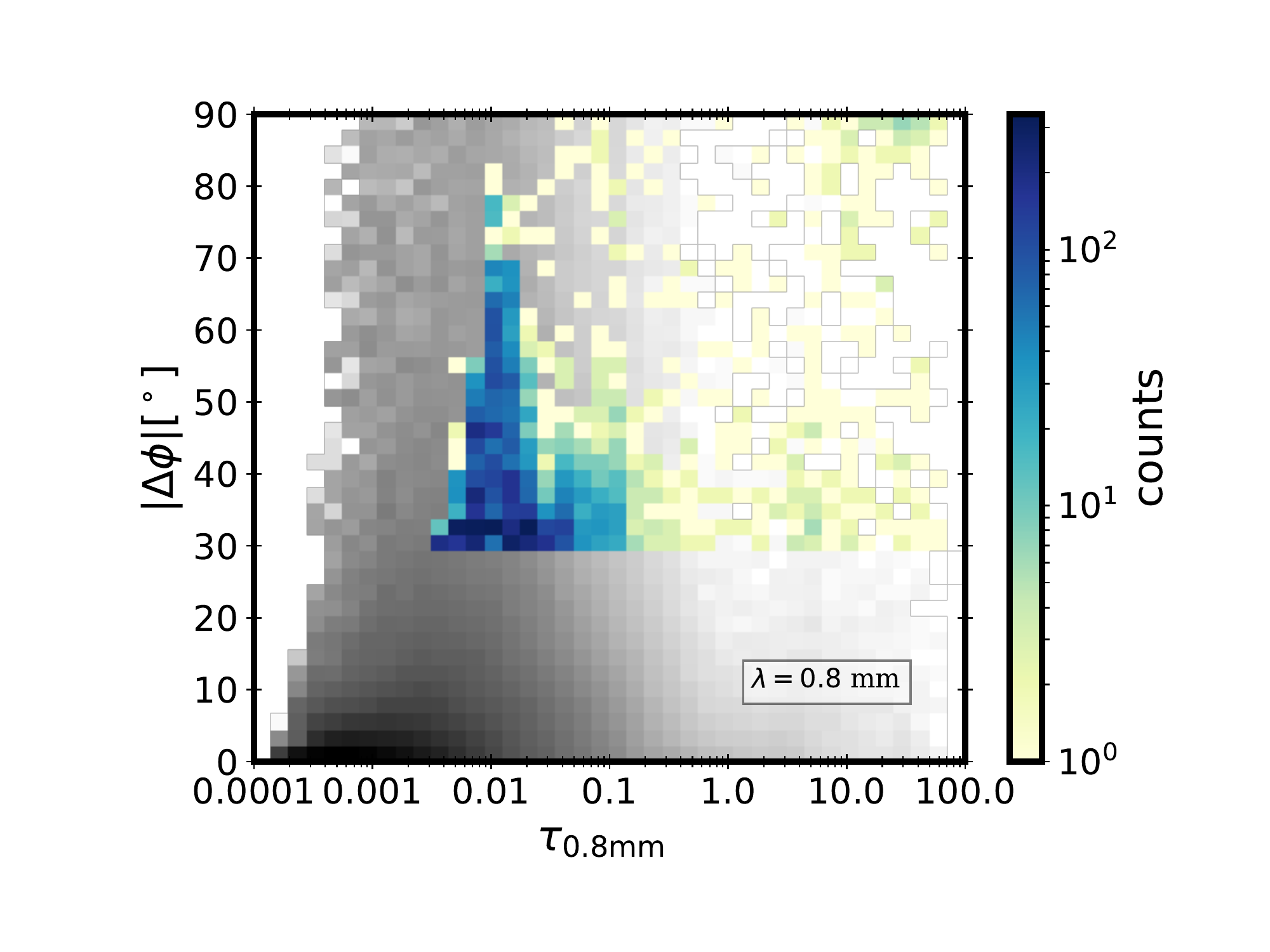}&
  \includegraphics[width=0.25\textwidth, trim={1.9cm 1.4cm 1.7cm 1.5cm},clip]{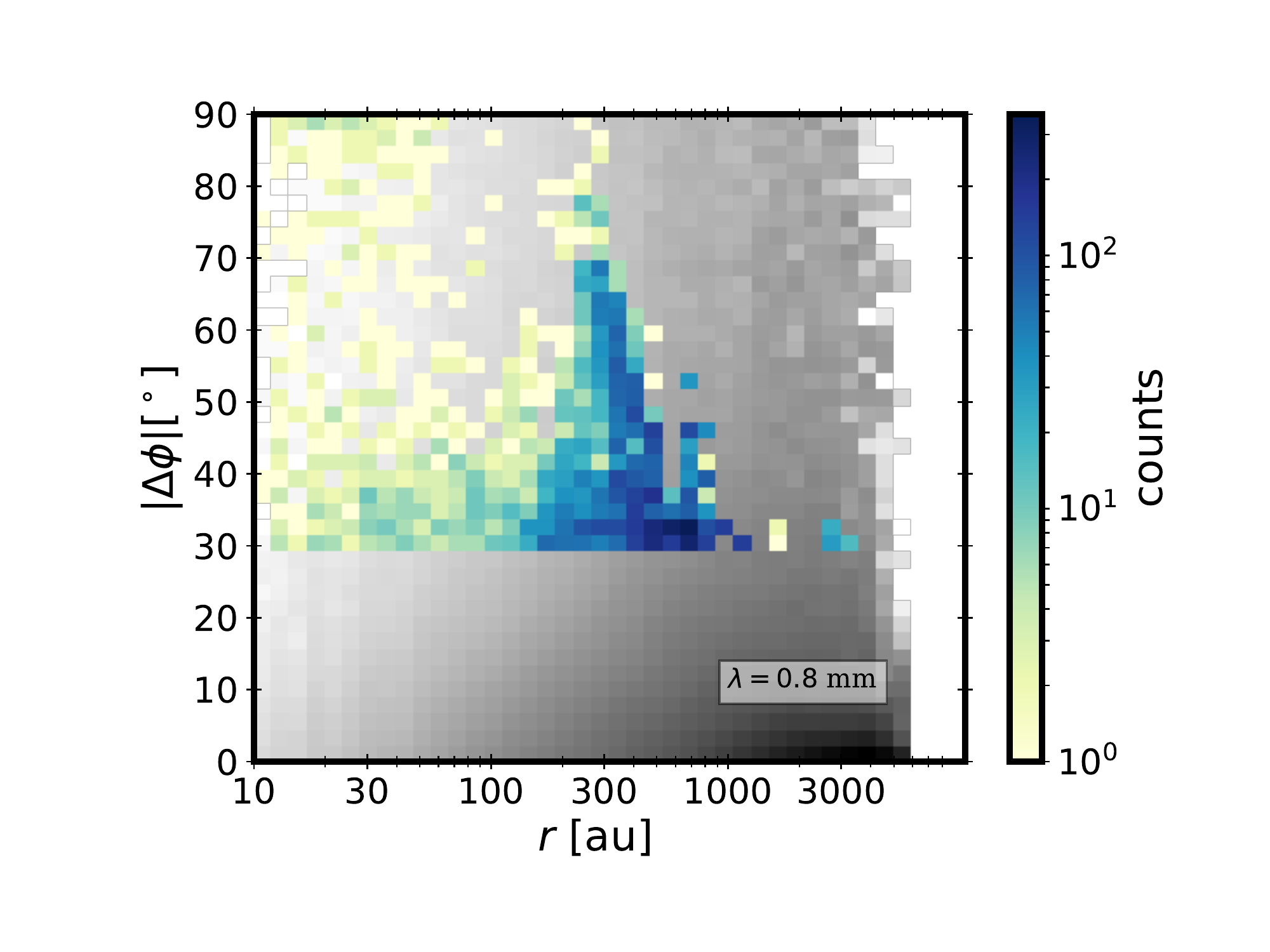}\\
  \includegraphics[width=0.25\textwidth, trim={1.9cm 1.4cm 1.7cm 1.5cm},clip]{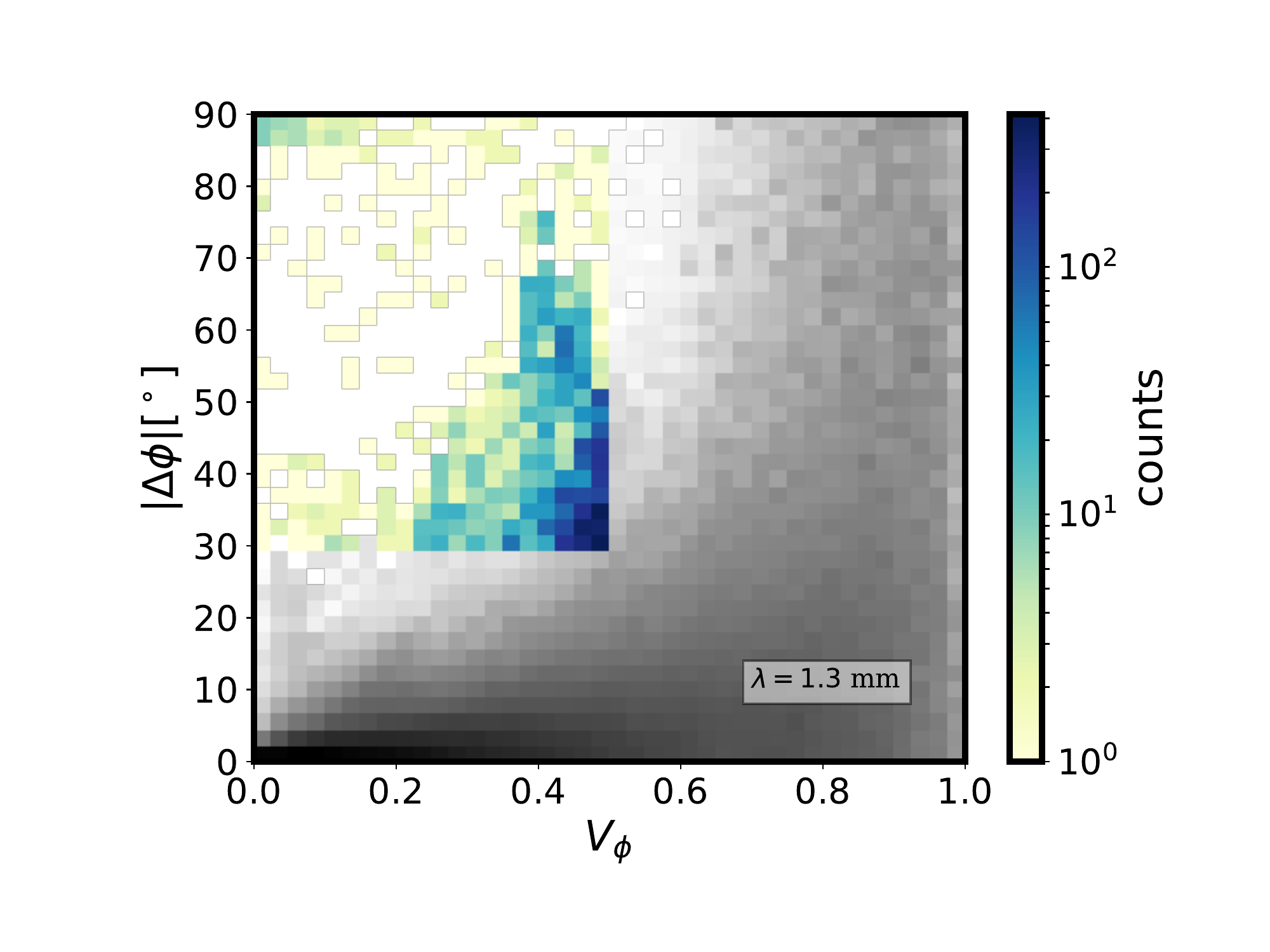}&
  \includegraphics[width=0.25\textwidth, trim={1.9cm 1.4cm 1.7cm 1.5cm},clip]{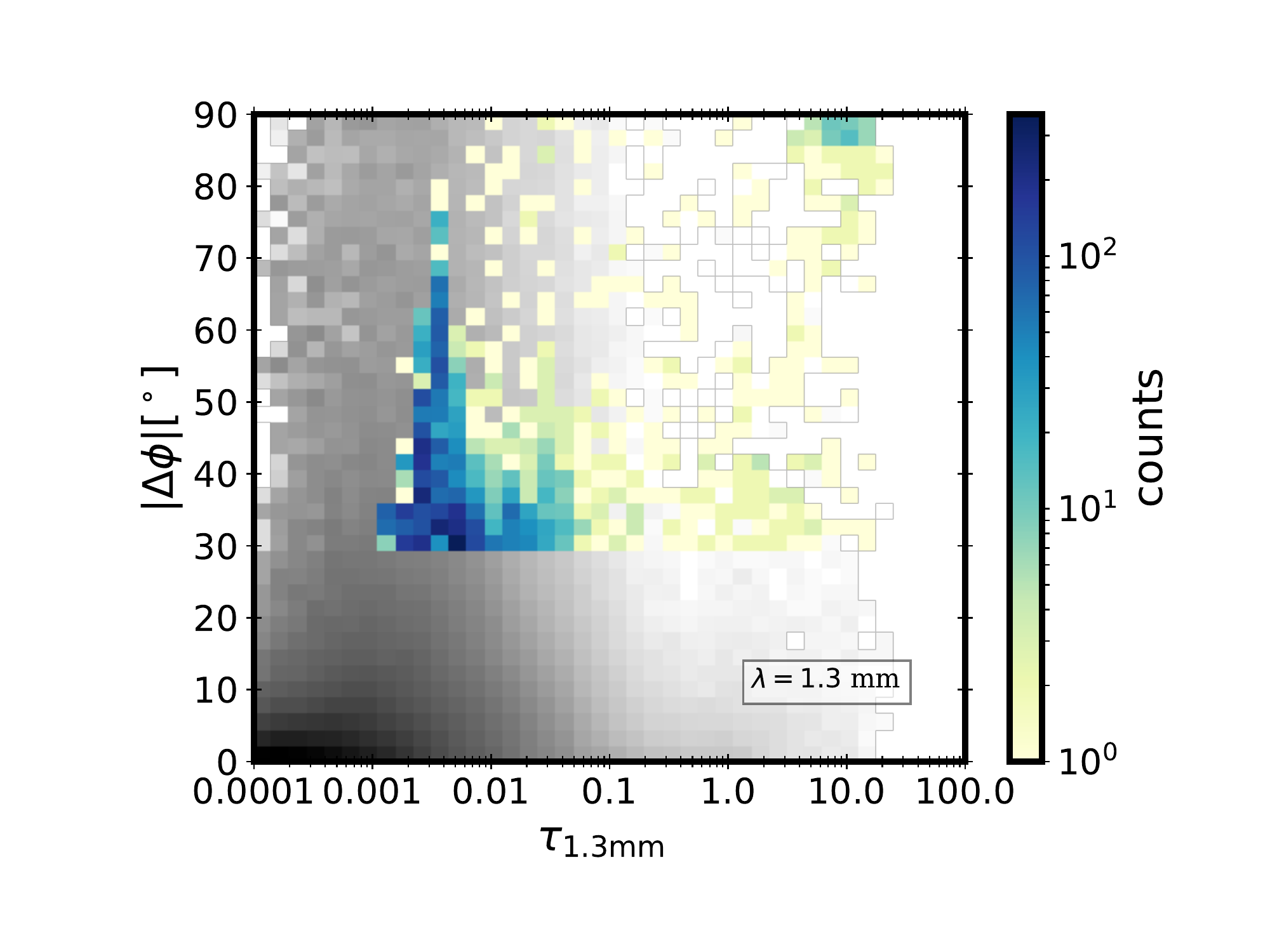}&
  \includegraphics[width=0.25\textwidth, trim={1.9cm 1.4cm 1.7cm 1.5cm},clip]{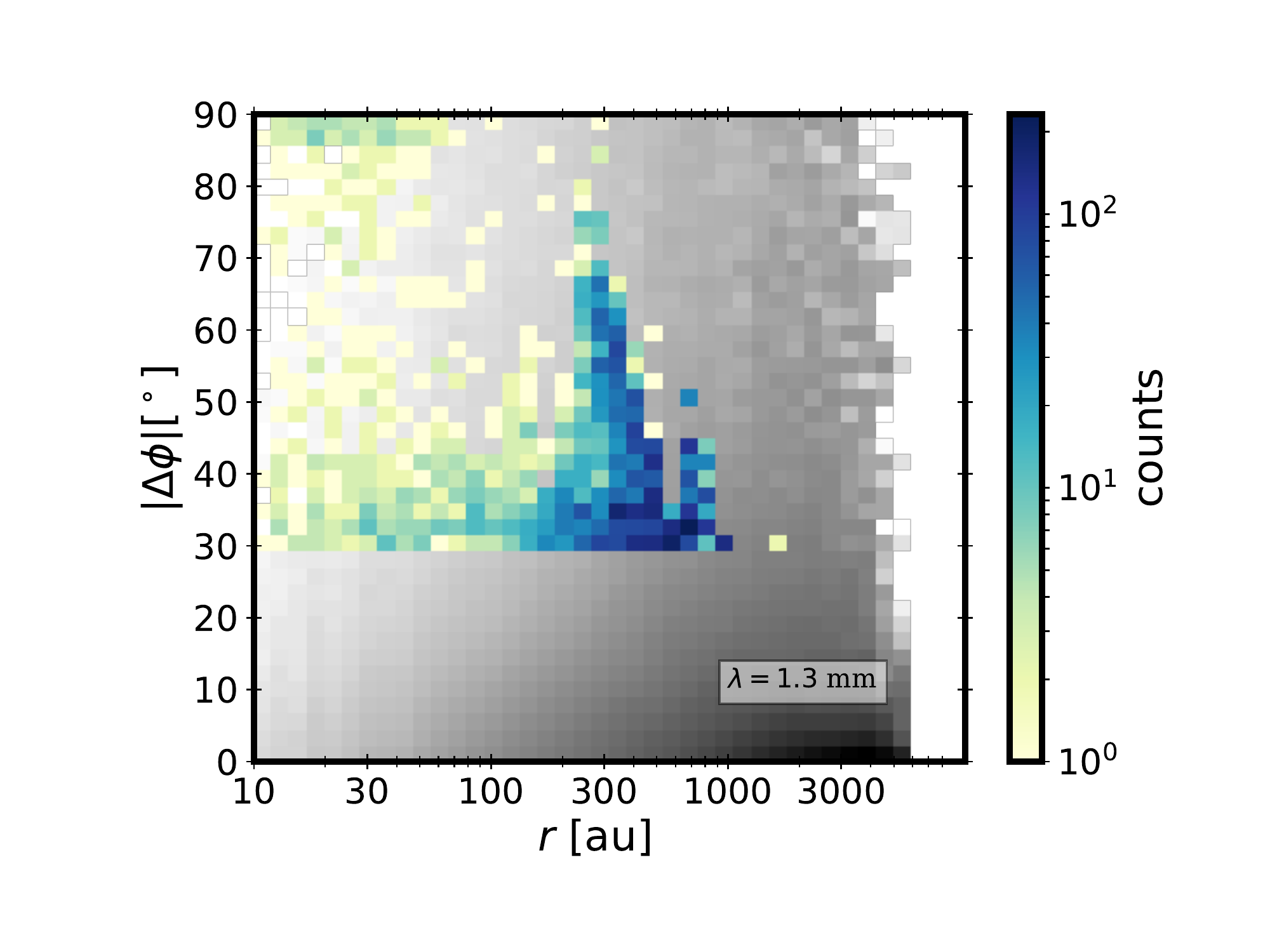}\\
  \includegraphics[width=0.25\textwidth, trim={1.9cm 1.4cm 1.7cm 1.5cm},clip]{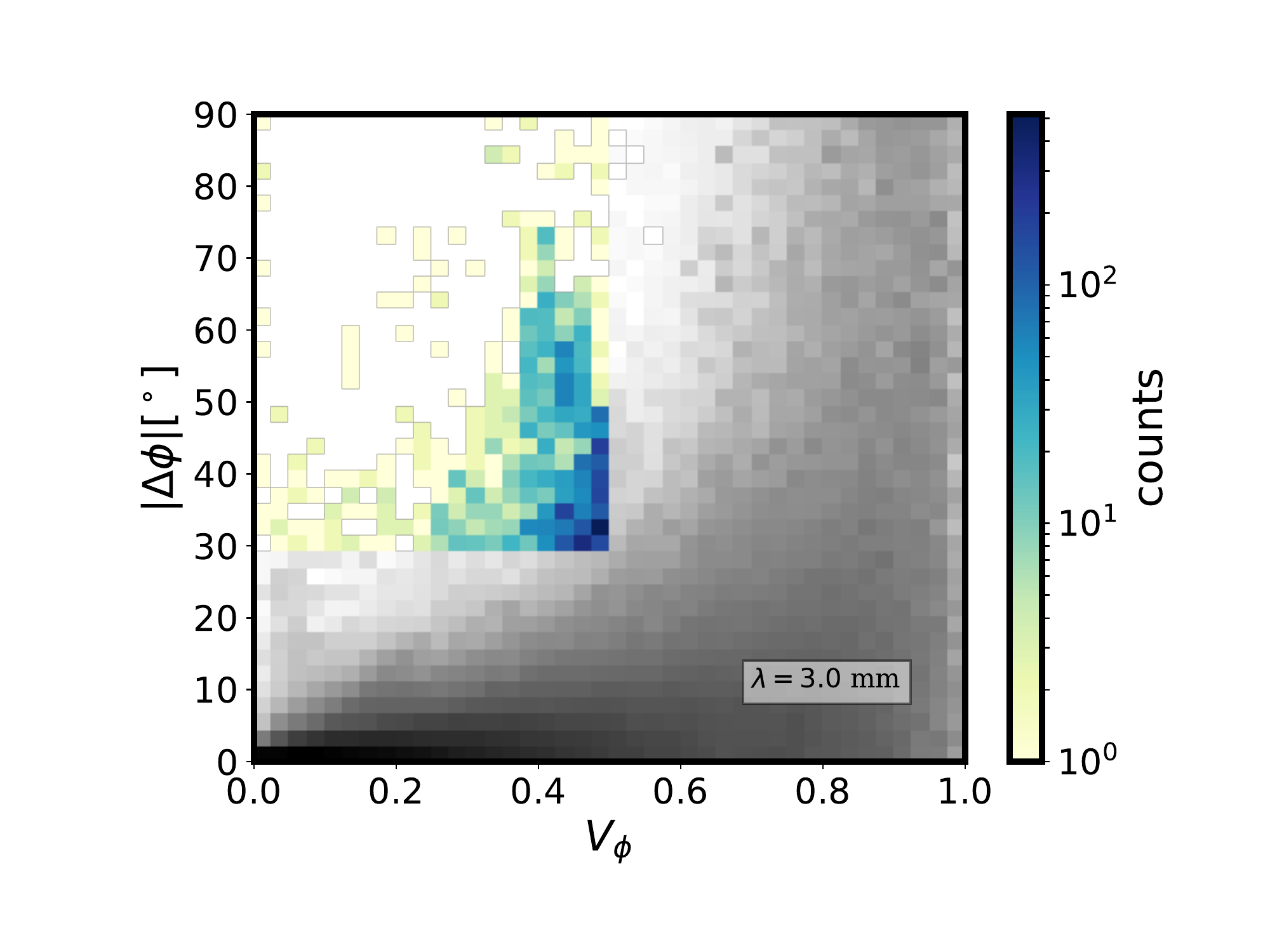}&
  \includegraphics[width=0.25\textwidth, trim={1.9cm 1.4cm 1.7cm 1.5cm},clip]{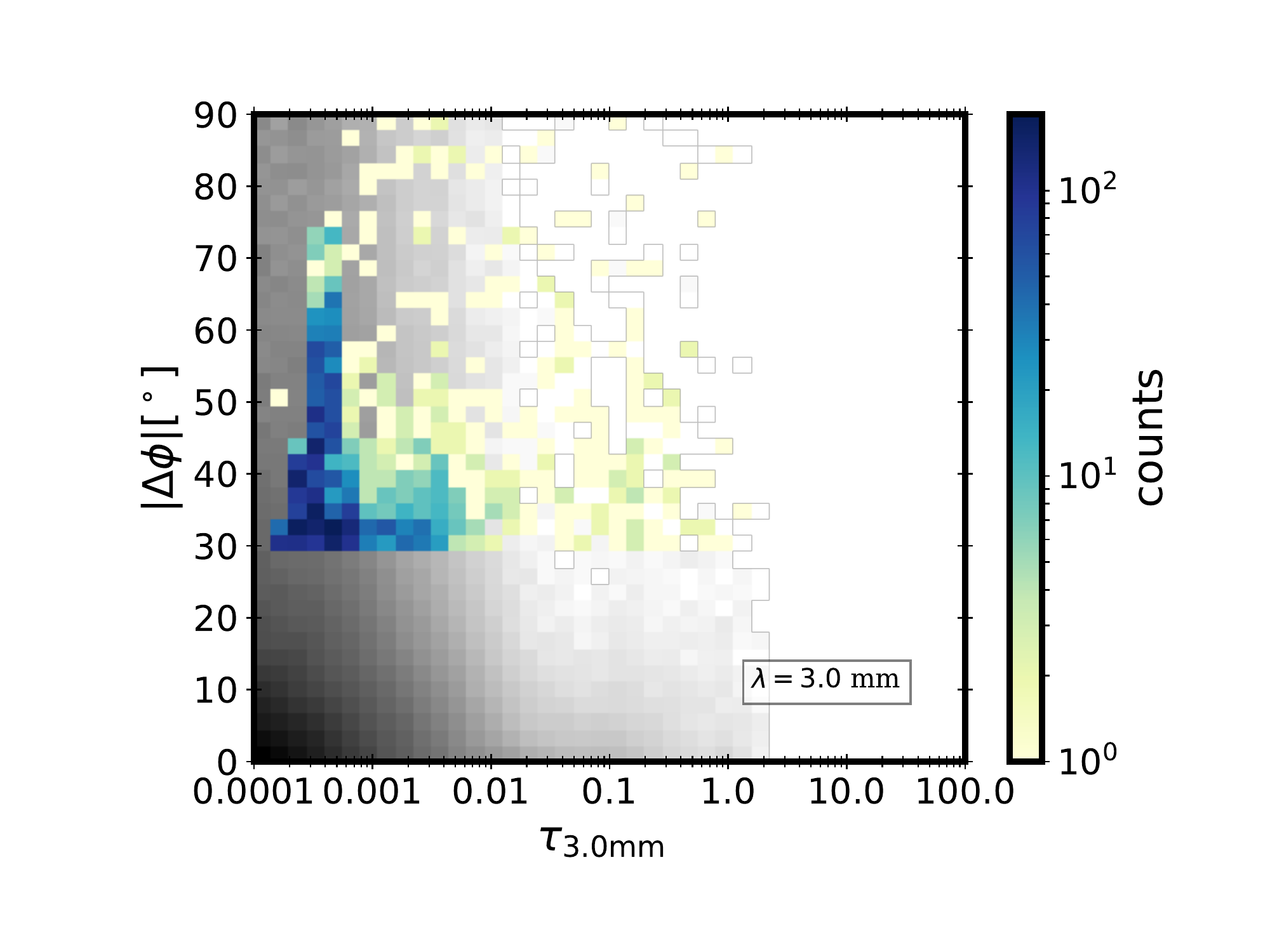}&
  \includegraphics[width=0.25\textwidth, trim={1.9cm 1.4cm 1.7cm 1.5cm},clip]{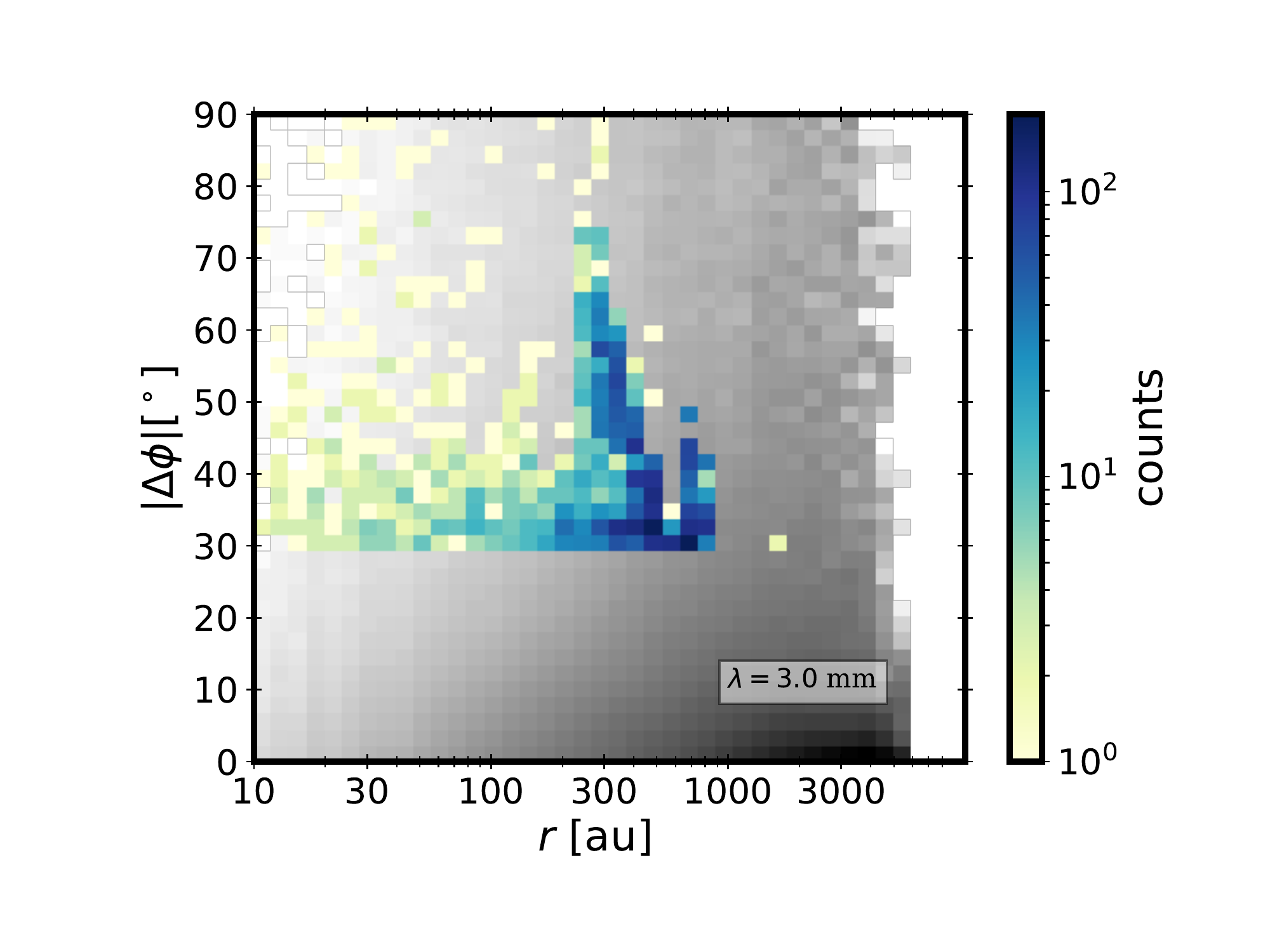}\\
\end{tabular}  
\caption{Same as Fig.~\ref{dphi_weird_RAT} but for the case where dust grains are perfectly aligned. }
\label{dphi_weird_PA}
\end{figure*}

Even though the great majority of discrepancies between the polarization vectors and the mean orientation of the magnetic field can be understood as a consequence of the intrinsic organization level, some points of high discrepancy ($|\Delta\phi|>30^\circ$) and low disorganization levels ($V_\phi<0.5$) remain unexplained. To better understand this issue 
we explore the effect of the opacity, which can cause a signal to be dominated by the foreground external layers, and thus not tracing the full line of sight, and the effect of the alignment efficiency, which if lowered can hinder the ability of polarization to properly trace the magnetic field.\\
The opacity effect can be studied by analyzing the distribution of $|\Delta\phi|$ as a function of the opacity and the projected radius at the available wavelengths. 
In Fig.~\ref{dphi_weird_RAT} we present the stacked (for all simulations, projections and time-steps) distribution of the discrepancy $|\Delta\phi |$ as a function of the intrinsic level of disorganization of the field lines ($V_\phi$), the opacity ($\tau$) and as a function of the radial distance ($r$). The colored region corresponds to the subset of points where $|\Delta \phi| > 30^\circ$ and $V_\phi \leq 0.5$, while the gray background corresponds to all the points (given for reference). This figure shows that the bulk of the points at $|\Delta\phi | \sim 30-40^\circ $ is still consistent with the intrinsic disorganization level ($\sigma_\phi \sim 30^\circ$ when $V_\phi\sim 0.5$), and that they are likely produced in regions disturbed by the outflows at a distance of $200-1000~\mathrm{au}$. Lines of sight with extremely high discrepancy levels $|\Delta\phi |>70^\circ$ are associated to regions located within a radius of roughly $100~\mathrm{au}$ with high opacities of about $2-100$. This might indicate that in these regions the signal recovered in the synthetic observations is just partial. This is coherent with the decrease of the number of problematic lines of sight for longer wavelengths, that have a lower optical depth. \\ 
To test the effect of the alignment efficiency we performed an extra set of identical synthetic observations, except for the alignment properties. The \textsc{Polaris} code allows us to simulate the polarized dust emission arising from perfectly aligned dust grains. When this hypothesis is used, all the dust grains susceptible to be aligned will contribute to the polarized dust emission, and the resulting polarized intensity and polarization fraction will correspond to the upper limits of these quantities. Figure~\ref{dphi_weird_PA} displays the same plots as  Fig.\ref{dphi_weird_RAT}, but for the perfectly aligned dust grains hypothesis.
From this figure we find that the number of lines of sight in the region $|\Delta\phi|>30^\circ$ and $V_\phi< 0.5$ is reduced by $25$, $22$, and $25\%$ at $\lambda=0.8, 1.3,$ and $3.0~\mathrm{mm}$, respectively, when using the perfect alignment hypothesis, indicating that a lower alignment efficiency is also contributing to a worse description of the mean magnetic field orientation, especially in the inner $200~\mathrm{au}$. Additionally, these results  highlight the role of the opacity effects, since even though dust grains are perfectly aligned, there are remaining patches of problematic lines of sight. This is likely due to the fact that the emission is dominated by the most external layers in the foreground, which is consistent with the fact that the number of conflicting lines of sight decreases with the observed wavelength (or equivalently with the opacity). 

\FloatBarrier
\section{Relation between the circular variance and the angular dispersion}\label{sec:sigmaV}

\begin{figure}
\centering
  \includegraphics[width=0.48\textwidth, trim={0.9cm 0.1cm 1.9cm 1.1cm},clip]{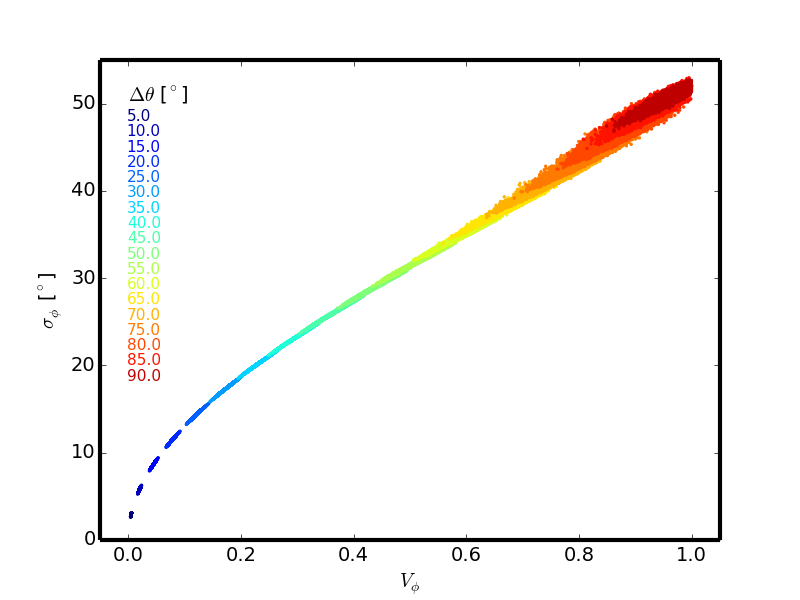}
\caption{Relation between the circular variance $V_\phi$ and the angular dispersion.
}
\label{Vphi2sigma}
\end{figure}

To understand how the circular variance ($V_\phi$), defined in Sect.~\ref{bidirstat}, and the angular dispersion ($\sigma_\phi$) are related we simulated several ensembles of random angles and computed the resulting  $V_\phi$ and  $\sigma_\phi$. 
To avoid any bias related to the position angle, we set $180$ regularly spaced central angles ($\phi_c$) from $-90^\circ$ to $89^\circ$. Secondly, 
around each central angle we defined $18$ opening angles $\Delta\theta$, varying  from $5^\circ$ to $90^\circ$ and we randomly generated an ensemble of $N=500$ uniformly distributed angles in the interval $[\phi_c - \Delta\theta, \phi_c + \Delta\theta)$. 
\\
As the mean angle does not represent the true orientation when dealing with bidirectional data, we define the mean orientation $\langle \phi \rangle$ as in Eq~\ref{Eq_mean_phiB}. The dispersion is thus defined as follows:\\ 
\raggedbottom
\begin{equation}
\sigma_\phi = \sqrt{\frac{1}{N} \sum_{i=1}^{N} {(\phi_i - \langle \phi \rangle)^2}},
\end{equation}  
\noindent where we use the smallest difference with respect to the mean mean orientation ($|(\phi_i - \langle \phi \rangle)|\leq 90^\circ$).
\\
We computed  $V_\phi$ and  $\sigma_\phi$ for each ensemble, and repeated the procedure $200$ times for each central angle and opening angle (totaling $648000$ ensembles). We present the results for the relation between the circular variance $V_\phi$ and the angular dispersion in Fig.~\ref{Vphi2sigma}. Each point of the figure shows the circular variance as a function of the angular dispersion of each individual simulated ensemble of $500$ data points. In this figure we color coded the opening angle  $\Delta\theta$ of each ensemble. We note that the angular dispersion is necessarily smaller than  $\Delta\theta$, which only represents the maximum allowed variation with respect to the central angle $\phi_c$. Figure~\ref{Vphi2sigma} shows that $\sigma_\phi$ increases monotonically with $V_\phi$ (and with  $\Delta\theta$), but its maximum value is limited to less than $55^\circ$.


\end{document}